%% file: lawther_ngc5548_BLR_modeling.tex
\documentclass[fleqn,usenatbib,useAMS]{mnras}

\usepackage{epic,eepic}
\usepackage{graphicx}
\usepackage{color}
\usepackage{epstopdf}
\usepackage{epsfig}
\usepackage{wasysym}

\usepackage{amsmath}

\usepackage[english]{babel}
\usepackage[latin1, applemac]{inputenc}
\usepackage{deluxetable}

\newcommand{\ncol}{\ensuremath{N_\mathrm{col}}}
\newcommand{\nh}{\ensuremath{n_\mathrm{H}}}
\newcommand{\phih}{\ensuremath{\Phi_\mathrm{H}}}
\newcommand{\logncol}{\ensuremath{\log[N_\mathrm{col}\,\mathrm{/cm}^{-2}]}}
\newcommand{\lognh}{\ensuremath{\log[n_\mathrm{H}\,\mathrm{/cm}^{-3}]}}
\newcommand{\logphih}{\ensuremath{\log[\Phi_\mathrm{H}\,\mathrm{/cm}^{-2}\mathrm{s}^{-1}]}}
\newcommand{\lline}{\ensuremath{L_{\mathrm{line}}}}
\newcommand{\lya}{\ensuremath{\mathrm{Ly}\alpha}}
\newcommand{\civ}{\ensuremath{\mathrm{C}\textsc{iv}}}
\newcommand{\halpha}{\ensuremath{\mathrm{H}\alpha}}
\newcommand{\hbeta}{\ensuremath{\mathrm{H}\beta}}
\newcommand{\heuv}{\ensuremath{\mathrm{He}\,\textsc{ii}\,1640\mathrm{\AA}}}
\newcommand{\heop}{\ensuremath{\mathrm{He}\,\textsc{ii}\,4686\mathrm{\AA}}}
\newcommand{\mgii}{\ensuremath{\mathrm{Mg}\,\textsc{ii}}}
\newcommand{\rin}{\ensuremath{r_{\mathrm{in}}}}
\newcommand{\rout}{\ensuremath{r_{\mathrm{out}}}}
\newcommand{\cloudy}{\textsc{Cloudy}}

\title{Quantifying the Diffuse Continuum Contribution of BLR Clouds to AGN Continuum Inter-band Delays}

\author[Lawther, D., Goad, M.R., Korista, K.T., Ulrich, O., Vestergaard, M.]{Lawther, D.,$^1$ Goad, M.R.,$^2$ Korista, K.T.,$^3$ Ulrich, O.,$^{3}$ Vestergaard, M.$^{1,4}$\\
	$^1$ Dark Cosmology Centre, Niels Bohr Institute, University of Copenhagen.\\
	$^2$ Department of Physics and Astronomy, Leicester Institute of Space and Earth Observation, University of Leicester, UK, LE1 7RH.\\
    $^3$ Western Michigan University, Department of Physics, 1120 Everett Tower, Kalamazoo, MI 49008-5252, USA.\\
    $^4$ Steward Observatory, University of Arizona, 933 N. Cherry Avenue, 85721 Tucson, AZ, USA}

\begin{document}
	
\maketitle
%

\begin{abstract}
Disk reverberation mapping of a handful of nearby AGN suggest 
accretion disk sizes which are a factor few too large for their luminosities, apparently at odds with the standard model.
Here, we investigate the likely contribution to the measured delay signature of diffuse continuum emission arising from broad line region gas. We start by constructing spherically symmetric pressure-law BLR models (i.e., $P(r)\propto r^{-s}$) that approximately reproduce the observed emission line fluxes of the strong UV--optical emission-lines in the best-studied source, NGC~5548. We then determine the contribution of the diffuse continuum to the measured continuum flux and inter-band delays, accounting for the observed variability behaviour of the ionizing nuclear continuum. Those pressure-law models that approximately reproduce the observed emission-line luminosities unavoidably produce substantial diffuse continuum emission. This causes a significant contamination of the disk reverberation signature (i.e., wavelength-dependent continuum delays). Qualitatively, the diffuse continuum delay signatures produced by our models resemble that observed for NGC~5548, including the deviation of the lag spectrum above that of a simple power-law in wavelength, short-ward of the Balmer and Paschen jumps. Furthermore, for reasonable estimates of the BLR covering fraction, the delay induced by diffuse continuum emission causes elevated inter-band delays over the entire UV--optical regime; for these pressure-law models, there are no `disk-dominated' wavelength intervals. Thus, the diffuse continuum contribution must be taken into account in order to correctly infer AGN accretion disk sizes based on inter-band continuum delays.

\end{abstract}

\begin{keywords}
galaxies: active -- methods: numerical -- galaxies : Seyfert
\end{keywords}

\input{sections/sec_introduction}

\input{sections/sec_theory}

\input{sections/sec_method}

\input{sections/sec_results}

\input{sections/sec_ccf}

\input{sections/sec_discussion}

\section{Conclusion}

Our conclusions are as follows.

\emph{(1)} For conditions thought relevant to the BLR of Seyfert 1 galaxies, simple radial pressure-law models with spherical geometry can broadly match the observed emission-line luminosities and variability behaviour of the majority of the strongest broad UV and optical emission lines. These models span 1--140 light-days in BLR radius, and require large (30--60\%) covering fractions.


\emph{(2)} Pressure-law models able to reproduce (or exceed) the observed emission-line luminosities also produce a significant amount of diffuse continuum emission over the entire UV--optical wavelength regime. Thus, the behavior of the diffuse continuum must be taken into account when interpreting observed continuum inter-band delays in terms of accretion disk sizes. In our models, the diffuse continuum is produced over a broad range in radii, and shows a strong wavelength dependence. The bulk of the diffuse continuum originates in high density, moderately ionized  gas, i.e., at radii {\em smaller\/} than the typical formation radius of H$\beta$.

\emph{(3)} After accounting for the fractional contribution of the wavelength-dependent diffuse continuum to the total continuum emission, we find additional lags of $\sim$a few days relative to the disk reprocessing model. The wavelength dependence of the delays produced by our models qualitatively resembles the measured inter-band continuum delays in NGC~5548. In particular, our models produce excess delays above the underlying disk power-law in the vicinity of the Balmer continuum, and a sharp drop in the delay signature long-ward of the Balmer jump, in agreement with the inferred delay spectra \citep{Edelson2015,Fausnaugh2016}. While we do not attempt to match the observed inter-band delays in detail, the elevated delays due to the diffuse continuum can roughly account for much of the discrepancy between observations and the predictions of reprocessing models. Better constraints on the covering fraction for the BLR, and on its radial density distribution and geometry, would help constrain the expected diffuse continuum lag spectrum.

\paragraph*{Acknowledgments: }

We thank the anonymous referee for helpful comments and suggestions which improved the quality of this manuscript. DL and MV gratefully acknowledge support from the Independent Research Fund Denmark (DFF) via grant no. DFF 4002-00275. 

\bibliographystyle{mnras}
\bibliography{lawther_ngc5548_BLR_modeling}

\clearpage
\input{tables/line_lags_table}

\end{document}

%% file: sections/sec_introduction.tex
\section{Introduction}\label{sec:introduction}

The immense energy outputs of active galactic nuclei (AGN) are ultimately fueled by gas accretion onto supermassive black holes \citep[e.g.,][]{LyndenBell1971}, which reside at the centres of most - perhaps all - massive galaxies \citep[e.g.,][]{Kormendy2001}. Thus, supermassive black holes in the local Universe probably achieved their current masses through bouts of AGN activity. The details of these accretion episodes are not well-understood. The in-falling gas is thought to form an accretion disk that heats up and emits much of the observed continuum radiation; \citet{Shakura1973} develop the standard $\alpha$-disk model of steady-state disk accretion, originally in the context of X-ray binaries. The nuclear continuum in turn excites fast moving gas (clouds) near the central engine, which emit the observed broad emission lines (BEL); these clouds reside in the so-called broad line region (BLR). On larger scales, the nuclear continuum also excites the dusty obscuring structure \citep[e.g.,][]{Antonucci1993} to produce the infrared dust emission feature \citep[e.g.,][]{Barvainis1987}, and excites gas in the host galaxy bulge to produce the narrow emission lines, the kinematics of which are dominated by the host galaxy potential \citep[e.g.,][]{Nelson1996}.

The large distances to AGN, coupled with the small sizes of their central regions  means that the  central regions of AGN remain unresolved, even for the nearest objects, defying scrutiny using conventional techniques (e.g, direct imaging). Indirect methods must therefore be used to probe their internal structure. Reverberation mapping (hereafter, RM) \citep{Blandford1982} is one such method, and has proven a particularly powerful probe of the central regions of AGN. In its traditional form,  RM  utilizes correlations between continuum and emission-line variations to map the spatial distribution and (with high quality velocity-resolved data) the kinematics of the broad emission-line gas \citep[e.g.,][]{Bentz2010,Denney2010,Skielboe2015}. This has revealed a compact, yet spatially extended BLR, sitting deep within the gravitational potential of the central supermassive black hole.  RM studies of multiple emission lines, spanning a broad range in ionization state in a handful of individual sources, suggest that the BLR is radially stratified, with strong gradients in gas density and/or ionization. Importantly, the gas appears to be largely virialized \citep{PW1999}, a property that when coupled to measured BLR sizes \citep{PET2004} has been usefully exploited to measure the mass of the central supermassive black hole in $\approx 60$, mostly nearby, AGN \citep{Bentz2015}. On-going multi-object spectroscopic surveys \citep[e.g.,][]{Shen2015b,King2015} should vastly increase the number of reverberation mapped AGN, albeit at reduced fidelity, and importantly, extend the number of objects for which black hole mass estimates are available to higher luminosity and redshift.

\subsection{Disk reverberation mapping}\label{sec:intro_diskreverb}
More recently, correlated multi-wavelength inter-band continuum variations have been used to test the standard model for accretion \citep[][see also \S \ref{sec:intro_5548}]{Collier2001,Collier2001b,Sergeev2005,Cackett2007,Troyer2016,Edelson2017}, with some surprising results. With a few assumptions (a standard $\alpha$-disk irradiated from above by a compact variable X-ray emitting source), the wavelength-dependent variations reveal the disk radial temperature profile $T(R)$ and, if the black hole mass is known, the mass accretion rate through the disk.  Formally, in the standard model the inter-band continuum delays $\tau(\lambda)$ increase with increasing wavelength $\lambda$ according to

\begin{equation}\label{eq:delayeqtn}
\tau(\lambda) \propto (M \dot{M})^{1/3}\lambda^{4/3}
\end{equation}

\noindent \citep[e.g.,][]{Fausnaugh2016}, where $M$ is the black hole mass, and $\dot{M}$ the mass accretion rate through the disk.
However, results from the most intensive multi-wavelength monitoring campaigns (most notably NGC~5548, see \S \ref{sec:intro_5548}) reveal continuum inter-band delays which are too large (by a factor of a few) for an $\alpha$-disk emitting at the observed luminosity \citep{Edelson2015,Fausnaugh2016,Edelson2017,Starkey2017,Cackett2017,Fausnaugh2018}. That is, the standard model for accretion developed for accreting binaries may be incorrect when applied to AGN. Similarly, quasar microlensing studies also point to larger than expected disk sizes \citep{Poindexter2008,Morgan2010,Mosquera2013}. 

So, are AGN accretion disks non-standard? In order to address this question it is crucial that all known variable contributions to the measured continuum bands are properly accounted for. Since the nuclear regions are unresolved, there will be additional contributions to the measured disk continuum emission, some of which may vary. Known contaminants include: the non-variable contributions to the observed continuum emission from stars in the host galaxy, and the variable broad emission lines. The latter are particularly problematic for broad band photometric data. Their contributions to the observed delays may be estimated given an estimate of the BEL delays, together with the relative flux contribution to the filter bandpass, as determined from single epoch spectra \citep[e.g.,][]{Fausnaugh2016}. However, a factor often neglected from these studies is the possibly substantial contribution to the continuum bands by variable continuum emission arising from the same gas responsible for emitting the broad emission-lines, the so-called diffuse continuum component (hereafter, DC)\footnote{The narrow line region will also emit a diffuse continuum component, though the size  scales for the NLR would imply that their contribution, as for the narrow emission-lines, is non-variable on the timescales relevant to RM campaigns.} \citep[e.g.,][]{Korista2001}. As noted by \citet{Korista2001}, the DC component, if significant, introduces a delay signature which broadly mimics the disk continuum interband delays, with delays generally increasing with increasing wavelength. Importantly, these authors show that the DC emission introduces a delay signature over and above that arising from simple disk reprocessing, particularly in the Balmer and Paschen continua. This contaminant may be partially responsible for the larger than expected disk sizes. Thus, interpreting the measured continuum interband delays (and thus disk sizes) {\em solely\/} in terms of a simple disk reprocessing scenario is formally incorrect.

\subsection{AGN STORM Monitoring of NGC~5548: a Test-bed for AGN Accretion Disk Physics}\label{sec:intro_5548}

NGC 5548 is a low-redshift ($z=0.01718$) Seyfert 1.5 galaxy that has been a target of multiple RM campaigns, including the 13-year \emph{AGN Watch} campaign \citep{Clavel1991,Krolik1991,Korista1995,Peterson2002}. In 2014, the \emph{AGN Space Telescope and Optical Reverberation Mapping} (AGN STORM; Peterson, PI) campaign observed NGC 5548 using the \emph{Hubble} Space Telescope, obtaining 171 usable epochs with $\approx$ daily cadence \citep{DeRosa2015}. The target was observed concurrently in the UV-optical to X-ray regime using \emph{Swift} \citep{Edelson2015} and in the optical from ground-based observatories spanning a range in longitude \citep{Fausnaugh2016,Pei2017}. In terms of cadence and spectral coverage, the AGN STORM program is arguably the most intensive RM campaign to date. Key findings of the AGN STORM campaign on NGC 5548 include:

\begin{itemize}
\item The response of the BELs appear to `decouple' from the continuum variations during a substantial portion of the 2014 campaign, showing very little response to the UV continuum variations \citep{DeRosa2015,Goad2016}.
\item The lags obtained during the `coupled' portion of the campaign suggest BLR size scales substantially smaller than those expected given the global BLR radius-luminosity relation \citep{Pei2017}, or indeed, the single-object radius-luminosity relation for NGC 5548 \citep{Eser2015}.
\item High-cadence \emph{Swift} UVOT monitoring reveals inter-band continuum delays \citep{Edelson2015,Fausnaugh2016}. These delays generally increase with increasing wavelength and
are broadly consistent with a $\tau(\lambda)\propto\lambda^{4/3}$ dependency (Equation \ref{eq:delayeqtn}). However, they are larger (by a factor of a few) than those predicted given the estimated $M$ and $\dot{M}$ for NGC~5548 \citep{Fausnaugh2016,Starkey2017}. Thus, if the measured delays are solely due to disk reprocessing, the disk in NGC~5548 is larger than predicted by the standard $\alpha$-disk model.
\item The continuum lag for the \emph{Swift} $U$ band, which samples the continuum in the vicinity of the Balmer jump, is elevated relative to the best-fit $\tau(\lambda)\propto\lambda^{4/3}$ delay model  \citep{Edelson2015,Fausnaugh2016}.
\end{itemize}


\subsection{Outline of this Work}

Here, we determine the likely contribution of the diffuse continuum emission from BLR gas to the measured continuum inter-band delays for AGN. We use the well-studied source NGC~5548 (\S \ref{sec:intro_5548}) as a test case. Of particular relevance to this work is the larger than expected UV--optical inter-band continuum delays measured during the AGN STORM 2014 campaign. These delays may be affected by DC emission, as evidenced by the elevated lag signal near the Balmer jump. Our intention is not to model the AGN STORM 2014 data in detail. Rather, by constructing a straw-man model which approximately reproduces the observed emission-line luminosities of the strongest UV and optical emission lines in NGC~5548, we aim to quantify the DC contribution arising from its BLR, both in terms of flux and variability behaviour, and thereby assess its influence on the measured inter-band continuum delays. This prepares the necessary groundwork for more detailed models of NGC~5548 and other AGN at comparable luminosities. 

While studies of inter-band continuum delays exist even for quasar-luminosity objects \citep{Mudd2017}, the highest-cadence RM data available are for lower-luminosity AGN in the local Universe. Thus, we expect our modeling results to be broadly relevant to current and future studies of disk inter-band delays. We proceed as follows: in \S \ref{sec:theory} we construct a model BLR for which the run of gas physics with radius is completely specified by a simple radial pressure-law, $P \propto r^{-s}$  \citep{Rees1989,Netzer1992,Goad1993,Kaspi1999}. We then use photoionization calculations to determine the emergent emission-line fluxes $\epsilon(r)$ as a function of radial distance $r$, integrating over the cloud distribution to determine the total emission-line luminosities (\S \ref{sec:method}). We investigate both constant ionization and constant density models, and assess their ability to match the gross properties (i.e., BEL luminosities, their ratios, and their variability timescales) of the strongest UV--optical broad emission-lines (\S \ref{sec:results_steadystate}). From these same models, we then compute the wavelength-dependent flux and delay distribution of the DC emission (\S \ref{sec:results_cont}), and its dependence on the density and ionization state of the BLR gas (\S \ref{sec:results_cont_comparison}). Finally, we perform Monte Carlo calculations, driving our model BLR with simulated continuum lightcurves, in order to estimate the measured delays of both the BEL and the diffuse continuum bands, and their dependence on the characteristics (amplitude and variability timescale) of the driving continuum lightcurve (\S \ref{sec:ccf}). We make a rough estimate of the total continuum lag spectrum (including that due to reprocessing in an $\alpha$-disk), and compare with the measured delays for NGC~5548, in \S \ref{sec:discussion}. In what follows we adopt a luminosity distance $D_{\rm L}=72.5$~Mpc to NGC~5548 \citep{Bentz2015}, assuming a $\Lambda$ CDM cosmology with $\Omega_{\Lambda}=0.7$ and $\Omega_{M} = 0.3$.

%% file: sections/sec_theory.tex
\section{Pressure-law Broad-line Region Models}\label{sec:theory}
%

We refer to models for which the pressure $P$ depends on the radial distance $r$ from the central continuum source,

\begin{equation}
P(r)\propto r^{-s} \, ,
\end{equation}

\noindent as pressure-law models. In this work we examine two limiting cases, $s=0$ and $s=2$, representing constant density and constant ionization parameter models, respectively. We here adopt a spherically symmetric BLR geometry spanning more than two decades in radial extent. This model is chosen for (i) its simplicity, and because (ii) we can compare our radial pressure-law models with the Local Optimally emitting Cloud model for this source presented by \citet{Korista2001}, which also adopts spherical symmetry. Here, we summarize the radial dependencies of various physical quantities for spherically symmetric pressure-law models; the derivations in this section follow \citet{Goad1993} and \citet{Rees1989}. We make the simplifying assumption that the cloud temperature does not vary with radius; for solar composition, photoionization equilibrium is achieved at temperatures $T\sim10^4$ K across a wide range of ionization parameter (Equation \ref{eq:ur}), and the gas temperature will therefore vary weakly with radius \citep[e.g.,][]{Netzer1990_book,Ilic2009}. 
For constant cloud temperatures, the cloud hydrogen gas density \nh\,is proportional to the pressure, P, and so 

\begin{equation}\label{eq:nh}
\nh(r)\propto r^{-s} \, .
\end{equation}

\noindent Thus, $s=0$ corresponds to a constant \nh\,throughout the BLR. The ionization parameter $U$ is defined as:

\begin{equation}
U(r)=\frac{Q_\mathrm{H}}{\nh(r)4\pi r^2c} \, ,
\end{equation}

\noindent where $Q_\mathrm{H}$ is the number of hydrogen ionizing photons emitted by the central continuum source per second. Thus, $s=2$ corresponds to a constant ionization parameter model, since:

\begin{equation}\label{eq:ur}
U(r)\propto r^{s-2} \, .
\end{equation}

\noindent The surface area per cloud, $A_c$, is proportional to $R_c^2$, where $R_c$ denotes the radius of a cloud. In general, $R_c$ depends on the pressure $P$, and is therefore constant for $s=0$. If we demand that the mass of each cloud is conserved as the clouds move radially outwards (i.e., clouds do not break up or coalesce within our region of interest), mass conservation implies that $R_c^3n_H=constant$. Thus, we obtain the relation

\begin{equation}
A_c(r) \propto R_c^2(r) \propto r^{2s/3} \, .
\end{equation}

\noindent The column density of each cloud, \ncol, depends on the gas density and cloud radius:

\begin{equation}\label{eq:ncol_r}
\ncol(r)\propto R_c\nh\propto r^{-2s/3} \, .
\end{equation}

\noindent The above relations determine the local physical conditions, as parameterized by \ncol, \nh\,and incident ionizing photon flux \phih, at any radius in a spherically symmetric pressure-law BLR model. These conditions determine the local surface emissivity $\epsilon(r)$ of a BLR cloud at radius $r$ (as determined via \cloudy\,modeling, \S \ref{sec:method_cloudy}). The total luminosity for an emission line is then found by integrating over the distribution in cloud properties such that:

\begin{equation}\label{eq:lline}
\lline=4\pi\int_{\rin}^{\rout}\epsilon(r)A_c(r)n_c(r)r^2dr \, ,
\end{equation}

\noindent where \rin\,and \rout\, are the inner and outer BLR radii respectively, $A_c$ is the surface area of a single cloud, and $n_c$ is the local number density of clouds. As dust grains strongly absorb UV photons, \rout\, is chosen to approximately coincide with the distance at which dust grains can form and survive ($\approx$~140 light days for NGC~5548). \\

\noindent Finally we determine the local cloud surface area, $\mathrm{d}(A_cn_c)$. Given our assumption that clouds do not break up or coalesce, mass conservation implies that the term $n_cvr^2$ is constant, for cloud velocity $v$. We make the additional simplifying assumption that the clouds are in virial motion ($v(r)\propto r^{-1/2}$). In that case, a cloud number density distribution $n_c(r)\propto r^{-3/2}$ fulfills the requirement of mass conservation, and the differential covering factor of the BLR obeys the relation

\begin{equation}
dC(r) \propto A_c(r)n_c(r) dr \propto r^{2s/3-3/2} dr \, .
\end{equation}

\noindent This allows us to determine a normalization $K(s,\rin,\rout)$ for the total line-emitting surface of the clouds, for a given BLR covering fraction. The luminosity integral becomes:

\begin{equation}\label{eq:lumin}
L=4\pi K\int_{\rin}^{\rout}\epsilon(r)r^{2s/3+1/2}dr \, ,
\end{equation}

\noindent while the normalization, for $\Omega=4\pi$ steradians at \rout, is given by

\begin{multline}
K=\left[\frac{2}{\sqrt{\rin}}-\frac{2}{\sqrt{\rout}}\right]^{-1}\,(s=0)\\
\mathrm{or}\,K=\frac{5}{6}\left[\rout^{5/6}-\rin^{5/6}\right]^{-1}\,(s=2)\, .
\end{multline}

%% file: sections/sec_method.tex
\section{Method: from Photoionization Grids to BLR Models}\label{sec:method}

\begin{figure}
	\centering
	\includegraphics[width=0.999\linewidth]{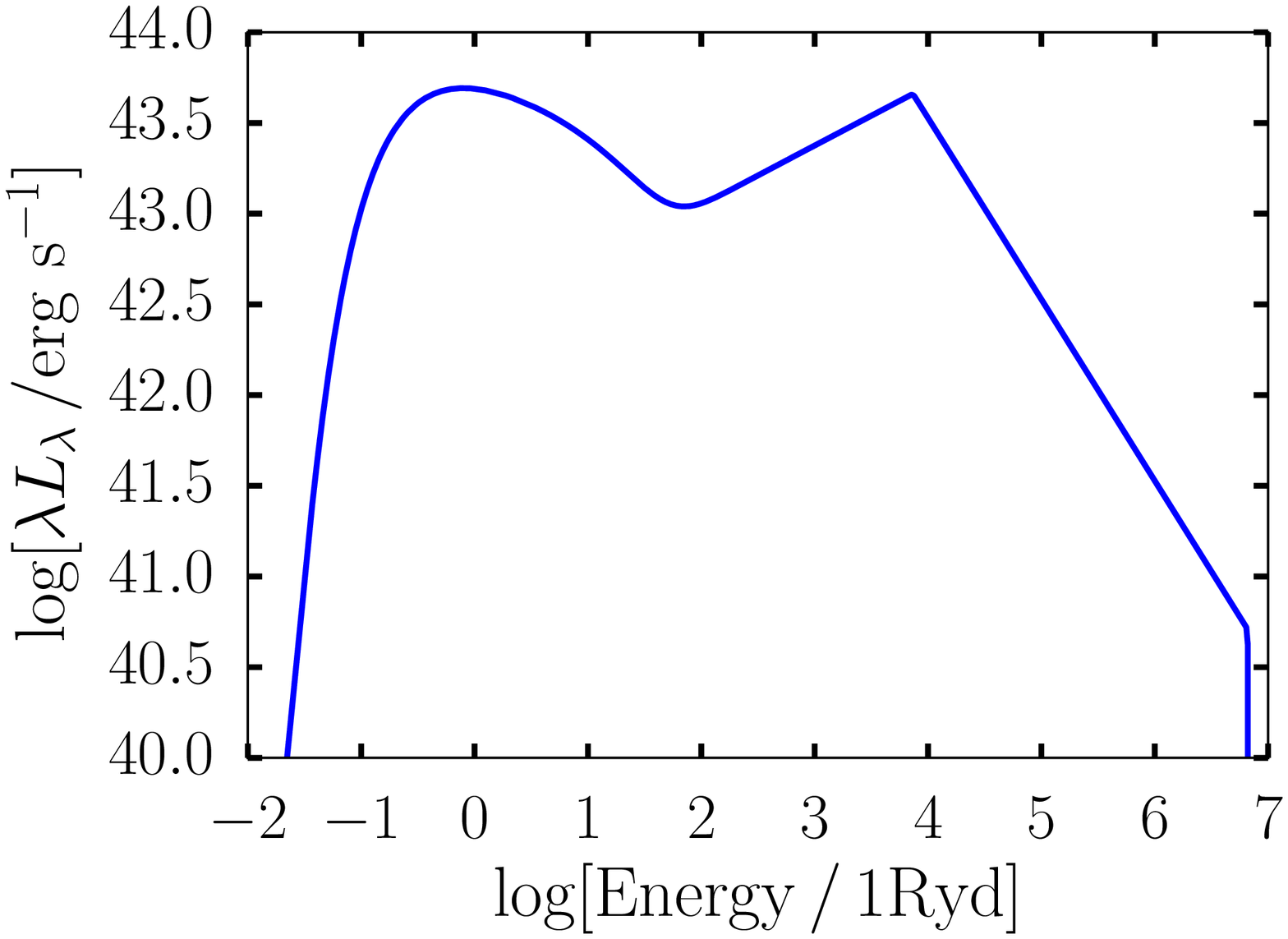}
	\caption{The NGC 5548 continuum SED presented by \citet{Mehdipour2015}, which we use to generate the \cloudy\,photoionization grids. The SED is scaled to the source-frame luminosity based on the observed mean continuum flux for the 2014 AGN STORM campaign, 
$F_{\lambda}$(1367\AA)$ = 42.64 \pm 8.6 \times 10^{-15}$~erg~s$^{-1}$~cm$^{-2}$~\AA$^{-1}$, applying the reddening curve presented by \citet{Cardelli1989} with $E(B-V)=0.03$ and $R(v)=3.1$ (as discussed in \S \ref{sec:results_steadystate}), and assuming a luminosity distance of $D_{\rm L}=72.5$~Mpc 
\citep{Bentz2015}. }
	\label{fig:cloudy_sed}
\end{figure}

\begin{figure*}
	\centering
	\includegraphics[width=0.99\linewidth]{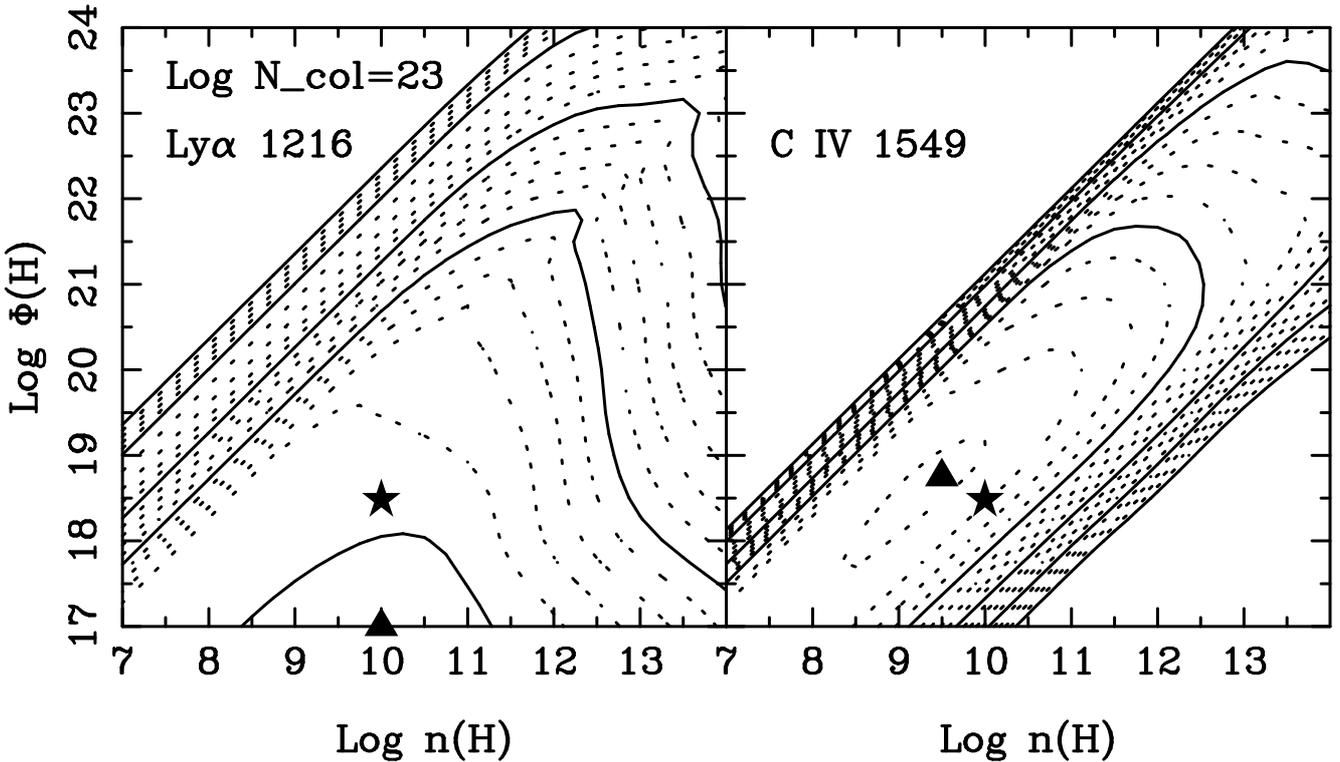}
	\caption{Example \cloudy\,photoionization grids for \lya\,\emph{(left panel)} and \civ\,\emph{(right panel)}, for \logncol=23. The contours display the total line equivalent width, i.e., the sum of the emission from the directly-illuminated and the outward-facing cloud faces, normalized to the incident continuum flux at $\lambda$1215\AA. The solid curves represent increments of 1 dex in equivalent width, while the dashed curves show 0.25 dex increments; the lowest contour represents an equivalent width of one. Black triangles show the maximum value of the line fluxes; black stars indicate the location in (\nh,\phih) space of a representative BLR cloud, that approximately reproduces the emission-line strengths of Ly$\alpha$ and C~{\sc iv} in a typical AGN (\citet{Davidson1979}). Vertical `slices' through these grids represent constant-density ($s=0$) pressure law models, while diagonal slices (at a 45$^{\circ}$ angle to the horizontal) represent constant ionization parameter ($s=2$) models; we note that the latter require an interpolation across multiple \cloudy\,grids, as \ncol\,is constant only for $s=0$ pressure laws. The \civ\, emission line emits most efficiently along a `ridge' centered on an ionization parameter $\log(U)\sim-1.5$, while the \lya\, emission line emits efficiently at low incident ionizing photon fluxes.}
	\label{fig:cloudy_grids}
\end{figure*}

\begin{figure*}
	\centering
	\includegraphics[width=0.99\linewidth]{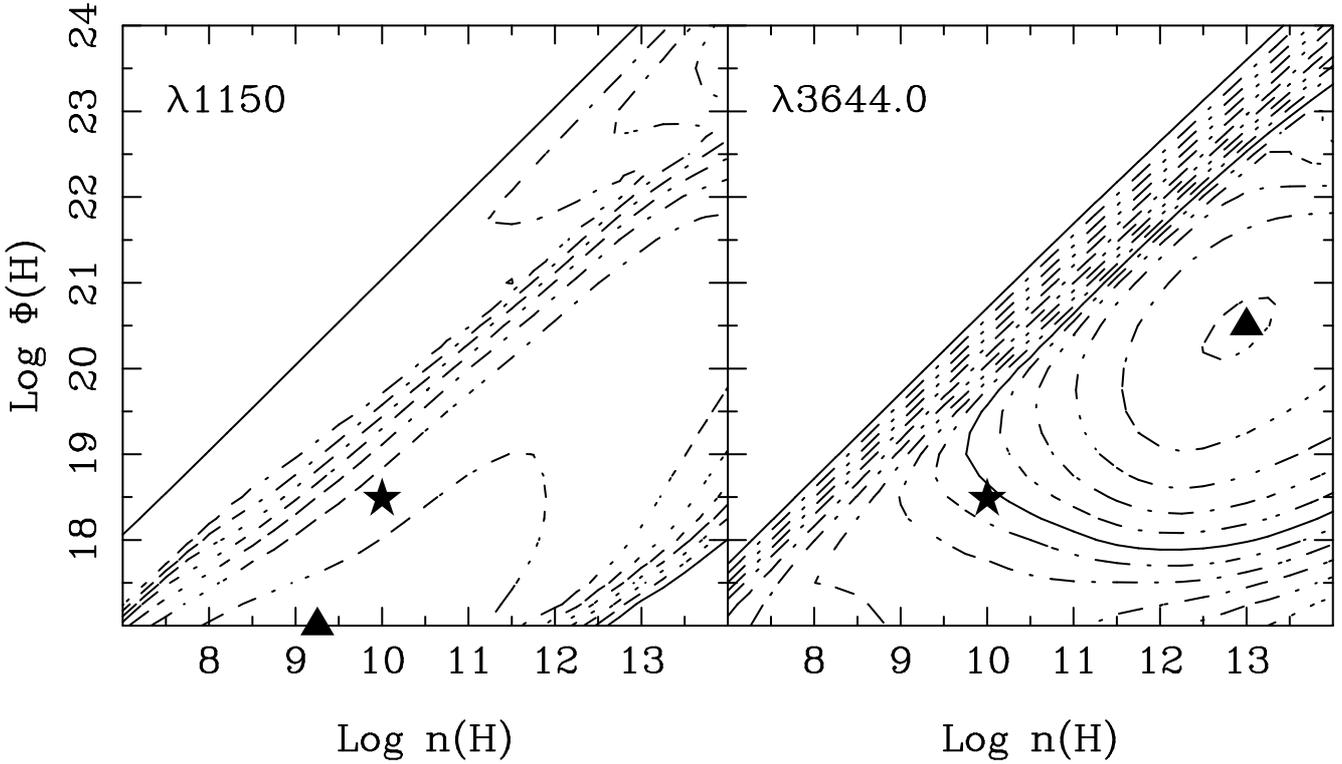}
	\caption{Example \cloudy\,photoionization grids for the diffuse continuum at various UV-optical wavelengths, for \logncol=23. The contours display the sum of the emission $F_\lambda$ from the directly-illuminated and the outward-facing cloud faces, normalized to the incident continuum flux at $\lambda$1215\AA. The solid curves represent increments of 1 dex in equivalent width, while the dashed curves show 0.1 dex increments; the lowest contour represents $\log[\lambda F_\lambda/\lambda F_\lambda(1250$ \AA$)]=-1$. Symbols as Figure \ref{fig:cloudy_grids}. At short wavelengths $(\lambda\apprle2000$ \AA) the brightest DC component is electron scattering, mostly free-free, which is most effective at low densities and low ionizing fluxes. Immediately bluewards of the Balmer break $(\lambda=3646$ \AA), the DC is dominated by Balmer continuum emission, which emits most effectively at high densities and fairly high ionizing fluxes.}
	\label{fig:cloudy_DC}
\end{figure*}

\subsection{Photoionization Models for Individual Clouds}\label{sec:method_cloudy}

To obtain the local surface emissivity $\epsilon(r)$ for a cloud located at radius $r$, we generate a grid of photoionization models using \cloudy\,version C13.05 \citep{Ferland2013}.  \cloudy\,performs photoionization modeling of individual spherical clouds of gas with radius $R_c$, situated at a radial distance $r$ from the central ionizing continuum source. We generate the photoionization models assuming that the covering fraction for the BLR is small, so that only a single reprocessing occurs\footnote{Such geometries are referred to as `open' in the \cloudy\,modeling parlance.}, and with $r$ sufficiently large that the clouds are effectively plane parallel slabs. In that case, the emission from both the inwards-facing (i.e., directly illuminated by the continuum) and outwards-facing surfaces eventually escapes to infinity. The ratio of inwards-facing to total emission does however, affect the form of the emission-line response functions (\S \ref{sec:method_anisotropy}).

For the incident ionizing continuum, we utilize the continuum SED for NGC~5548 as presented by \citet{Mehdipour2015} (Figure \ref{fig:cloudy_sed}) based on multi-wavelength (X-ray--UV--optical--IR) observations performed 2013 June -- 2014 February. This SED is comprised of: an accretion disk emission feature, a power-law hard X-ray continuum component, and a reflection feature producing the Fe K emission complex. The hard X-ray component has a photon index $\Gamma\approx1.8$, a high-energy cutoff set to $T_e=400$~keV and a low-energy cutoff at $\sim1$~keV. The accretion disk component includes Compton up-scattering of seed photons at temperature $T_\mathrm{seed}\approx0.8$~eV in a corona with $T_c\approx150$~keV; this up-scattering produces the observed soft X-ray excess. Much of the intrinsic continuum emission responsible for powering the broad emission lines is produced in the unobservable extreme-UV spectral region (photon energies~$\geq1$~Ryd), which for this model is dominated by the Compton up-scattered accretion disk component.

Our model grids span two decades in hydrogen gas column density, $22\le\logncol\le24$, seven decades in hydrogen gas density, $7\le\lognh\le14$, and seven decades in incident ionizing photon flux, $17\le\logphih\le24$, with a resolution of 0.25 in the logarithm of each of these quantities. For each grid point, we use \cloudy\,to obtain the local surface emissivities not only for an extensive list of atomic transitions, but also for reprocessed continuum emission (the diffuse continuum, DC) at UV-optical and infra-red wavelengths emitted from the same gas. For the $s=2$ models, given the steady-state luminosity and BLR size for NGC 5548, we subsequently found that a substantial emission line flux is produced by gas with \logncol$\sim21.5$. For the particular $s=2$ BLR we study in-depth (Model 2, \ref{sec:results_steadystate}), we therefore generate additional \cloudy\,models with $\logncol\ge21$, so as to obtain accurate emissivity and responsivity functions in the outer BLR.

In Figure~\ref{fig:cloudy_grids} we present representative \cloudy\,emission line equivalent widths (hereafter, EW), measured relative to the incident continuum at $\lambda$1215\AA\, for two of the strongest observed emission lines, Ly$\alpha$ and C~{\sc iv}, for a fixed hydrogen gas column density \logncol=23 cm$^{-2}$. The \lya\, emission line emits most efficiently at low ionizing photon fluxes; indeed, its efficiency  increases towards \logphih$<17$, the lower limit of the parameter space probed by our photoionization grids. This implies that the Hydrogen lines could in principle be produced very efficiently at low ionizing fluxes. However, such conditions are unlikely to be relevant here: the outer edge of the BLR is most likely determined by the dust sublimation radius (at which \logphih$>17$), as the ionizing continuum is efficiently absorbed by dust grains. In contrast, high-ionization lines such as \civ\,tend to be produced efficiently along a diagonal `ridge' in (\phih,\nh) space, corresponding to a particular ionization parameter, $U$; for \civ\, the optimal ionization parameter, i.e., one that passes through the peak EW, is $\log(U)\sim-1.5$. We present photoionization grids for two representative DC bands in Figure \ref{fig:cloudy_DC}. These may be compared to the more extensive model grids of DC bands presented in \citet{Korista2001} spanning the entire UV--optical continuum.
While their results differ from ours in detail, mainly due to the use of different ionizing continua, the gross shapes of their contour plots are similar to those used here.

In general, the gas becomes overionized at high ionization parameters, thus the DC does not emit efficiently at high \phih\,and low \nh\,(upper left regions of the photoionization grids), while the DC is not strongly sensitive to \phih\,and \nh\,(hence the widely spaced contour lines) once the ionization parameter becomes sufficiently low.

\begin{figure}
	\centering
	\includegraphics[width=0.999\linewidth]{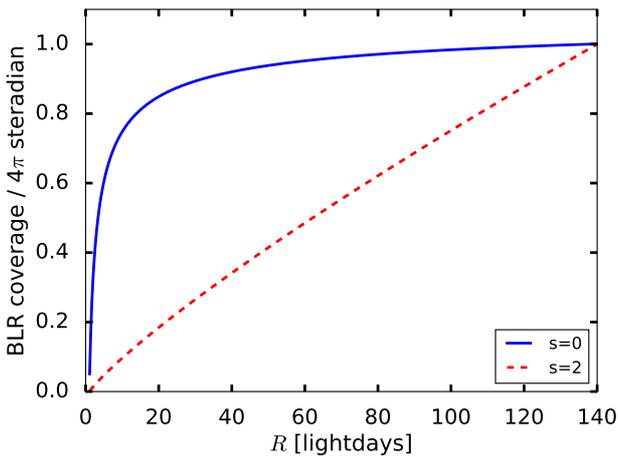}
	\caption{Cumulative sky coverage (as seen from the continuum source), as a function of radial distance $r$, for BLR models with a total cumulative covering fraction of $\Omega=4\pi$ steradians at the outer radius. For a given $\Omega$, the constant-density ($s=0$) models distribute the clouds at smaller radii relative to the $s=2$ models.}
	\label{fig:coverage}
\end{figure}

\subsection{Extent and Geometry of the Model BLR}

In the absence of a clear consensus on the broad line region geometry, we limit this investigation to spherical BLR geometries only. \citet{Perez1992} showed that, in the absence of additional constraints, a well-determined 1-d transfer function cannot {\em uniquely\/} determine the geometry of the BLR, while even 2-d transfer functions reveal some ambiguity. For NGC 5548, the highest fidelity reverberation maps for the strong UV and optical emission lines from the AGN STORM campaign suggest that the variable BLR may occupy a more flattened configuration (Horne, private communication). Our objective in this paper, however, is simply to illustrate the dependencies of the flux and lag spectra of the DC on a range of BLR cloud physical conditions and radial distributions, as a prelude to more detailed models for which constraints on the geometry are also included. In this context, a spherical BLR geometry allows a more intuitive interpretation of the model flux and delay spectra in terms of the cloud distribution and local photoionization physics. Alternative BLR geometries would introduce additional assumptions for which we currently lack strong independent constraints, e.g., the BLR viewing angle and opening angles. Adoption of a spherical BLR geometry also aids comparison with the spherically symmetric locally optimally-emitting cloud model for NGC 5548 presented by \citet{Korista2001}. We note that deviations from spherical symmetry may prove necessary in any future work aimed at reproducing the observed flux and delay spectra in detail (\S \ref{sec:discussion}).


For a spherical BLR, the integrated emission-line luminosities produced by our models depend on our choice of inner and outer BLR radii (\rin\,and \rout), the total covering fraction at the outer radius (equation \ref{eq:lumin}), and our chosen normalization, i.e., for a given $s$, the value of $\Phi_{\rm H}$, $n_{\rm H}$ and $N_{\rm col}$ at some fixed radial distance $r$. The constant pressure models ($s=0$) distribute a large fraction of the BLR clouds close to the inner radius (Figure \ref{fig:coverage}). This causes the emission line luminosities to depend strongly on \rin\,for $s=0$ models: much of the BLR becomes over-ionized as \rin\,decreases, and the emission line luminosity decreases sharply. The $s=2$ models are less sensitive to this effect, as they are constant in ionization parameter, $U$. In this work, we adopt \rin=1~light-day, and \rout=140~light-days (the approximate dust sublimation radius, given the luminosity of NGC 5548). We note that due to the central concentration of BLR clouds for $s=0$, \rin\,cannot be much smaller than 1~light-day if our constant-density models are to produce the required emission line luminosities for reasonable values of the total covering fraction. In their phenomenological BLR model for NGC 5548, \citet{Pancoast2014} find $\rin=1.39^{+0.80}_{-1.01}$ light-days, consistent with our adopted value. 

\subsection{Radial Distribution of Clouds:} 

Our \cloudy\,modeling assumes that the BLR clouds do not self-shadow, and that there is no intrinsic reddening of the nuclear continuum as seen at the inward-facing cloud surface. In that case, the ionizing continuum at the inward-facing surface of a given cloud scales as $r^{-2}$. We parameterize the radial dependence of \logphih\,as

\begin{equation}\label{eq:phi_to_radius}
\log(r)=-0.5(\logphih-20)+15.413+r_{20} \, .
\end{equation}

\noindent Here, $r_{20}$ is the radius (in light-days) at which \logphih=20. The value of $r_{20}$ depends on the continuum SED and its luminosity. For NGC 5548, $r_{20}\approx14.8$ light-days, as determined using the observed continuum flux, $F_\lambda(1367$ \AA$)=42.6(\pm 8.6)\times 10^{-15}$~erg cm$^{-2}$ s$^{-1}$ \AA$^{-1}$ \citep{DeRosa2015}, corrected for Galactic extinction adopting $E(B-V)=0.03$ (\citealt{Korista2000} - we discuss this choice in \S \ref{sec:results_steadystate}), and assuming a luminosity distance to NGC~5548 of $D_\mathrm{L}=72.5$~Mpc \citep{Bentz2015}. 

In combination with the pressure-law relations (\S \ref{sec:theory}), $r_{20}$ allows us to convert the emergent surface emissivities $\epsilon(\phih,\nh,\ncol)$ extracted from our photoionization grids into radial emissivity functions $\epsilon(r)$. For $s=0$ models, both \ncol\,and \nh\,are constant throughout the BLR (\S \ref{sec:theory}). Therefore, the relevant emissivities for such models comprise a vertical `slice' through a single \nh, \phih\, grid. We perform a linear interpolation in log-space over \phih\,to obtain $\epsilon(r)$. For $s=2$, the ionization parameter $U$ is constant, $n_{\rm H}\propto r^{-2}$, and $N_{\rm col}\propto r^{-4/3}$. Such models correspond to diagonal lines in (\nh, \phih) space, with \ncol\, decreasing radially as per Equation \ref{eq:ncol_r}. To obtain $\epsilon(r)$ for these models, we perform a linear interpolation in log-space over \phih, \nh\, and \ncol.

With the emissivity distribution $\epsilon(r)$ in hand, we numerically integrate equation \ref{eq:lline} from \rin\,to \rout\,to obtain the total luminosity for a given BEL (or diffuse continuum band), assuming full coverage ($\Omega=4\pi$ steradian) as seen from the continuum source, at the BLR outer radius. A reduced covering fraction (i.e., $< 4\pi$~steradian) then corresponds to a downwards linear scaling of the emission-line luminosity.\\

\subsection{Determining the Response Functions}\label{sec:method_radii}

The observed BLR emission signal is due to the integrated emission of individual BLR clouds located at radii $\rin<r<\rout$. Thus, the BLR responds to continuum variations over a distribution of time delays $0\le\tau\le2\rout/c$; the detailed shape of the emission line response function will depend on BLR geometry and on viewing angle \citep[e.g.,][]{O'Brien1994}, as well as on the detailed gas physics. \citep[e.g.,][]{Goad1993}. In addition, the \emph{measured} emission-line response will depend on the variability behavior (i.e., amplitude and characteristic timescale) of the ionizing continuum source \citep[see also \S \ref{sec:method_driving}]{Goad2014}.

For a given BEL, the luminosity-weighted effective radius of the BLR is given by the centroid of the differential luminosity as a function of radius:

\begin{equation}\label{eq:emissivity_radii}
r_{\epsilon}=\frac{\int_{\rin}^{\rout}r\lline (r)\mathrm{d}r}{\int_{\rin}^{\rout}\lline (r)\mathrm{d}r}.
\end{equation}

\noindent A similar relation can be defined for the diffuse continuum bands, replacing $L_{\rm line}$ with $\lambda F_{\lambda}$, the monochromatic diffuse continuum emission at $r$.

\subsubsection{Why Responsivity Matters}\label{sec:method_responsivity}

If the emission of each BLR cloud varies linearly (though not necessarily 1:1) with respect to the nuclear continuum variations, and the clouds emit isotropically, $r_\epsilon$ would also represent the observed effective reverberation radius. However, the responsivity and isotropy of realistic BLR clouds depend on the local photoionization physics of the individual clouds. We define the local responsivity $\eta(r)$ for the contribution \lline$(r)$ to a given emission line from radius $r$ as:


\begin{equation}
\eta(r)=\frac{\Delta\log[\lline(r)]}{\Delta\log[\phih(r)]}.
\end{equation}

\noindent If $0<\eta<1$ for all $r$, the line equivalent width decreases when the continuum luminosity increases (i.e., an intrinsic Baldwin effect, \citealt{Goad2004}), while $\eta<0$ produces an inverse correlation between line flux and continuum luminosity.

For $s=0$ models, given our assumption that there is no internal extinction or cloud--cloud shadowing in the BLR, a change in ionizing flux corresponds to an instantaneous `shift' of the clouds along the radial axis, allowing us to measure $\eta(r)$ directly for each \cloudy\,($n_\mathrm{H}$,$\Phi_\mathrm{H}$) grid as

\begin{equation}\label{eq:responsivity_calc}
\eta(r)=-\frac{\delta\log(\epsilon)}{2\delta\log(r)}; \mathrm{\,\,(s=0)},
\end{equation}

\noindent where the factor 2 is due to the radial dependence of the ionizing flux. For a constant steady-state ionization parameter $U$ (i.e., $s=2$ models), as both \nh\,and \phih\, follow inverse-square laws in radius, a small increase in flux corresponds to a constant increase in $U$ at all radii:

\begin{equation}\label{eq:responsivity_calc_s2}
\eta(r)=\frac{\delta\log(\epsilon)}{\delta\log(U)}; \mathrm{\,\,(s=2).}
\end{equation}

\noindent RM measures the effective radius of those BLR regions that respond to luminosity variations. In the specific case of $\eta(r)=1$ for all $r$, the line equivalent widths remain constant as the continuum luminosity varies, and the effective variability radius of the BLR is equal to $r_{\epsilon}$. Otherwise, the measured time delays will approximately correspond to the responsivity-weighted BLR radius,

\begin{equation}\label{eq:responsivity_radii}
r_{\eta}=\frac{\int_{\rin}^{\rout}r\eta(r)\lline (r)\mathrm{d}r}{\int_{\rin}^{\rout}\eta(r)\lline (r)\mathrm{d}r}.
\end{equation}

\noindent In general, $\eta(r)$ is luminosity-dependent. In the linear regime (i.e., small fluctuations about some mean level), the responsivity is simply the logarithmic slope of the $\Phi_\mathrm{H}$ versus $\epsilon$ relationship at a given $r$. This linear approximation will break down for large changes in continuum luminosity. Formally, under these circumstances the responsivity corresponding to a particular continuum level is best-determined directly from a grid of photoionization model calculations.

\subsubsection{Accounting for Anisotropic Emission}\label{sec:method_anisotropy}

If the BLR clouds emit anisotropically, the expected values of both $r_{\epsilon}$ and $r_{\eta}$ will be modified accordingly \citep[e.g.,][]{Ferland1992,O'Brien1994}. For example, consider a BLR cloud located along the line of sight to the observer. If this cloud emits isotropically, it will contribute to the response function at a time delay $\tau=0$ relative to the ionizing continuum. However, if this cloud emits only from its inwards-facing surface, it is not observed, and does not contribute to the response at $\tau=0$. On the other hand, an identical cloud located on the opposite side of the broad-line region will contribute to the response function at a time delay $\tau=2\,r/c$. Thus, while the steady-state line luminosity is not affected by anisotropy (assuming spherical symmetry), the line response is shifted towards larger delays if the BLR clouds preferentially emit from their illuminated face.\\

Following \citet{O'Brien1994}, we define an anisotropy factor, $F(r)=\epsilon_{\mathrm{inwd}}(r)/\epsilon_{\mathrm{tot}}(r)$, where $\epsilon_{\mathrm{inwd}}$ and $\epsilon_{\mathrm{tot}}$ are the inwards-facing emissivity and effective total emissivity for the cloud, respectively. Isotropic emission corresponds to $F=0.5$. The value of $F$ depends upon the local photoionization physics for individual clouds, and thus are determined directly from our \cloudy\,modeling. The `observed emissivity' of a given cloud then depends on the fraction of the inwards-facing and outwards-facing surfaces we observe. Since the cloud geometry is unknown we choose an emission line radiation pattern which approximates the phases of the moon:

\begin{equation}
\epsilon_{\mathrm{obs}}(r,\theta)=\epsilon_{\mathrm{tot}}(r)\left[1-(2F(r)-1)\cos(\theta)\right].
\end{equation}

\noindent Here, $\theta$ denotes the angle between the line of sight and the cloud radial vector. Given that there is a linear mapping between $\cos(\theta)$ and the time delay $\tau$ at a given radius, we numerically integrate out the angular dependence, and reformulate Equations \ref{eq:emissivity_radii} and \ref{eq:responsivity_radii} in terms of $\tau$ to obtain an emissivity-weighted effective delay, $\tau_{\epsilon}$, and a responsivity-weighted effective delay, $\tau_{\eta}$. These quantities correspond to $r_\epsilon/c$ and $r_\eta/c$, respectively, for isotropically-emitting clouds. The total luminosity can then be recovered by integrating $\mathrm{d}L(\tau)$ over $0<\tau<2\,\rout/c$, where $\mathrm{d}L(\tau)$ is the 1-d response function. We note that, as the responsivity of a given emission line may be negative at some or all radii, the response function may not have a well-defined responsivity-weighted centroid $\tau_{\eta}$; in practice, we find that $s=2$ models tend to display negative responsivity at large radii.

%% file: sections/sec_results.tex
\section{Steady-State Models}\label{sec:results}

As the available parameter space for our pressure-law BLR models spans several orders of magnitude in \phih, \nh\,and \ncol, we first determine which steady-state models are roughly consistent with the observed luminosities of the strongest UV and optical BEL in NGC~5548. This allows us to exclude models for which the BEL emission is too weak to be relevant for this test case. We then select two representative models, one each for $s=0$ and $s=2$ (\S \ref{sec:results_steadystate}). For these models we investigate the BEL responses (\S \ref{sec:results_linelags}) and, of particular relevance here, the relative contribution and wavelength-dependent behavior of the diffuse continuum (\S \ref{sec:results_cont}). Lastly, we examine the dependence of the diffuse continuum luminosity, response functions, and measured lags, on the parameters \nh, \phih, and $U$, for the broader class of $s=0$ and $s=2$ pressure laws (\S \ref{sec:results_cont_comparison}).

\subsection{BEL Luminosities}\label{sec:results_steadystate}

\begin{figure*}
	\centering
	\includegraphics[width=0.49\linewidth]{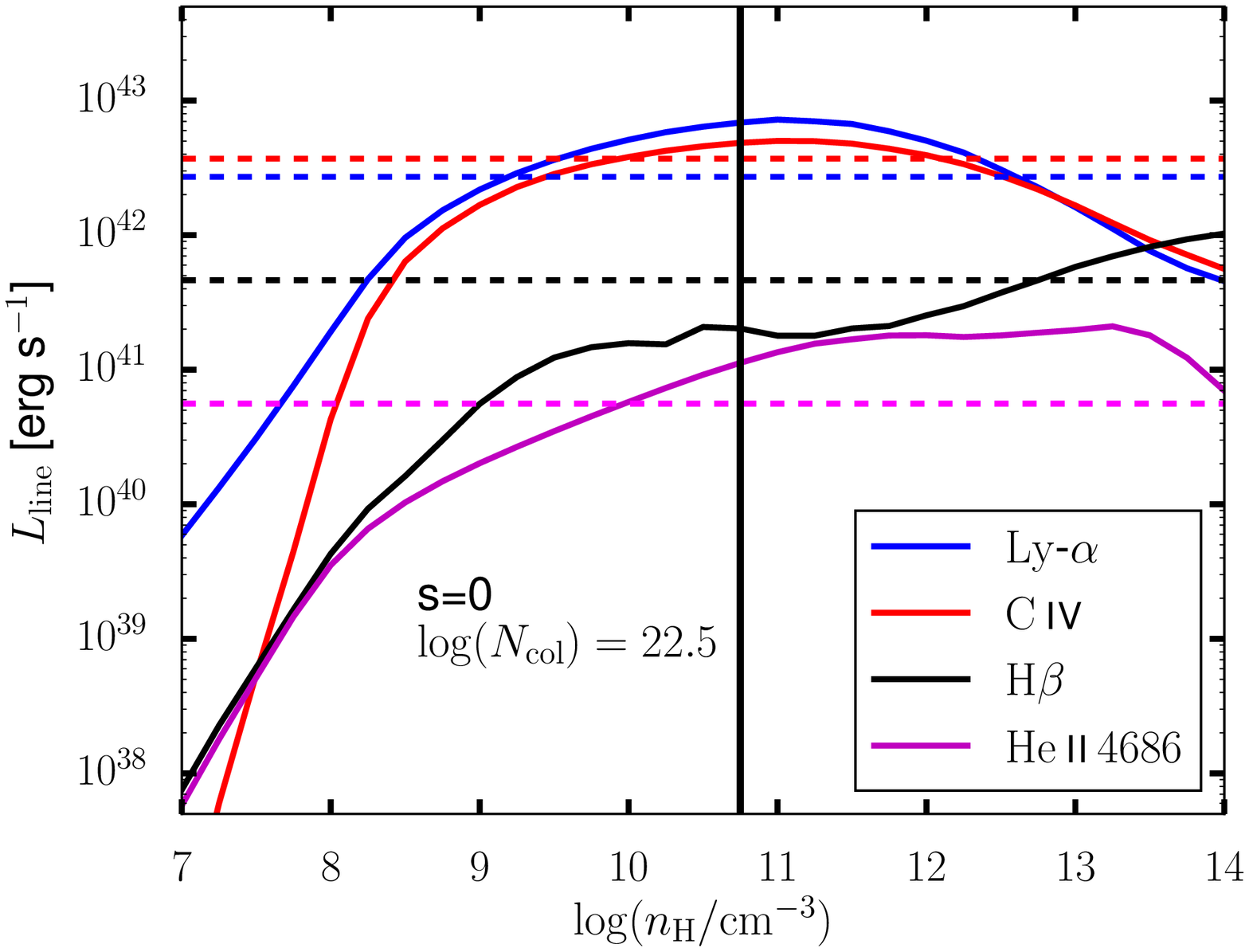}
	\includegraphics[width=0.49\linewidth]{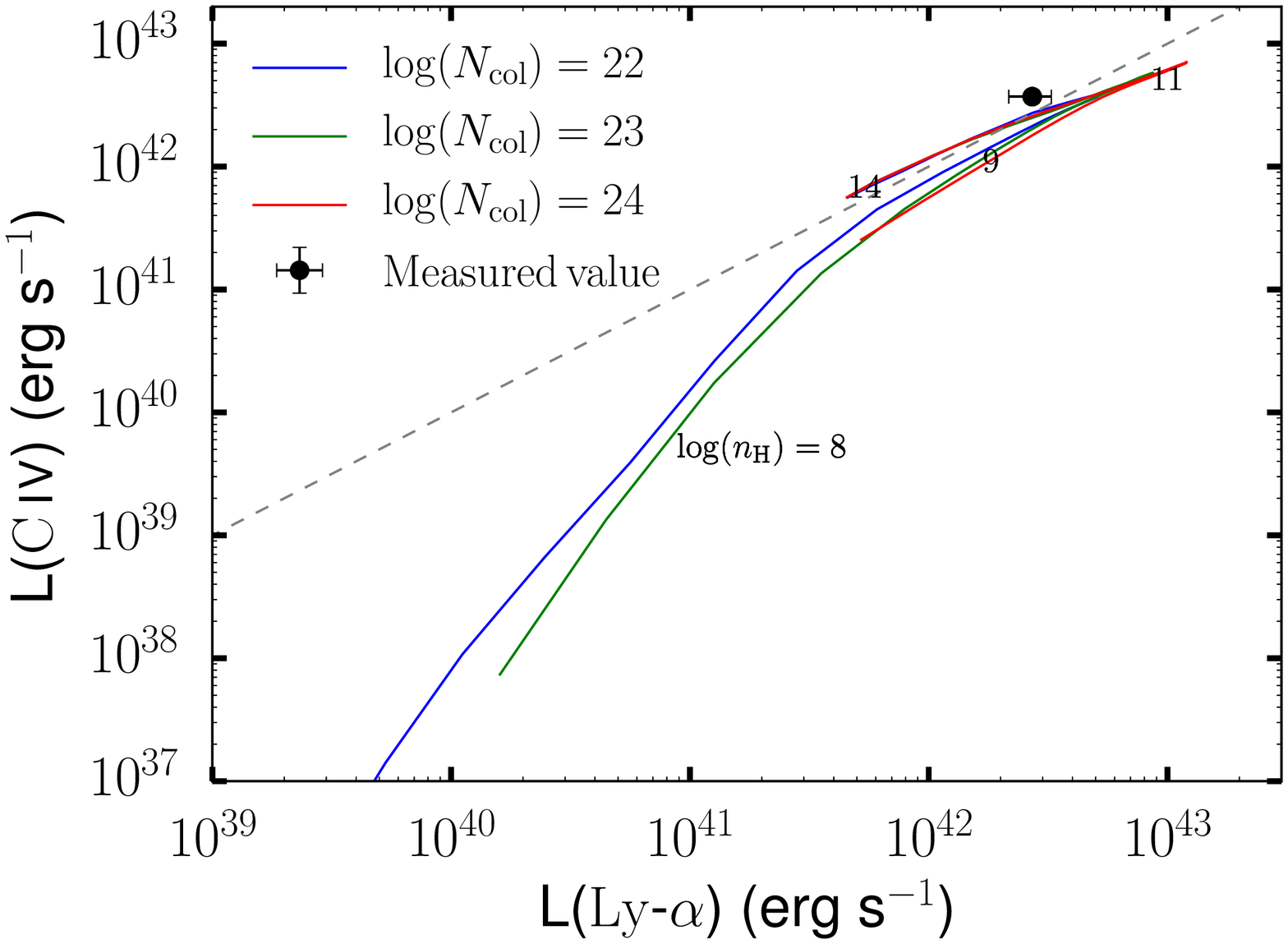}
	\includegraphics[width=0.49\linewidth]{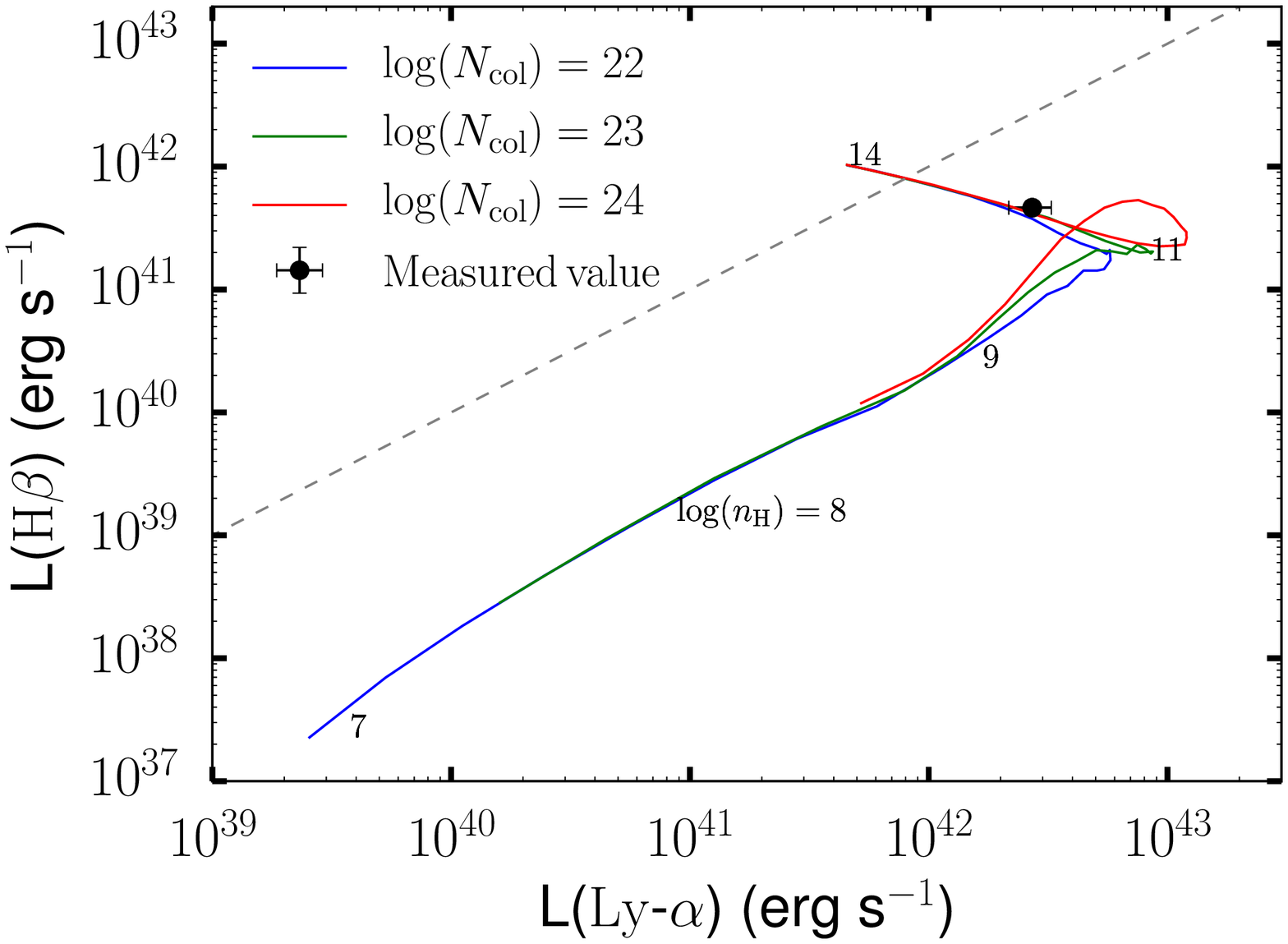}
	\includegraphics[width=0.49\linewidth]{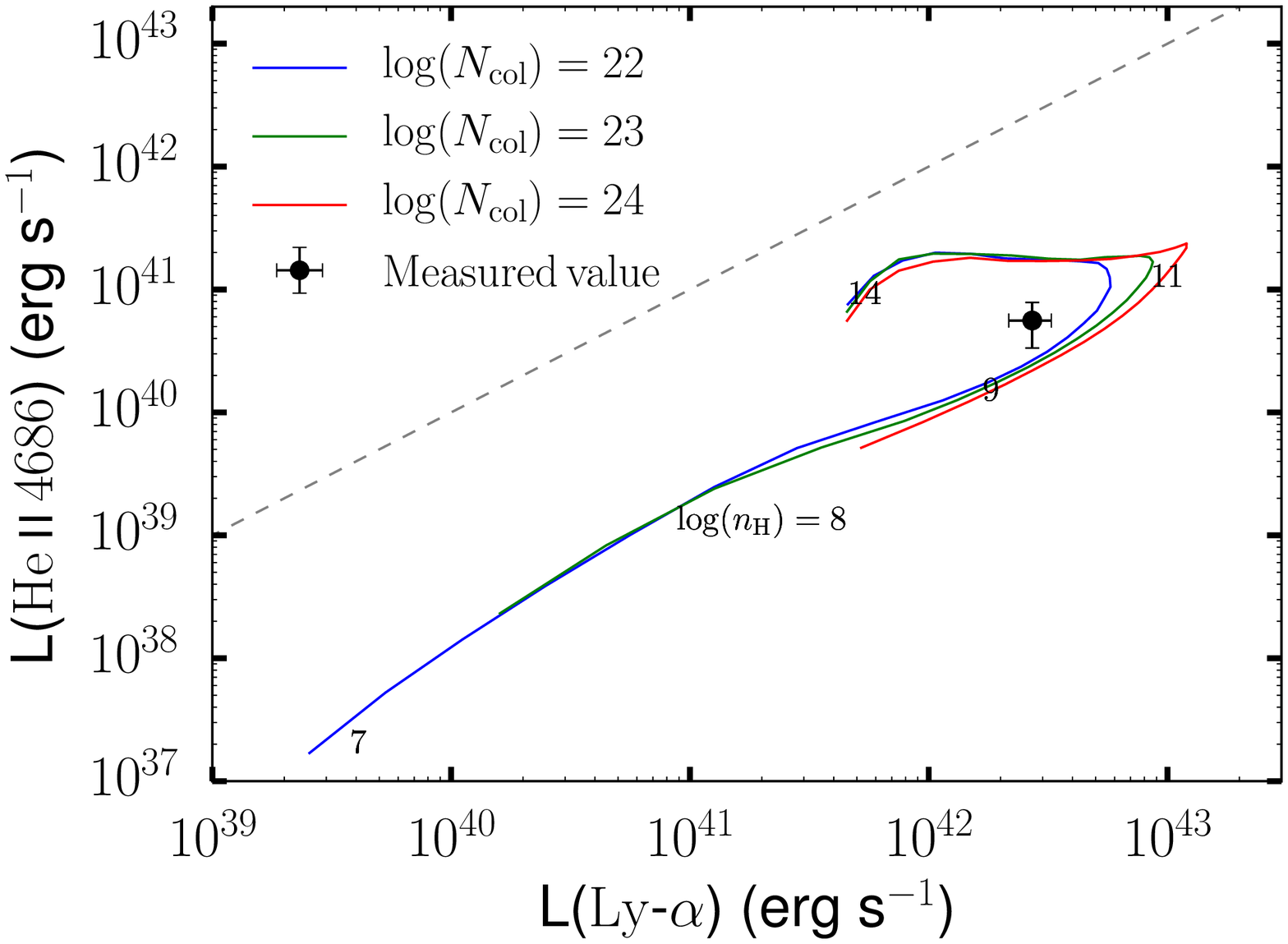}	
	\caption{\emph{Top left:} Model BEL luminosities as a function of \nh\, (solid curves). Here we assume that the BLR covers $4\pi$ steradian of the continuum source (see discussion in \S \ref{sec:results_steadystate}). The measured broad emission-line luminosities for NGC~5548 \citep{DeRosa2015,Pei2017} are shown as dashed lines. For the observed \civ\,and \hbeta\,luminosities, we subtract a narrow emission-line component (see \S \ref{sec:results_steadystate} for details). The black vertical line shows \lognh=10.75, corresponding to our Model 1. The main conclusion to be drawn here is that the measured emission-line luminosities for \lya, \civ\,and \heop\,are exceeded for $9.75\apprle\lognh\apprle12.25$, while higher densities are required to reproduce the \hbeta\ luminosity (\S \ref{sec:results_steadystate}). \emph{Other panels:} Predicted BEL luminosities for \civ\,\emph{(top right)}, \hbeta\,\emph{(bottom left)} and \heop\,\emph{(bottom right)}, relative to \lya, for a grid of our $s=0$ BLR models. Each of the three `tracks' represents a single value of \logncol; the density \nh\, ranges over $7\le\lognh\le14$, and increases in a counterclockwise direction. We indicate the location of selected values of \nh\,for the \logncol$=22$ track; the tracks with larger column densities have higher line luminosities at low \nh\,but are similar at high gas densities. As a visual aid, the dashed black lines indicate 1:1 line ratios. Black circles represent the observed line luminosities for NGC 5548, after correcting for reddening and narrow-line contamination. In general, we note that gas densities of \lognh$\sim11$ tend to maximize the line intensities, and that the relative intensities of the BELs investigated here are not strongly sensitive to \ncol\, (as the `tracks' are not widely separated) for the range of column densities explored here.}
	\label{fig:line_lumins}
\end{figure*}

Here, we determine the range of pressure law models able to roughly reproduce the observed BEL luminosities. Ideally we require our model integrated BEL luminosities (as calculated for a covering factor of $4\pi$ steradian) to exceed their corresponding measured values. This is because our model does not account for BLR self-shadowing, making it unreliable at high covering factors. Reducing the covering factor corresponds to a linear scaling of the BEL luminosities, e.g., for a covering fraction of $4\pi/3$ steradian, the model luminosities given by Equation \ref{eq:lumin} should exceed observations by a factor 3.

\subsubsection{Observed Line Luminosities:}

We compare our models to the observed UV BEL \citep{DeRosa2015} and optical BEL \citep{Pei2017} luminosities for NGC 5548 during the 2014 AGN STORM campaign. While \citet{Schlafly2011} find a Galactic reddening of $E(B-V)=0.02$ towards NGC 5548, \citet{Kraemer1998} find that the narrow-line region in NGC 5548 has an additional intrinsic reddening of $E(B-V)\approx0.04$. For consistency with previous photoionization studies of this AGN \citep[e.g.,][]{Korista2000}, and given the uncertainty on the total (Galactic plus intrinsic) reddening of the BLR, we adopt $E(B-V)=0.03$. We de-redden the observed luminosities using the reddening curve presented by \citet{Cardelli1989}, with $R(V)=3.1$. For \civ\,and \hbeta, we subtract a narrow emission-line component from the observed BEL luminosities, as determined by \citet{Korista2000} and \citet{Peterson1991}, respectively; while the narrow emission line strengths for NGC 5548 have shown some variation on multi-year timescales \citep{Peterson2013}, we only require a first-order correction for the purposes of this study. The inferred emission-line luminosities, corrected for reddening and for narrow-line emission, are listed in Table \ref{tab:observedlags}.

\subsubsection{Model 1, Constant Density $(s=0)$:} We calculate the BEL luminosities produced by constant-density ($s=0$) models spanning the full range of hydrogen gas densities $7\le\lognh\le14$ and gas column densities $22\le\logncol\le24$. A density of $\lognh>10$ is required in order to produce sufficient \civ\,luminosity to match the 2014 observations (Figure \ref{fig:line_lumins}), and to produce stratification in the emission-line delays (Figure \ref{fig:line_lags_model1}). For a column density of \logncol$=22.5$, our model emission-line luminosities can not simultaneously exceed the measured \lya, \hbeta, \civ\,and \heop\, BEL luminosities (Figure \ref{fig:line_lumins}, upper left). In terms of observed line ratios, the model \hbeta\,line is under-produced relative to \lya\,except at very high gas densities (\logncol$\apprge13$). While we do find one model that marginally exceeds the observed luminosities of all four lines (namely, that with $\log[n_\mathrm{H}/\mathrm{cm}^{-3}]=9$ and \logncol$=24$), this extreme case would imply a Compton-thick yet low-density BEL at all $r$. The diameters of individual BLR clouds are $\sim0.5$ light-day for this extreme case; the smoothness of observed BEL velocity profiles excludes such large sizes \citep[e.g.,][]{Laor2004}. Thus, if the underlying photoionization modeling is correct, a single $s=0$ pressure-law component cannot account for the measured BEL strengths in NGC 5548. At least one additional high-density ($\lognh\apprge12.5$) component would be required to produce sufficient luminosity in all BELs. However, the underproduction of \hbeta\,in our models may instead be an issue with the treatment of radiative transfer in our photoionization modeling. The \cloudy\,algorithm uses the local escape probability formalism, which becomes unreliable at the very high hydrogen optical depths typical of BLR clouds \citep{Netzer1990_book,Kaspi1999}. In the absence of an exact treatment of radiative transfer, it is difficult to determine whether the low \hbeta\,line emission for single-component models actually implies the existence of additional BLR emission regions with higher densities. All else equal, the simplicity of a single-component model is attractive in the context of studying the diffuse continuum component (\S \ref{sec:results_cont}). 

We therefore define our \textbf{Model 1} as an $s=0$ pressure-law BLR with \lognh=10.75 and \logncol=22.5; this model fulfills our luminosity criterion for all of the strongest UV and optical BEL apart from \hbeta. Our choice of \ncol\, here is somewhat arbitrary; the emission-line luminosities and response function centroids are only strongly sensitive to \ncol\,for low column densities, $\logncol<22.5$. Specifically, the observed \civ, \lya\,and \heop\,luminosities are exceeded for $10.25<$\lognh$<11.25$ for $22.5\leq\ncol\leq24$.

\subsubsection{Model 2, Constant Ionization Parameter $(s=2)$:}

\begin{figure}
	\centering
	\includegraphics[width=0.99\linewidth]{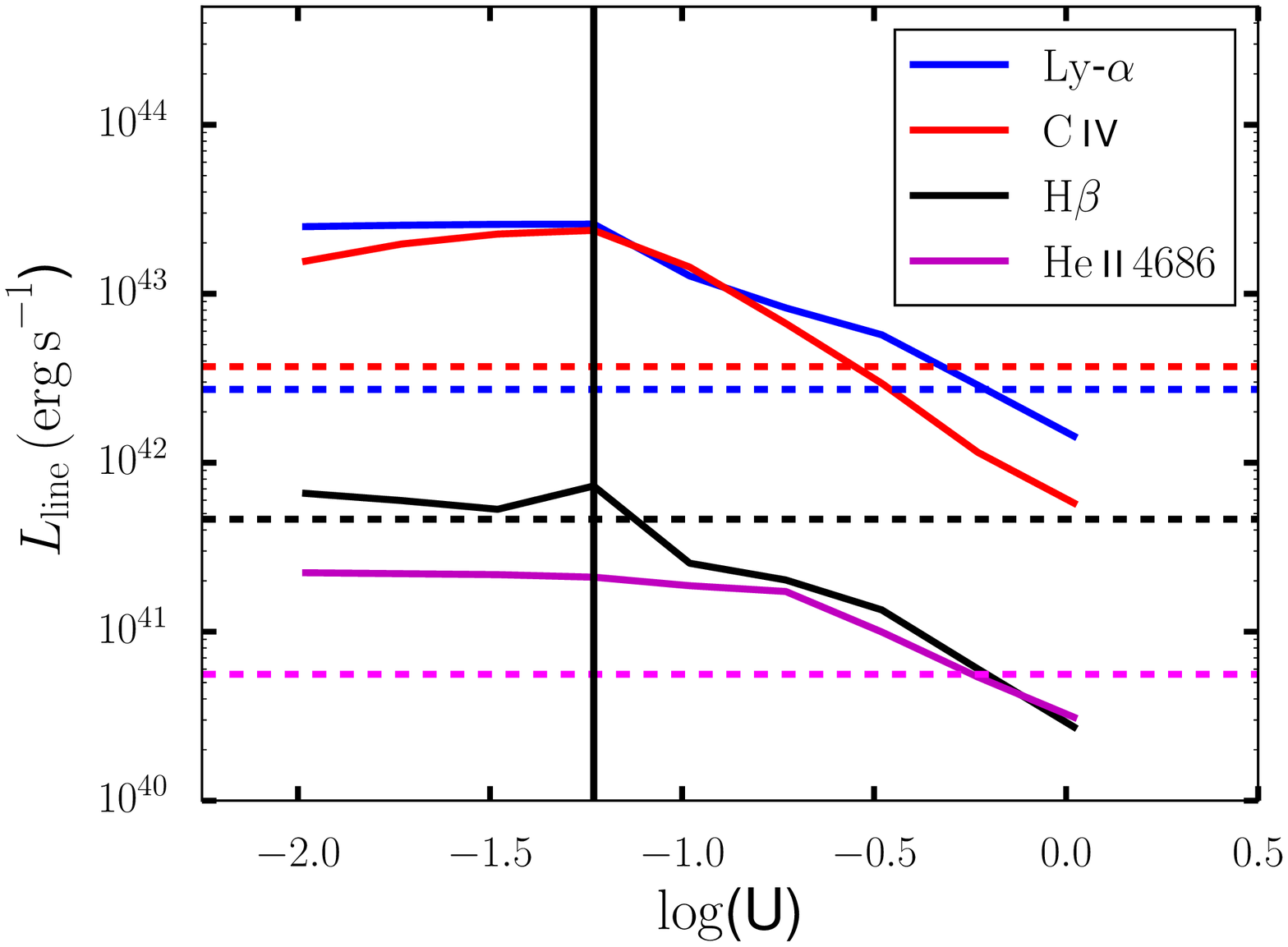}
	\caption{Model BEL luminosities for \lya, \civ, \hbeta, and \heop, for the $s=2$ (constant ionization parameter) steady-state model. These values of \lline\,assume that the BLR covers $4\pi$ steradian of the continuum source (see discussion in \ref{sec:results_steadystate}). The measured emission-line luminosities (dashed lines) are exceeded for ionization parameters $\log(U)\apprle-1.1$. The solid vertical line indicates $\log(U)=-1.23$, corresponding to our Model 2; for our chosen normalization of \ncol, this value approximately maximizes the line intensities.}
	\label{fig:line_lumins_s2}
\end{figure}

The family of constant ionization parameter ($s=2$) models described here display an increasing column density and increasing density towards the centre (Equations \ref{eq:nh}, \ref{eq:ncol_r}). The $s=2$ models generally produce more BEL luminosity than do the $s=0$ models because they distribute a larger fraction of the BLR clouds at large $r$ (Figure \ref{fig:coverage}), i.e., at lower incident ionizing photon flux, at which the optical recombination lines (H$\alpha$ and H$\beta$) and Mg~{\sc ii} tend to have higher surface emissivities. In contrast to our chosen  $s=0$ model, constant $U$ models are capable of exceeding the measured emission line luminosities of all of the strong UV--optical emission lines (including \hbeta) for $\log(U)\apprle-1.1$ (Figure \ref{fig:line_lumins_s2}). We do not explore models with $\log U < -2.1$, as for these models $n_{\rm H}$ lies outside the bounds of our model grids for radii $\leq r_{\rm in}$. We define \textbf{Model 2} as an $s=2$ pressure law BLR with constant ionization parameter $\log(U)=-1.23$. This is the $s=2$ model for which the \hbeta\,and \civ\,BEL luminosities are maximized (Figure \ref{fig:line_lumins_s2}). By construction, Model 2 has locally identical steady-state  physical conditions (i.e., $n_\mathrm{H}$, $N_\mathrm{col}$ and $U$) to Model 1 at $r_{20}$, where $r_{20}\approx14.8$ lightdays for NGC 5548. For \rin=1 lightday and \rout=140 lightdays, Model 2 spans $21\apprle\logncol\apprle24$, and $8.8 < \lognh < 13.1$. 

\subsection{BEL Response Functions}\label{sec:results_linelags}

\begin{figure*}
	\centering
	\includegraphics[width=0.49\linewidth]{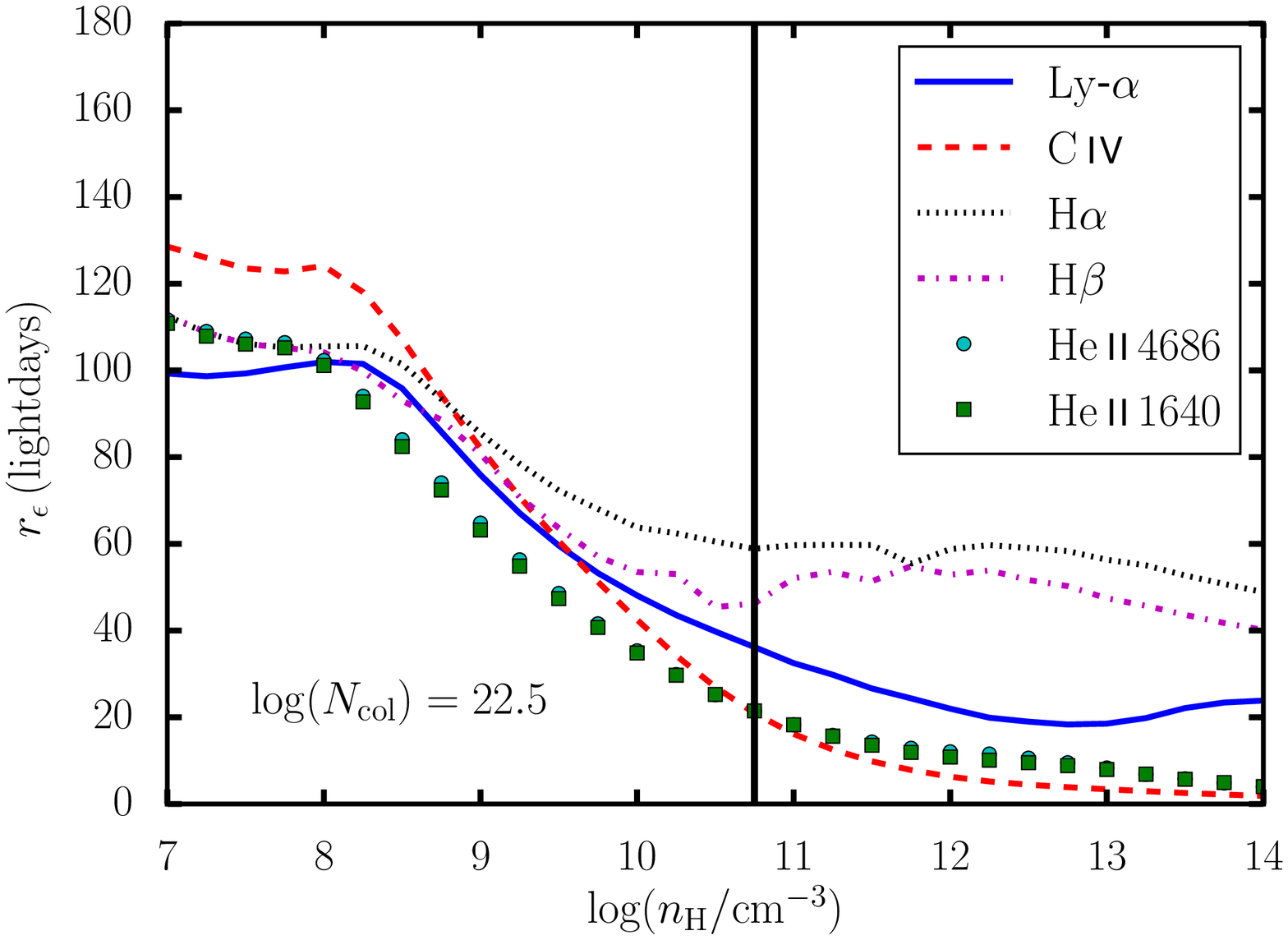}
	\includegraphics[width=0.49\linewidth]{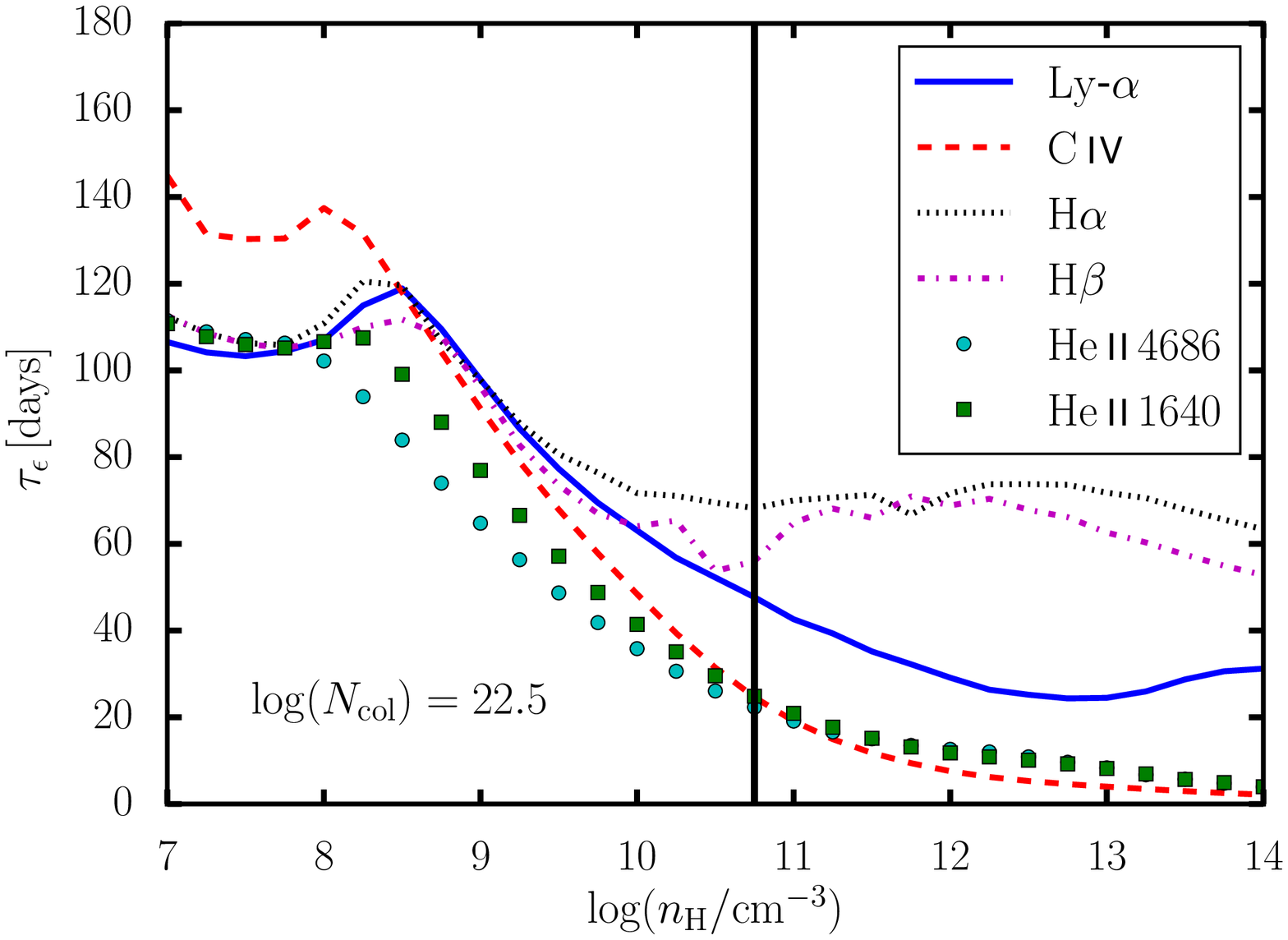}
	\caption{The emissivity-weighted effective radii, $r_\epsilon$ (left panel), and emissivity-weighted delay $\tau_\epsilon$ (right panel), as a function of \nh, for $s=0$ models with \logncol=22.5. The delays shown in the right panel accounts for anisotropic emission from the BLR clouds, which acts to increase the time delays relative to the isotropic case. In general, lower values of \nh\,\,result in a BLR that emits most efficiently at larger radii, increasing the emissivity-weighted delay. The black solid lines show the location of \lognh=10.75\,, corresponding to our Model 1.}
	\label{fig:line_lags_model1}
\end{figure*}


We summarize the measured broad emission line delays for the strong UV and optical emission lines for the AGN STORM 2014 monitoring campaign in Table \ref{tab:observedlags}, and present the response function centroids generated by our models in Table \ref{tab:modellags}. For Model 1, the lines tend to be most responsive at lower ionizing photon flux, i.e., in the outer BLR. Thus, we see $\tau_\eta>\tau_\epsilon$ for Model 1. Model 2 tends to produce larger emissivity-weighted response function centroids than Model 1; this is a direct consequence of the $s=2$ pressure law distributing more coverage (and, therefore, more line-emitting gas) at larger radii relative to the $s=0$ models (Figure \ref{fig:coverage}). However, Model 2 has small (and often slightly negative) line responsivities at large radii. Thus, the lines respond most strongly in the inner BLR, and we have $\tau_\eta<\tau_\epsilon$ for Model 2. For \lya, \civ\,and \halpha\,the response functions are positive in the inner regions but strongly negative at large $r$ for Model 2. They therefore lack a well-defined centroid; however, the effective radius that would be measured using RM observations can still be estimated by calculating the BEL response to continuum variations, as we explore via Monte Carlo modeling (\S \ref{sec:ccf}).

For lines with a well-determined response function centroid $\tau_\eta$, both models produce values of $\tau_\eta$ that are somewhat larger than the measured lags reported for the strong UV and optical broad emission lines in NGC~5548 prior to the AGN STORM campaign \citep{Cackett2015,Bentz2010b,Denney2010,Peterson2002,Korista1995,Clavel1991}. For example, for Model 1, we find a \hbeta\,response function centroid of $\tau_\eta\approx$62~days, which is significantly larger than the \hbeta\,lag predicted by either the empirically determined global radius-luminosity relationship \citep[][]{Bentz2013}, or by the single-object $L$(1350)\AA-\hbeta\,RL relationship for NGC~5548 as determined by \citet{Eser2015}. Much of the discrepancy with the pre-2014 RM results can be attributed to the variability behavior of the continuum source, as discussed in \S \ref{sec:ccf}.

The AGN STORM campaign detected anomalously short (given the source luminosity) BEL lags, of less than 10 days, for NGC 5548 in 2014. None of our $s=0$ models produce \hbeta\,response function centroids of less than $\sim50$ days (Figure \ref{fig:line_lags_model1}), nor do the $s=2$ models with $-2\le\log(U)\apprle-0.5$. While $s=2$ models with $\log(U)>-0.5$ do produce centroids at shorter delays ($\tau_\eta\approx10$ days for \hbeta), such models do not produce sufficient luminosity in \civ\,to match observations (Figure \ref{fig:line_lumins_s2}). 

\subsection{The Diffuse Continuum for Models 1 and 2}\label{sec:results_cont}

Using \cloudy, we generate radial surface emissivity distributions $\epsilon(r,\lambda)$ for the DC emission at more than 60 discrete wavelengths, sampling the UV to near-infrared regime from $\lambda\lambda$ 910 \AA\ -- 41137 \AA. This range encompasses the Lyman limit to just beyond the K-band, the longest wavelength band sampled from the ground in disk and dust reverberation mapping studies \citep[e.g.,][]{Kishimoto2007,Mandal2018}.

\begin{figure*}
	\centering
	\includegraphics[width=0.49\linewidth]{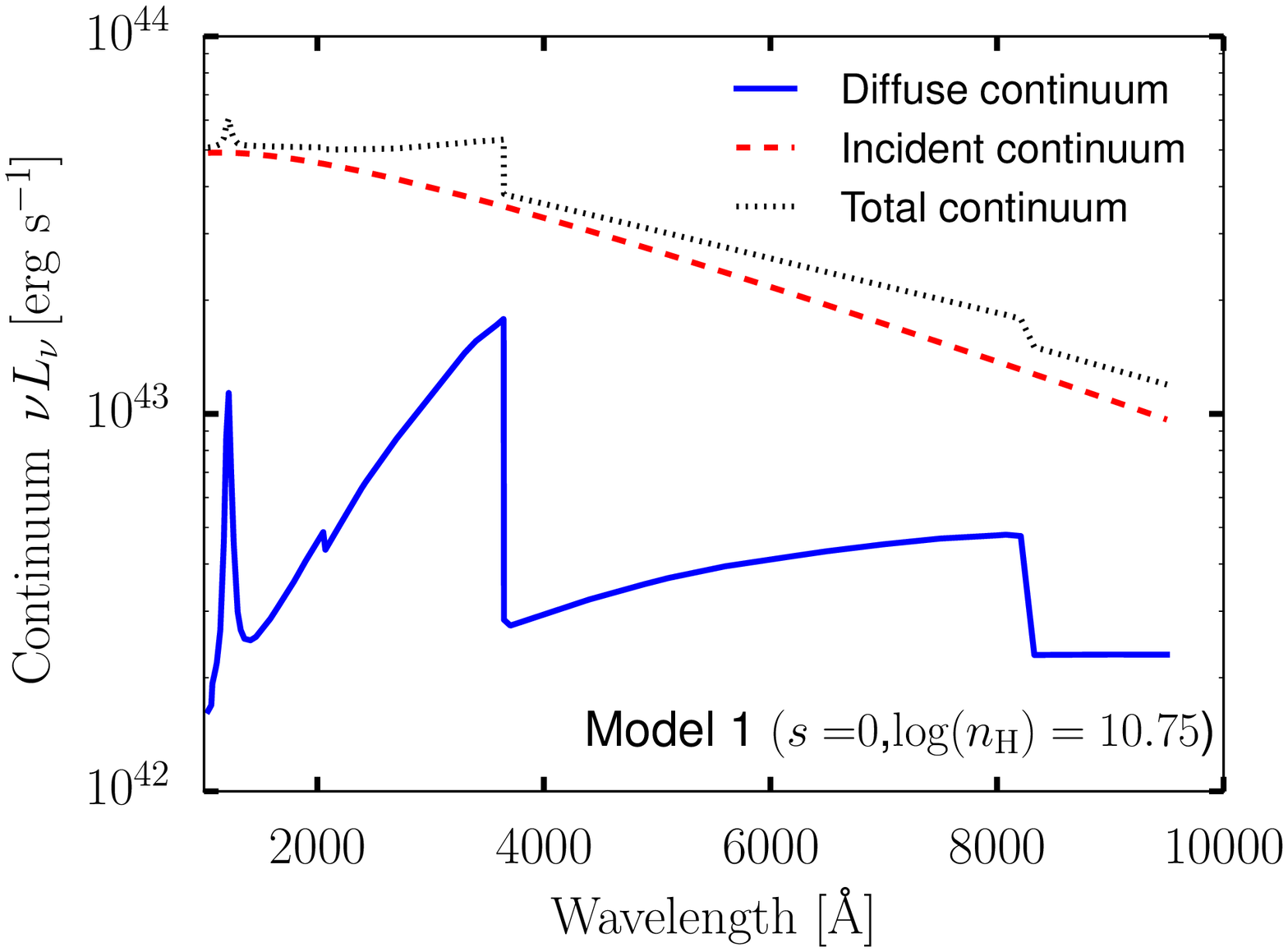}
	\includegraphics[width=0.49\linewidth]{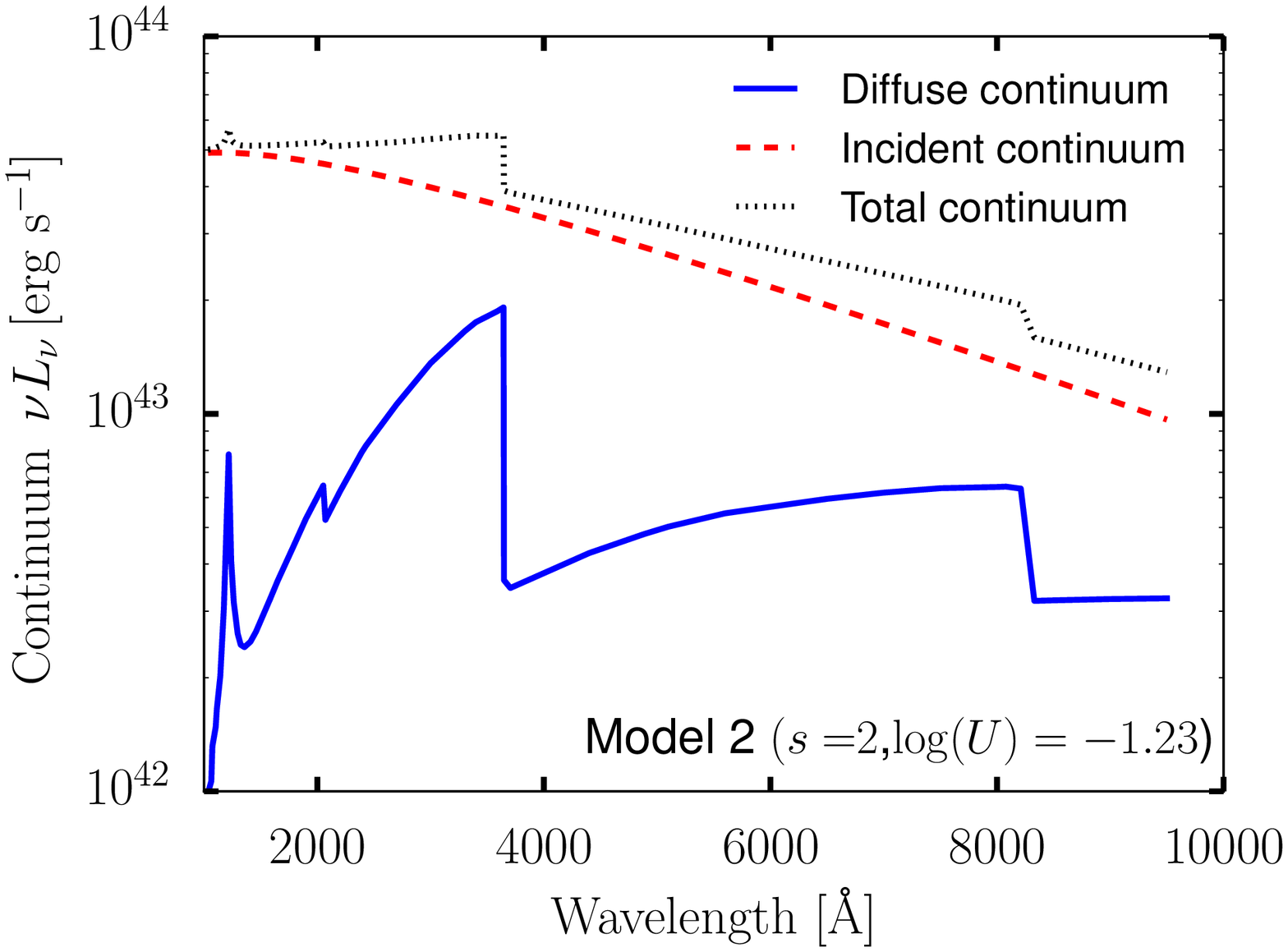}
	\includegraphics[width=0.49\linewidth]{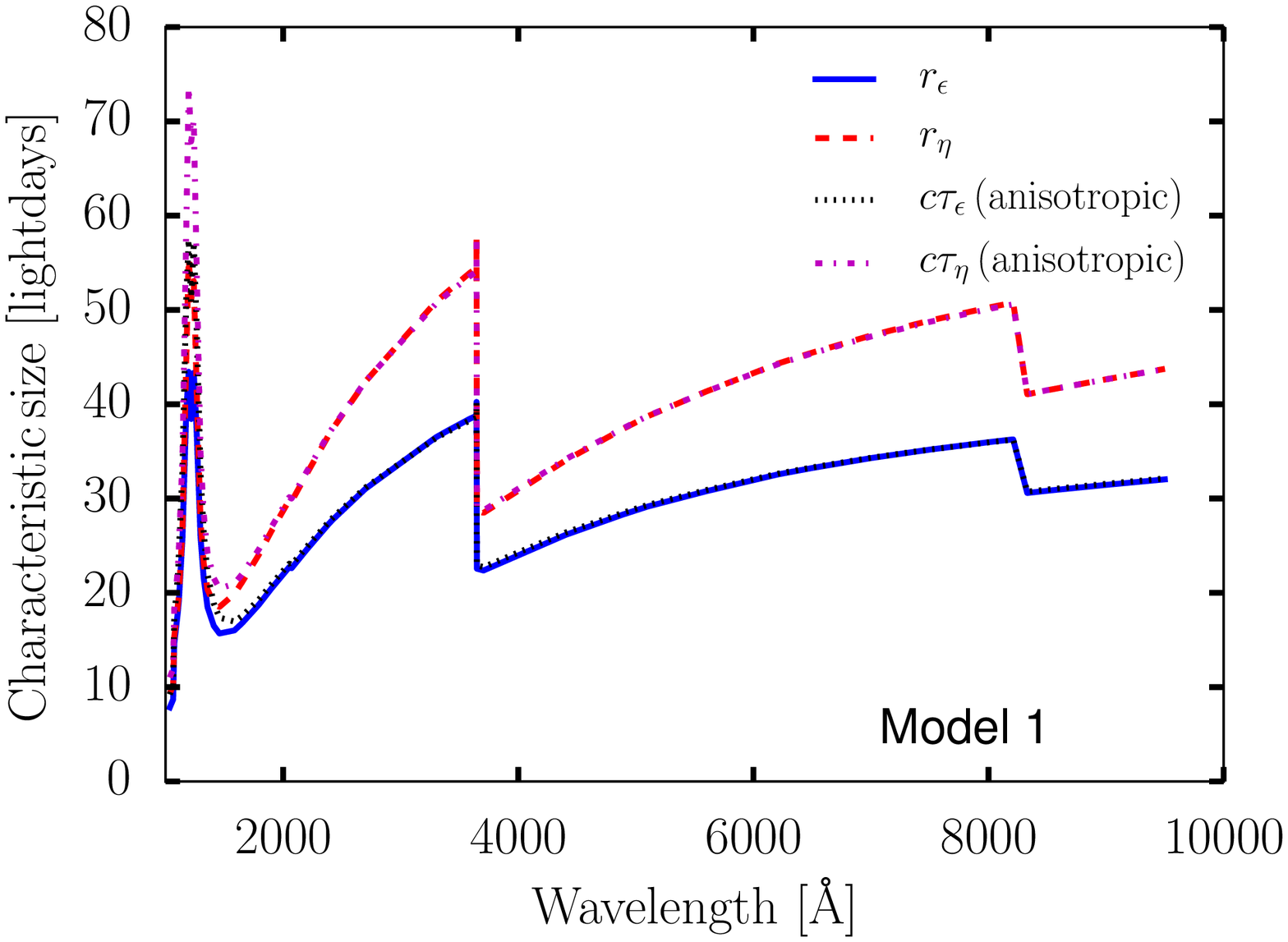}
	\includegraphics[width=0.49\linewidth]{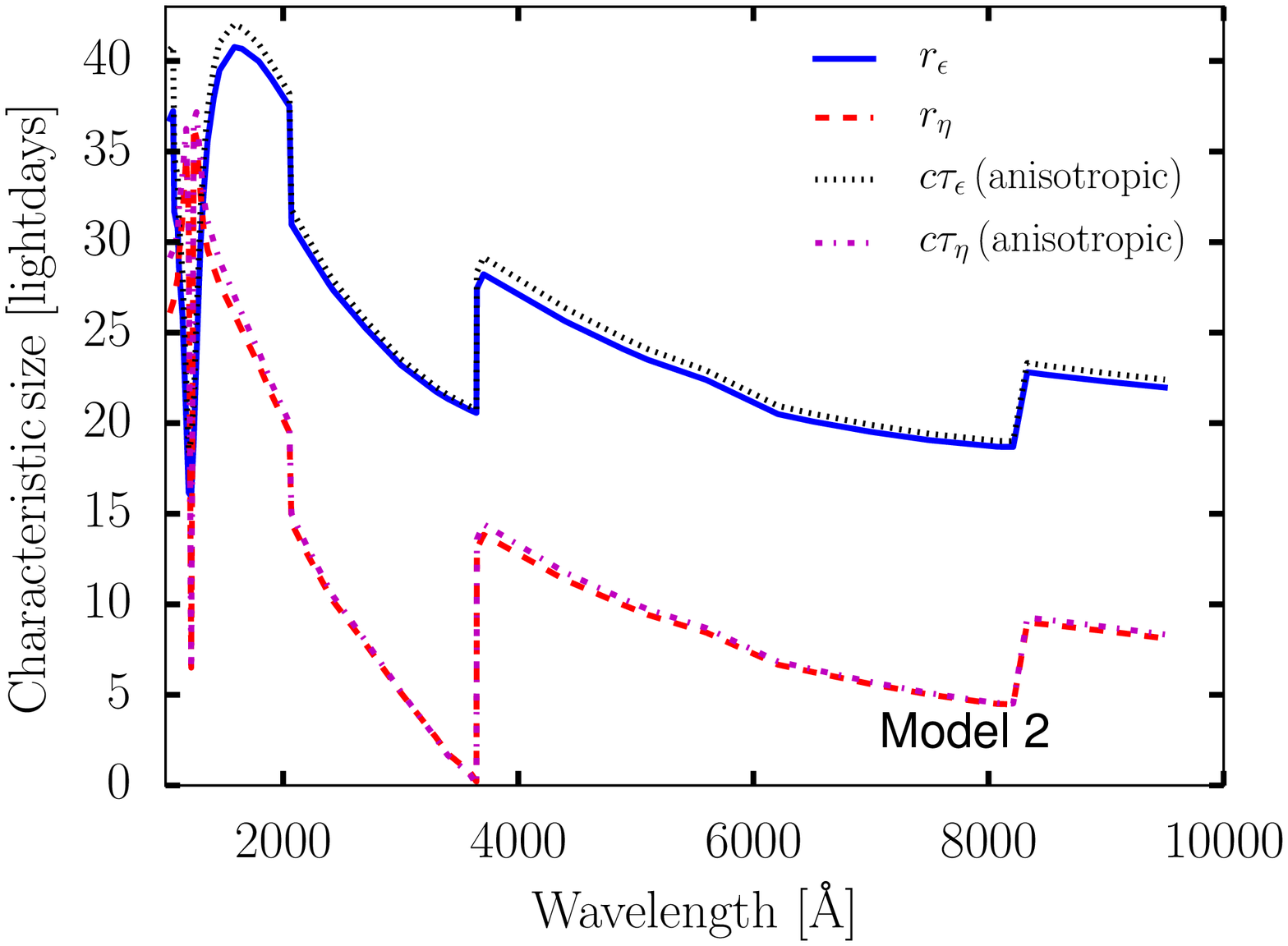}
	\caption{\emph{Top:} The specific luminosities $\mathrm{\lambda}L_{\mathrm{\lambda}}$ for the diffuse BLR continuum components, as a function of rest-frame wavelength, for Model 1 ($s=0$, left panel) and Model 2 ($s=2$, right panel). We also show the luminosities of the nuclear continuum as presented by \citet{Mehdipour2015}, and the total continuum that would be observed for these models, i.e., the sum of the nuclear and diffuse components.  \emph{Bottom:} The effective emissivity-weighted ($r_{\epsilon}$) and responsivity-weighted ($r_{\eta}$) radii in units of lightdays, and centroids of the emissivity--weighted ($\tau_{\epsilon}$) and responsivity-weighted ($\tau_{\eta}$) response functions in units of days, for the diffuse continuum response functions, as a function of wavelength. As the response functions describe the response of the BLR to a delta-function pulse in the continuum luminosity, their centroids do not take the variability properties of an AGN-like driving continuum into account; see Figure~\ref{fig:cont_lags} for the corresponding driven lags. We note that for both models $r_\epsilon$ and $\tau_\epsilon$ almost overlap, as do $r_\eta$ and $\tau_\eta$ - this implies that the diffuse continuum emits fairly isotropically at all radii.} 
	\label{fig:cont_models}
\end{figure*}

Each continuum wavelength is processed separately as per the methodology described in \S \ref{sec:method}. We first determine the DC luminosity $\nu L_\nu(\mathrm{diff.})$ at each wavelength, by integrating equation \ref{eq:lline} from \rin\,to \rout\, (Figure \ref{fig:cont_models}, top panels), replacing the emission line surface emissivity $\epsilon(r)$ with the monochromatic DC emissivity $\epsilon(r,\lambda)$. For both $s=0$ and $s=2$ models, the diffuse continuum represents a significant fraction of the total continuum (incident+diffuse), contributing up to $\sim$40\% of the measured continuum luminosity blue-wards of the Balmer jump, assuming that the nuclear source is fully covered by the BLR (more realistic covering fractions are explored in \S \ref{sec:discussion}). Furthermore, both models display strong Balmer and Paschen continua, rising towards longer wavelengths, with a significant drop in emission redwards of the Balmer ($\lambda$3648\AA) and Paschen ($\lambda$8204\AA) jumps.

Next we determine the wavelength-dependent response functions for each of the DC bands. We include the effects of responsivity (\S \ref{sec:method_radii}) and anisotropy (\ref{sec:method_anisotropy})  in the same fashion as for the BELs. The inclusion of anisotropic emission does not strongly affect the measured centroids for either model; this demonstrates that the DC emits fairly isotropically. The Model~1 ($s=0$) DC has a higher responsivity at large radii, so $\tau_\epsilon<\tau_\eta$ (Figure \ref{fig:cont_models}, bottom left). For Model 2 ($s=2$), the DC is more responsive in the inner regions, thus $\tau_\epsilon>\tau_\eta$ for that model (Figure \ref{fig:cont_models}, bottom right). We also highlight that the DC delays at wavelengths near $H\beta$ are \emph{ significantly smaller\/} (by a factor $\sim$2) than the delays predicted for the H$\beta$ BEL itself.

While the wavelength-dependence of the DC luminosities for the two models are broadly similar, their temporal behavior is markedly different.  For Model~1, the lags are larger blue-wards of the Balmer and Paschen jumps, while Model~2 displays the opposite behaviour, with elevated lags red-wards of the jumps. Whether such differences in temporal behaviour are ever realised in practice will depend on the fractional contribution of the DC emission to the total continuum emission $F_\mathrm{diff}$ (see \S \ref{sec:ccf_DC} for details). 

\subsection{Comparison of Diffuse Continuum Properties for Pressure-Law BLRs}\label{sec:results_cont_comparison}

As noted above, our single-component pressure-law models do not  reproduce the measured emission-line luminosities or their ratios for NGC~5548 in detail; multiple components (spanning a broad range in density and/or ionization) would be required. For this reason, we now extend the analysis of \S \ref{sec:results_cont} to encompass models spanning a broader range of \nh\,and $U$.

\subsubsection{Density Dependence}

\begin{figure*}
	\centering
	\includegraphics[width=0.985\linewidth]{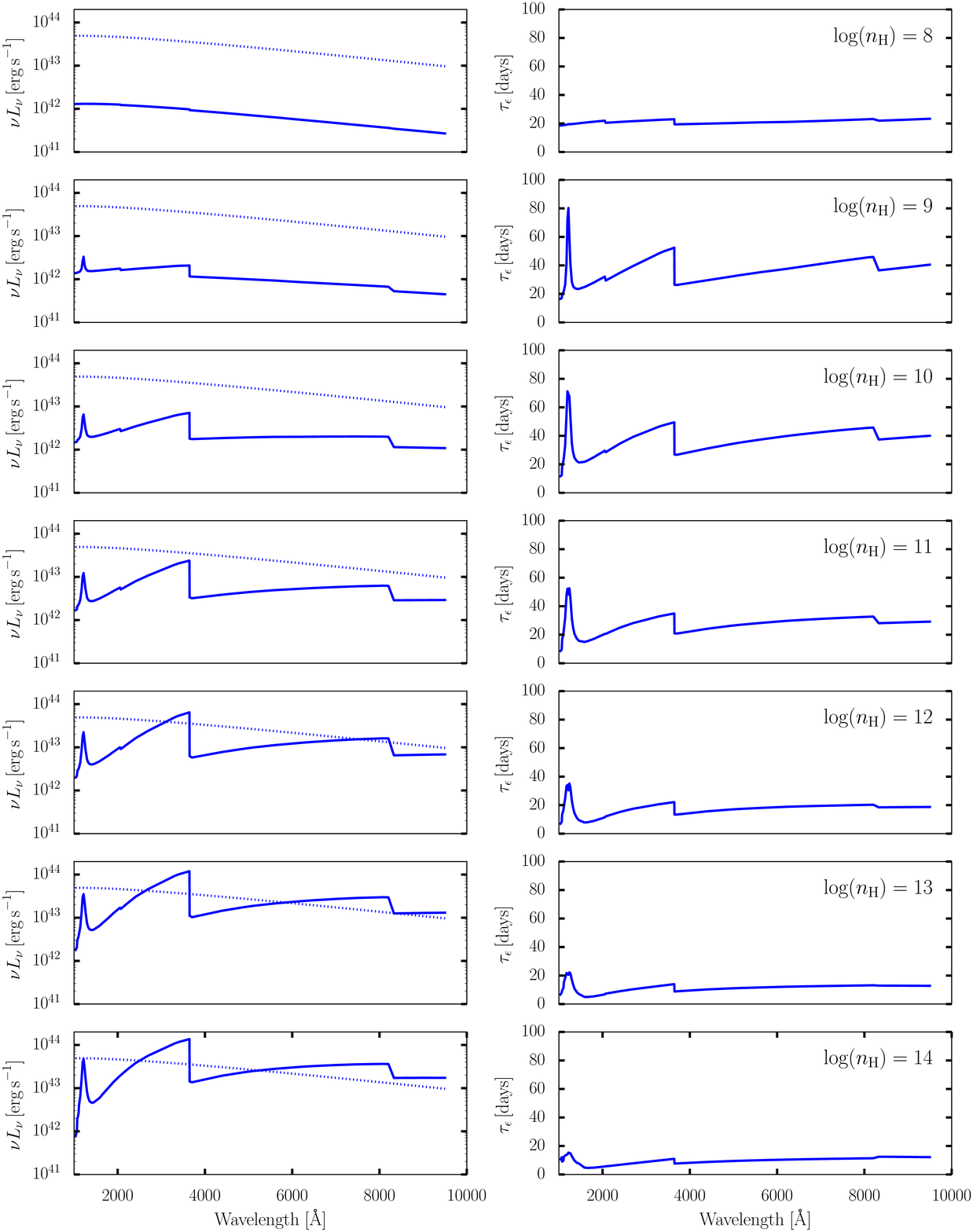}
	\caption{\emph{Left panels:} Specific luminosity of the diffuse (solid line) and incident (dashed line) continuum components, for $s=0$ models spanning a range of \nh, $8\le\lognh\le14$.The DC contribution is stronger, and the Balmer and Paschen features are more pronounced, at higher \nh. \emph{Right panels:} the corresponding wavelength-dependent response function centroids $\tau_{\eta}(\lambda)$ for the diffuse continuum, including the effects of anisotropy. For higher \nh, the DC is efficiently emitted closer to the black hole, producing shorter delays. As these quantities are measured from the response functions, they do not take the properties of the driving continuum into account; see Figure~\ref{fig:cont_nh_lags} for the corresponding driven lags.}
	\label{fig:cont_nh_lumin_resp}
\end{figure*}

We investigate the DC emitted by a range of $s=0$ models spanning $8\leq\lognh\leq14$, with other parameters identical to Model~1. The behaviour of the DC depends strongly on \nh. At low densities, the gas is highly ionized, and the reprocessed continuum is dominated by electron scattering, mostly free-free. For $\lognh=8$, the DC luminosity is almost two orders of magnitude smaller than that of the incident continuum, and is produced at radii of $\sim20$ light-days (Figure \ref{fig:cont_nh_lumin_resp}, top panels). As \nh\, increases, the DC emission becomes more prominent, its SED becomes dominated by the Balmer and Paschen continuum features (mostly free-bound recombination continuum), and it is produced at larger radii, $20\apprle\tau_{\eta}\apprle80$ light-days (Figure \ref{fig:cont_nh_lumin_resp}, middle panels). The delay centroid $\tau_\eta$ shows a strong wavelength-dependence, with longer lags blue-wards of the Balmer and Paschen jumps. At the highest \nh, ($\lognh > 10^{12}$~cm$^{-3}$) the BLR reprocesses so much of the ionizing continuum that, for full source coverage, the DC actually exceeds the brightness of the incident nuclear continuum over much of the UV--optical regime. At these high densities, unlike H$\beta$, the DC is dominated by gas at small BLR radii (high incident ionizing photon flux), and therefore responds on comparatively shorter timescales, $\tau_\eta<20$ days (see also Korista and Goad 2001, their Figures~1a, 1b).

\subsubsection{Ionization Parameter Dependence}

\begin{figure*}
	\centering
	\includegraphics[width=0.985\linewidth]{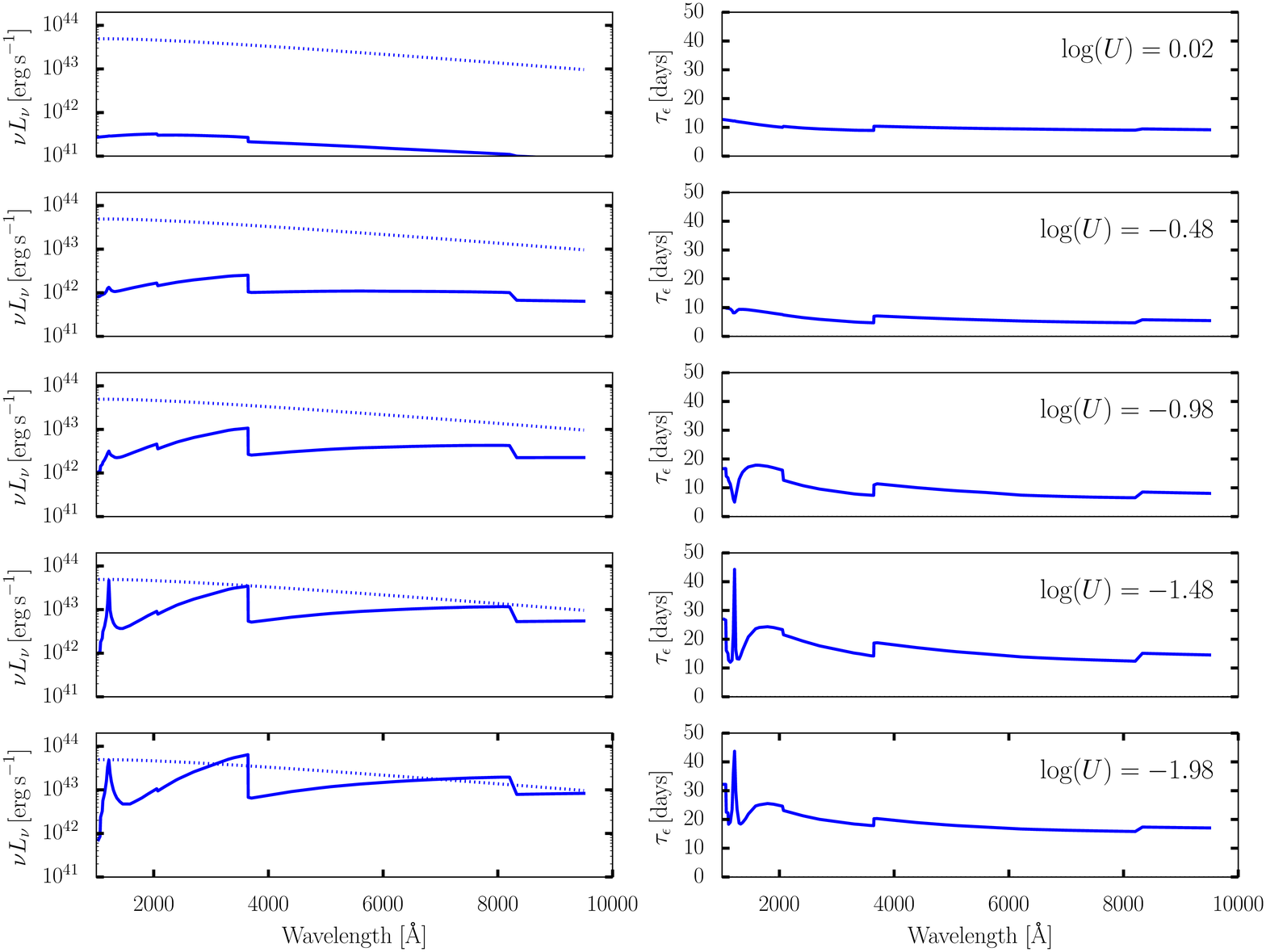}
	\caption{\emph{Left panels:} Specific luminosity of the diffuse (solid line) and incident (dashed line) continuum components, for $s=2$ models spanning a range of ionization parameter, $0.52\ge\lognh\ge-1.98$. The DC emits more brightly for models with lower $U$. \emph{Right panels:}  the corresponding wavelength-dependent responsivity-weighted centroid $\tau_{\eta}(\lambda)$ of the response function for the diffuse continuum, including the effects of anisotropy. The temporal behavior near the Balmer and Paschen breaks for the $s=2$ models is qualitatively different to the $s=0$ case (Figure \ref{fig:cont_nh_lumin_resp}). As these quantities are measured from the response functions, they do not take the properties of the driving continuum into account; see Figure~\ref{fig:cont_nh_lags_s2} for the corresponding driven lags.}
	\label{fig:cont_nh_lumin_resp_s2}
\end{figure*}

We investigate a range of $s=2$ models spanning $-1.98\leq\log(U)\leq0.52$, altering the density normalization at $r_{20}$, and keeping the \ncol\ normalization the same as before, i.e., a high $U$ model corresponds to a lower density normalization at $r_{\rm 20}$ (the highest $U$ model shown here has a density at $r_{\rm 20}$ of $\log n_{\rm h} = 9.0$~cm$^{-3}$).
%
For $s=2$ models, the response function centroids $\tau_{\eta}(\lambda)$ show the opposite trend in wavelength to $s=0$ models, that is, $\tau_\eta(\lambda)$ is slightly smaller blue-wards of the Balmer and Paschen jumps (Figure \ref{fig:cont_nh_lumin_resp_s2}). The DC contribution is not strongly sensitive to the ionization state of the gas over a broad range in $U$, as the gas is cooled primarily by line emission in this regime; the flux contours for the DC bands on the $n_{\rm H}$--$\Phi_{\rm H}$ plane are widely separated (e.g., Figure \ref{fig:cloudy_DC}). At high values of $U$ the reprocessed continuum is dominated by weak free-free continuum and electron scattering and thus resembles a power-law. This continuum is weak relative to the incident UV--optical continuum ($<$~1\% of the total), while the total line emission from such gas is insufficient to match the measured broad emission-line luminosities. 

%% file: sections/sec_ccf.tex
\section{Monte Carlo Modeling of BLR Lightcurves}\label{sec:ccf}

The values of $\tau(\lambda)$ presented thus far, for both the broad emission lines and reprocessed continuum, represent the "steady-state" values. In practice such values are seldom (if ever) realized because measured values depend not only on where the lines and continuum form, but also on the nature of the driving continuum (i.e. amplitude and characteristic variability timescale), and on the precise details of the monitoring program (e.g., campaign duration and sampling rate). See \citet{Goad2014} for details. A possible exception would be for a short sharp continuum event (i.e., a delta-function pulse in the continuum). Here, we simulate an ensemble of driving continuum light-curves that have similar statistical properties to those observed for NGC~5548, and combine these with the response functions of our BLR models to obtain simulated light-curves for the broad emission lines and diffuse continuum bands. We then measure the lags directly from the cross-correlation of our model light-curves with the driving continuum light-curve, mimicking a real observing situation. In general, this process results in measured delays that are significantly shorter than the "steady-state" values (as indicated by the centroid of the response function).

For comparison with recovered response functions, we here adopt a "locally linear" response approximation, estimating the marginal response of the emission lines (and diffuse continuum bands) to continuum variations about their steady-state (average) values. This approach is suitable (and computationally less expensive) for small continuum variations about the mean, but becomes progressively poorer as the amplitude of the continuum variations increases. Key assumptions underpinning this method include:

\begin{itemize}
\item the driving ionizing continuum source as seen from the BLR is point-like.
\item there is a linear relationship (though not necessarily 1:1) between the driving continuum variations and the emission-line (or DC) response.
\item the emission-lines (or DC) respond effectively instantaneously to local variations in the incident ionizing continuum flux.
\item no new material is added or destroyed.
\item the dominant timescale is the light crossing time.
\item as the driving continuum varies, only its amplitude changes, not its overall shape.
\end{itemize}

\noindent These assumptions are adopted for expediency only. For all but the highest ionization lines, the assumption of a point-like continuum source remains a valid approximation. SED variations can and do occur, but are generally secondary to changes in the ionizing continuum flux. A locally-linear response can be justified provided the amplitude of the continuum variations remain small. Under these conditions, the emission-line (or DC) light curve at time $t$, $L(t)$ can be represented by the convolution of the emission-line (or DC) transfer function $\Psi(\tau)$ with the driving ionizing continuum light-curve at previous times $C(t-\tau)$ : 

\begin{equation}
L(t)=\int_{-\infty}^{\infty} \Psi(\tau)C(t-\tau)d\tau \, .
\end{equation}

Typically, this equation is solved in its linearized form,

\begin{equation}
\Delta L(t) = \int_{-\infty}^{\infty} \Psi^{\prime}(\tau)\Delta C(t-\tau) d\tau \, ,
\end{equation}

\noindent with non-variable emission-line (DC) components consigned to a background term.

\subsection{Driving the BLR with a Model Continuum}\label{sec:method_driving}

On timescales of $\sim$ weeks to months, the continua of AGN vary approximately as a damped random walk (DRW) in the logarithm of the flux \citep[e.g.,][]{Kelly2009,Kozlowski2010,Zu2013,Edelson2014}. A DRW process is characterized by a variability amplitude $\sigma_{\mathrm{DRW}}$, along with a damping (or characteristic) timescale $T_{\mathrm{char}}$ upon which it tends to return to the mean. For model BLRs driven by DRW continua, \citet{Goad2014} perform an in-depth investigation of the influence of both the observational constraints (campaign length, observational cadence), and the DRW parameters, on the measured lags. The lag for emission line lightcurves produced by their model BLR, as measured using the interpolated cross-correlation function (ICCF) method \citep{White1994}, is strongly dependent on $T_{\mathrm{char}}$ of the input continuum, in the sense that short characteristic timescales lead to underestimated lags (relative to the centroid of the response function). The measured lags approach the expected (i.e., steady-state) values for $T_{\mathrm{char}}>\rout/c$.

The continuum variability displayed by NGC~5548 over the period 1989--1993 can be described as a DRW with characteristic timescale $T_\mathrm{char}\approx 40$~days and signal variance $\sigma_{DRW}=0.04$ \citep{Collier2001}. This damping timescale is considerably shorter than the maximum light travel times for our model BLR (i.e., $2\rout/c=280$ days). It is therefore important to take the continuum variability behavior into account when comparing our model lags to measured values. To this end, we generate DRW continuum light-curves with which to drive the BLR. For uniform sampling, the DRW is equivalent to a discrete, autoregressive AR(1) process \citep[][as detailed in their Appendix]{Kelly2009}; we generate light-curves with a cadence of 1 rest-frame day, using the AR(1) algorithm to obtain the logarithm of the flux in each subsequent time-step. Unless otherwise indicated, the DRW continuum light-curves used in this work are generated using $T_\mathrm{DRW}=40$ days and $\sigma_{DRW}=0.04$, with a campaign duration of 500 rest-frame days.

The BEL and DC light-curves, as driven by this ionizing continuum, are then generated by convolving the DRW light-curve with the relevant BLR response function (\S \ref{sec:method_radii}) representative of a particular emission line or diffuse continuum band. We show an example DRW continuum light-curve, along with the driven BEL light-curves for four emission line species (Ly$\alpha$, C~{\sc iv}, H$\beta$ and Mg~{\sc ii}), in Figure \ref{fig:drivelines1}. These indicate a broad range in amplitude and delay, i.e, a stratified and spatially extended BLR, as is commonly observed.

\subsection{Lag Determination via Cross-Correlation}\label{sec:method_measuring}

Given the continuum and the BEL (or DC) light-curves described in \S \ref{sec:method_driving}, we measure the \emph{as-observed} lag, following the interpolated cross-correlation function (ICCF) method \citep{White1994}, commonly used to determine lags for real RM campaigns \citep[e.g.,][]{Eser2015,Lu2016,Pei2017}. We measure both the peak value of the cross-correlation function ($\tau_\mathrm{CCF,peak}$) and the centroid ($\tau_\mathrm{CCF,cent}$), determining the centroid over the range in which the CCF coefficient exceeds 80\% of its peak value. We primarily work with $\tau_\mathrm{CCF,cent}$ in our analysis, as this measure is less susceptible to bias towards shorter lags \citep{Perez1992a}. The detailed shape of the CCF depends strongly on the specific realization of the DRW continuum (i.e., the continuum autocorrelation function, ACF), which is stochastic in nature. We therefore generate 10000 realizations of the DRW continuum for each of our model BLRs, and determine the CCF peak and CCF centroid for each realization. This allows us to determine the mean and standard deviation of $\tau_\mathrm{CCF,cent}$ for each BEL and each diffuse continuum waveband, holding the DRW parameters $T_\mathrm{char}$ and $\sigma_{DRW}$ constant. Quoted lags (Table 2, columns 7 and 8) represent the mean and standard deviation of the lag distributions (peak and centroid), given 10000 DRW continuum realizations. We assume measurement errors of 1\% for the driving continuum lightcurve, and of 5\% for the BEL (or DC) lightcurve.

\subsection{CCF Lags for the Emission Lines}\label{sec:ccf_BEL}

\begin{figure*}
	\centering
	\includegraphics[width=0.49\linewidth]{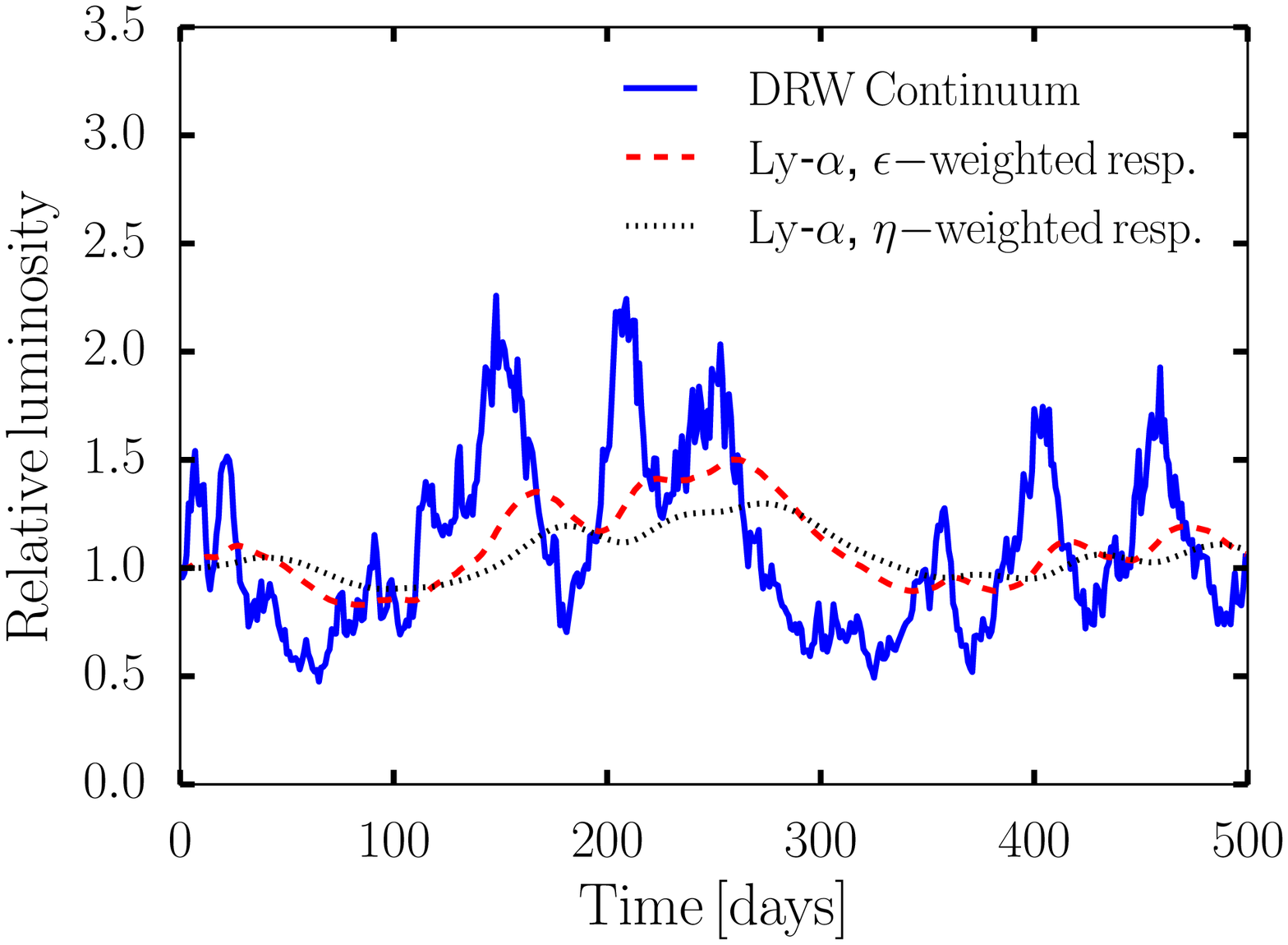}
	\includegraphics[width=0.49\linewidth]{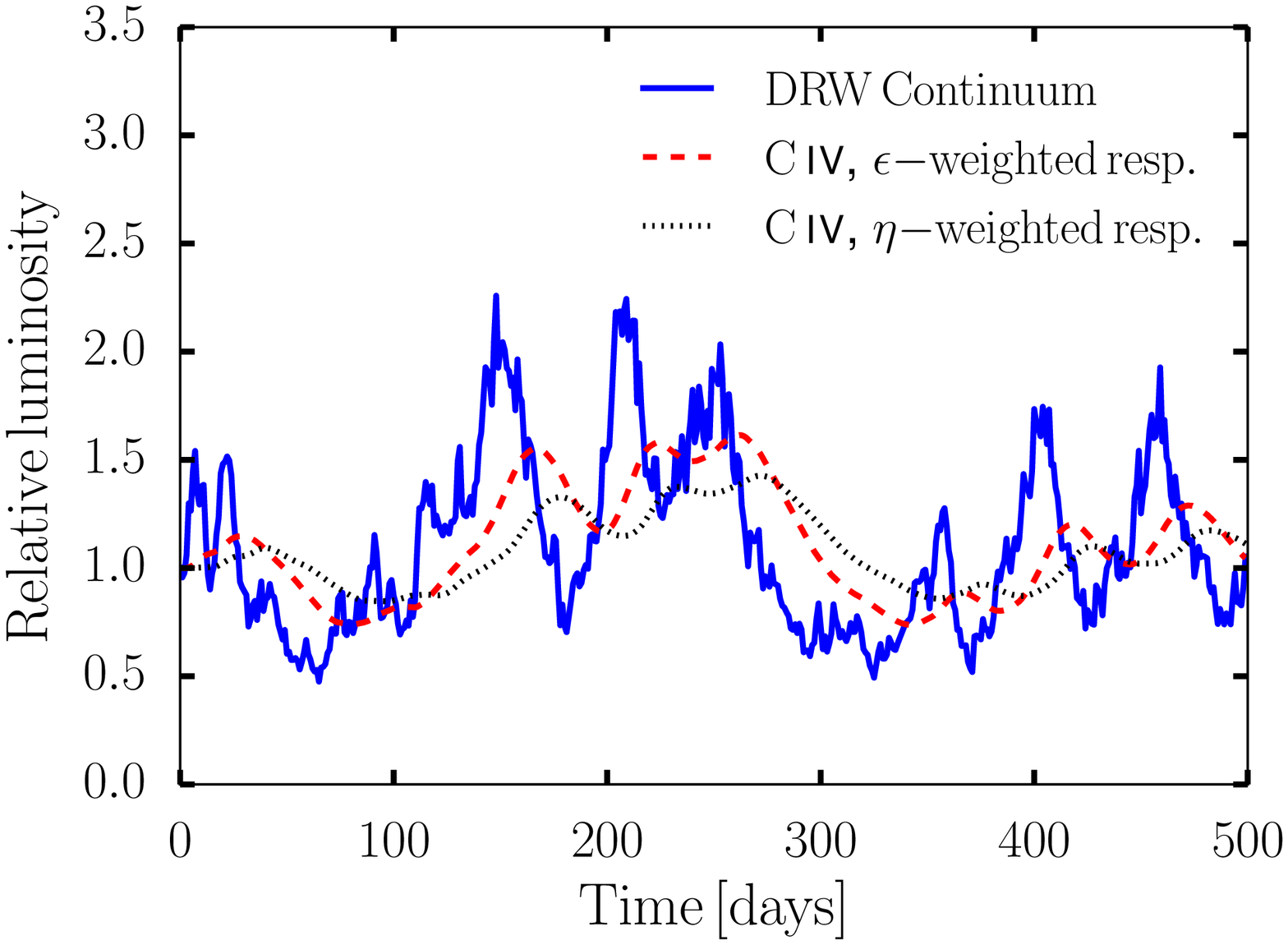}
	\includegraphics[width=0.49\linewidth]{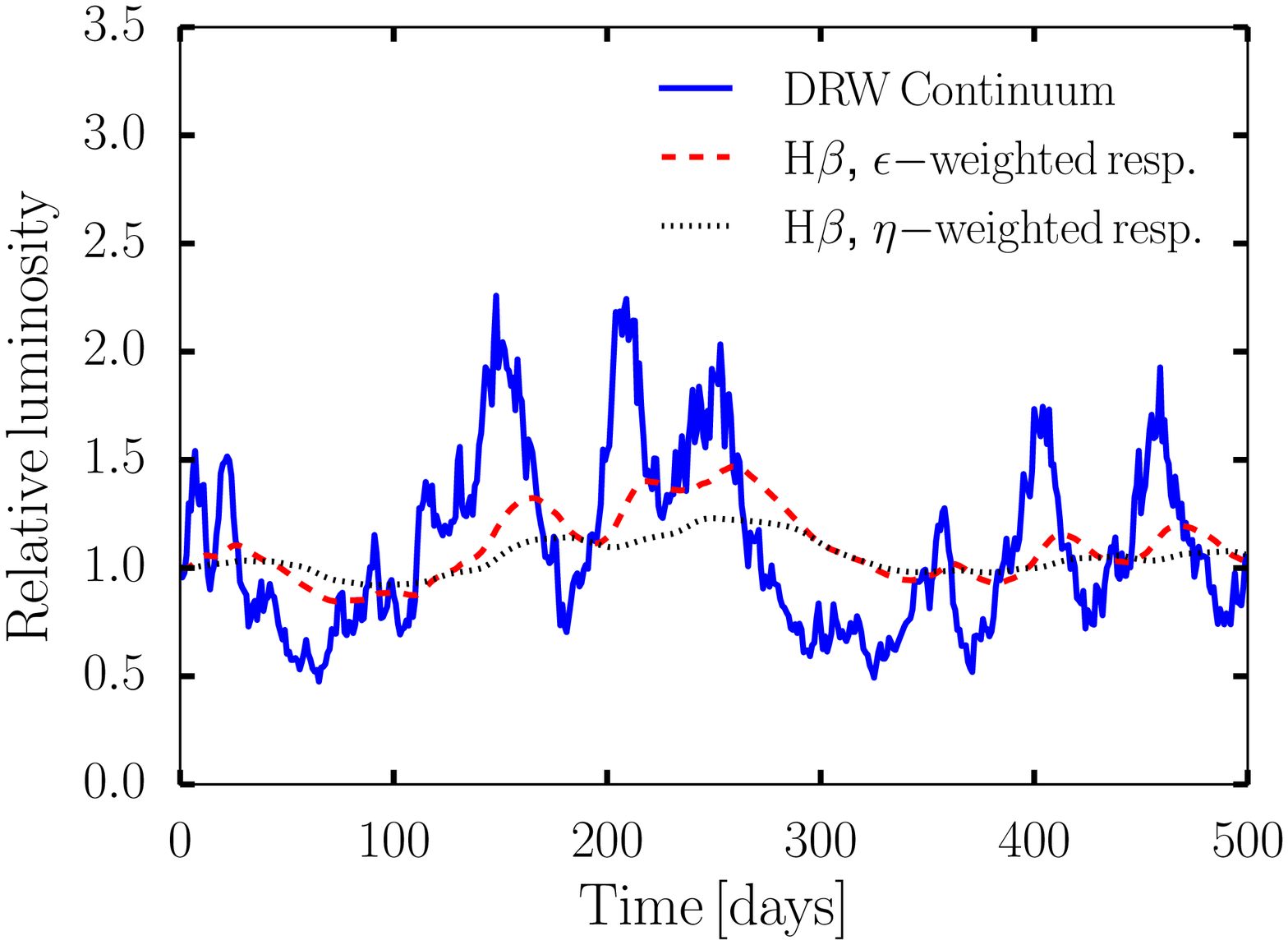}
	\includegraphics[width=0.49\linewidth]{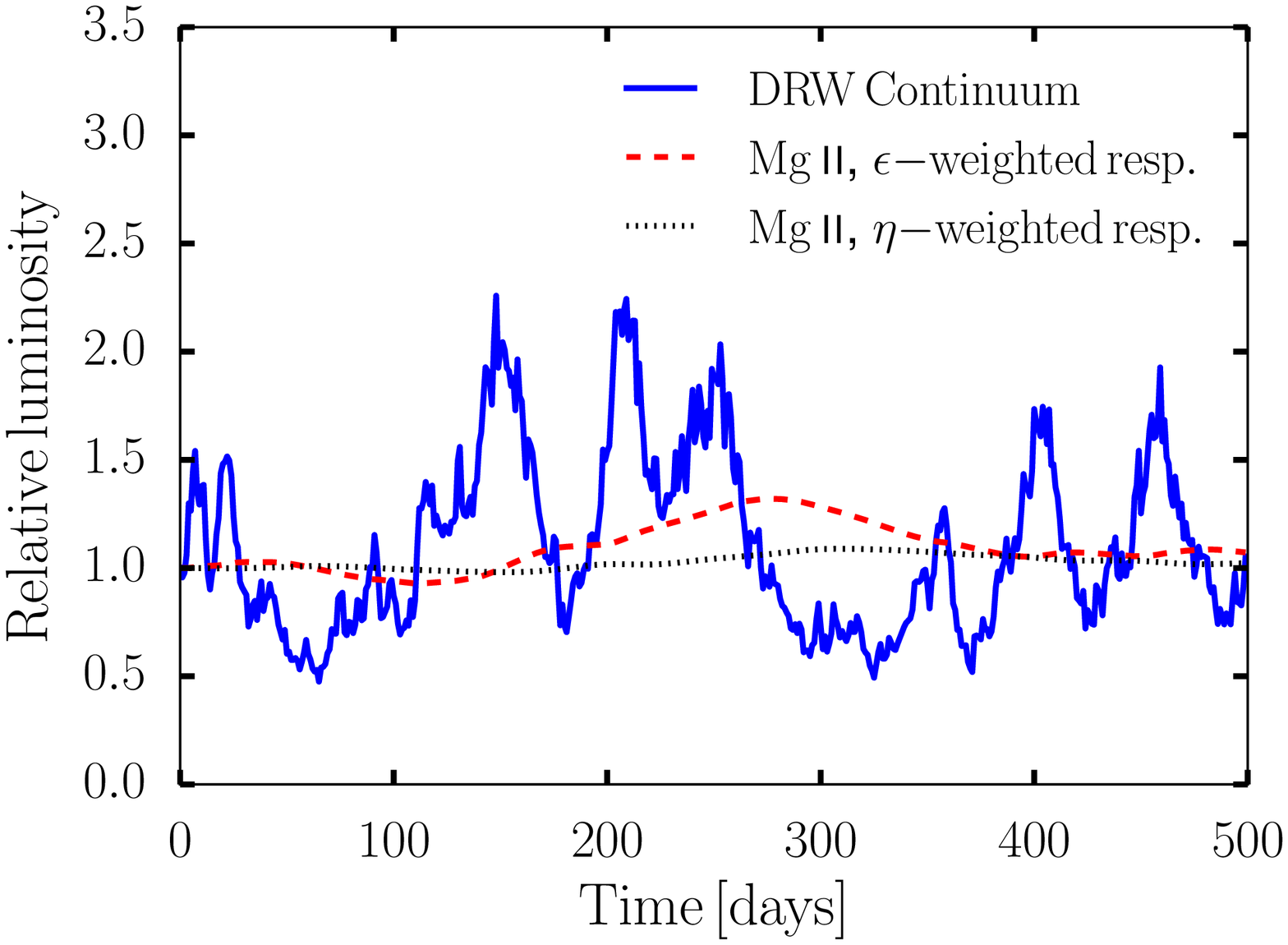}
	\caption{Driving the BELs for model~1  using a Damped Random Walk continuum with $T_{\mathrm{char}}=40$ days and $\sigma_{\mathrm{char}}=0.4$ (\S \ref{sec:method_driving}). Here, we show one example realization of the DRW continuum; in order to determine the average delays and their statistical uncertainties, we generate 10000 realizations of this continuum lightcurve. Including the effects of anisotropic emission generally increases the observed lag and decreases the response amplitude. As an extreme example, the \mgii\,BEL responds very weakly in our model when anisotropic emission is taken into account (i.e., the $\eta$-weighted lightcurve).}
\label{fig:drivelines1}
\end{figure*}

We show light-curves for four Model 1 BEL, as driven by a single realization of the DRW continuum, in Figure \ref{fig:drivelines1}. For each BEL we show two light-curves, one of which is generated for a response function that includes the effects of line responsivity (\S \ref{sec:method_responsivity});  these effects tend to delay and weaken the line response. Of the lines tested, the \civ\,BEL shows the strongest response, while the response of \mgii\,is negligible once the line responsivity is taken into account. A weak response for Mg~{\sc ii} appears to be a general property of photoionization model calculations \citep[e.g.,][]{Goad1993}. Observationally, \citet{Cackett2015} find for NGC 5548 that the Mg~{\sc ii} broad-line luminosity does not correlate with continuum variations. For a sample of 68 quasars with multiple SDSS observations, \citet{Zhu2017} find that Mg~{\sc ii} responds only weakly to continuum variations ($\eta\sim0.46$). These findings may preclude the use of this line for reverberation mapping studies, though we note that a handful of significant lag measurements have been reported for this line \citep[e.g.,][]{Metzroth2006,Shen2016}. 
Further, while Mg~{\sc ii} line widths have been used as proxies for the BLR velocity field in single-epoch $M_\mathrm{BH}$ estimates \mbox{\citep[e.g.,][]{Vestergaard2009,Shen2011}}, for a fraction of objects, the Mg~{\sc ii} line width is not representative of the H$\beta$ line width, as is often assumed \citep{VEST2011}.

We present the BEL CCF centroids for Models 1 and 2 in Table \ref{tab:modellags}. Much of the discrepancy between the response function centroids (\S \ref{sec:results_linelags}) and the pre-2014 observed lags can be attributed to the variability behaviour of the continuum source: for \hbeta\,we retrieve a lag of $\tau_{\mathrm{CCF,cent}}=33.3\pm7.2$ days (Model 1), or $\tau_{\mathrm{CCF,cent}}=27.6\pm12.8$ (Model 2). Both are formally consistent with the longest \hbeta\,delays measured historically for NGC 5548 \citep[$26.9^{+1.5}_{-2.2}$ days,][]{Zu2011}, which are observed at continuum luminosities similar to those of the 2014 campaign. 

The lags obtained using the CCF method are, however, still significantly larger than the values obtained by \citet{Pei2017} for the 2014 campaign. The damping timescale is considerably shorter during the 2014 campaign ($\approx$12 days, as measured from the Structure function by MG). We therefore test the sensitivity of the measured \hbeta\,lags to the damping timescale of our model continuum. While larger values of $T_{\mathrm{DRW}}$ do yield larger lags, approaching $\tau_\eta$ \citep[as also found by][]{Goad2014}, we find CCF centroids of $\tau_{\mathrm{CCF,cent}}\sim30$ days (albeit with a larger standard deviation of the lag distribution) even for small $T_\mathrm{DRW}\sim5$ days. Thus, our pressure law models cannot reproduce the very short lags observed in 2014, even allowing for an abnormally short characteristic timescale for the driving continuum.

\subsection{CCF Lags for the Diffuse Continuum}\label{sec:ccf_DC}

\begin{figure*}
	\centering
	\includegraphics[width=0.49\linewidth]{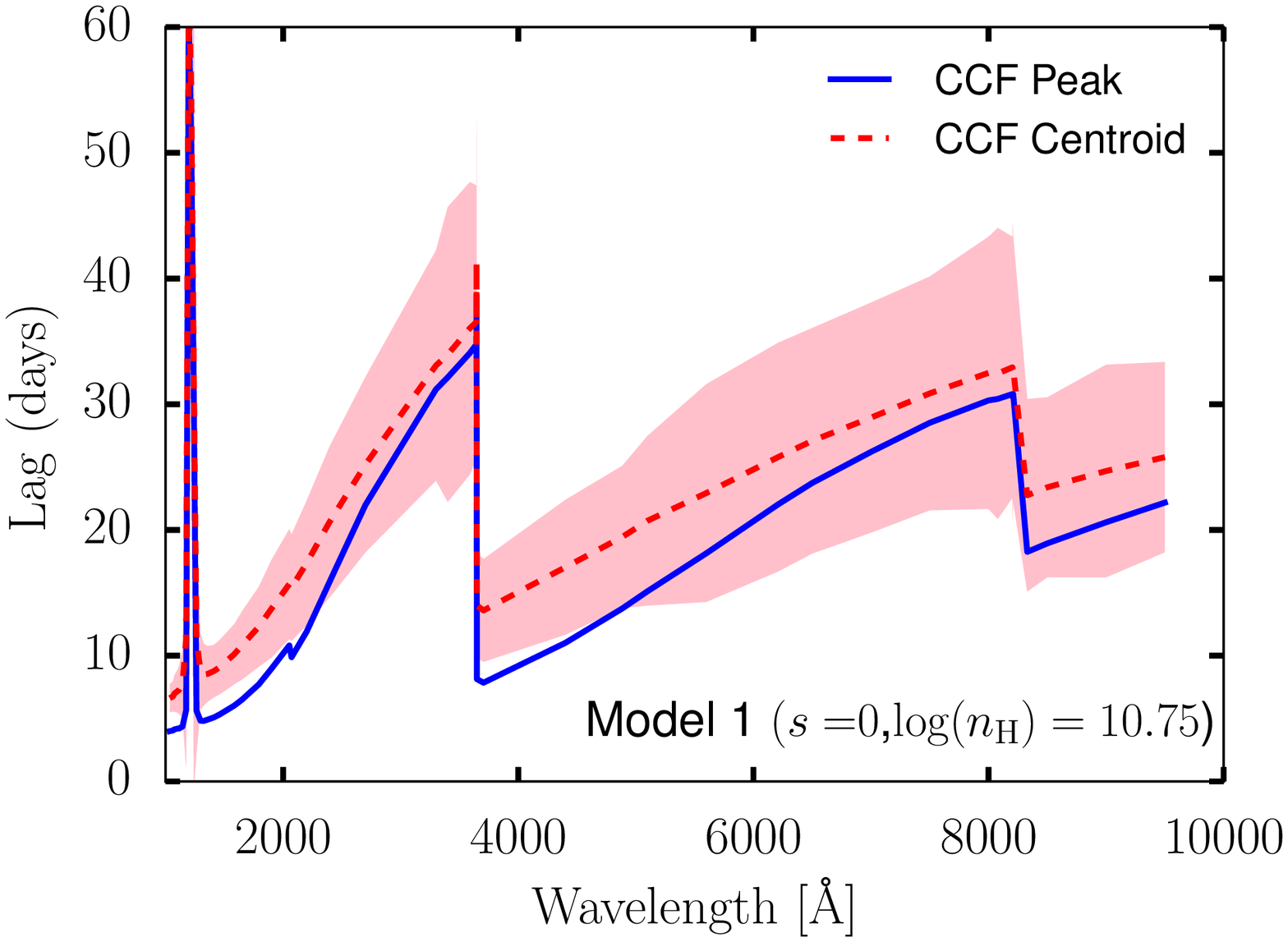}
	\includegraphics[width=0.49\linewidth]{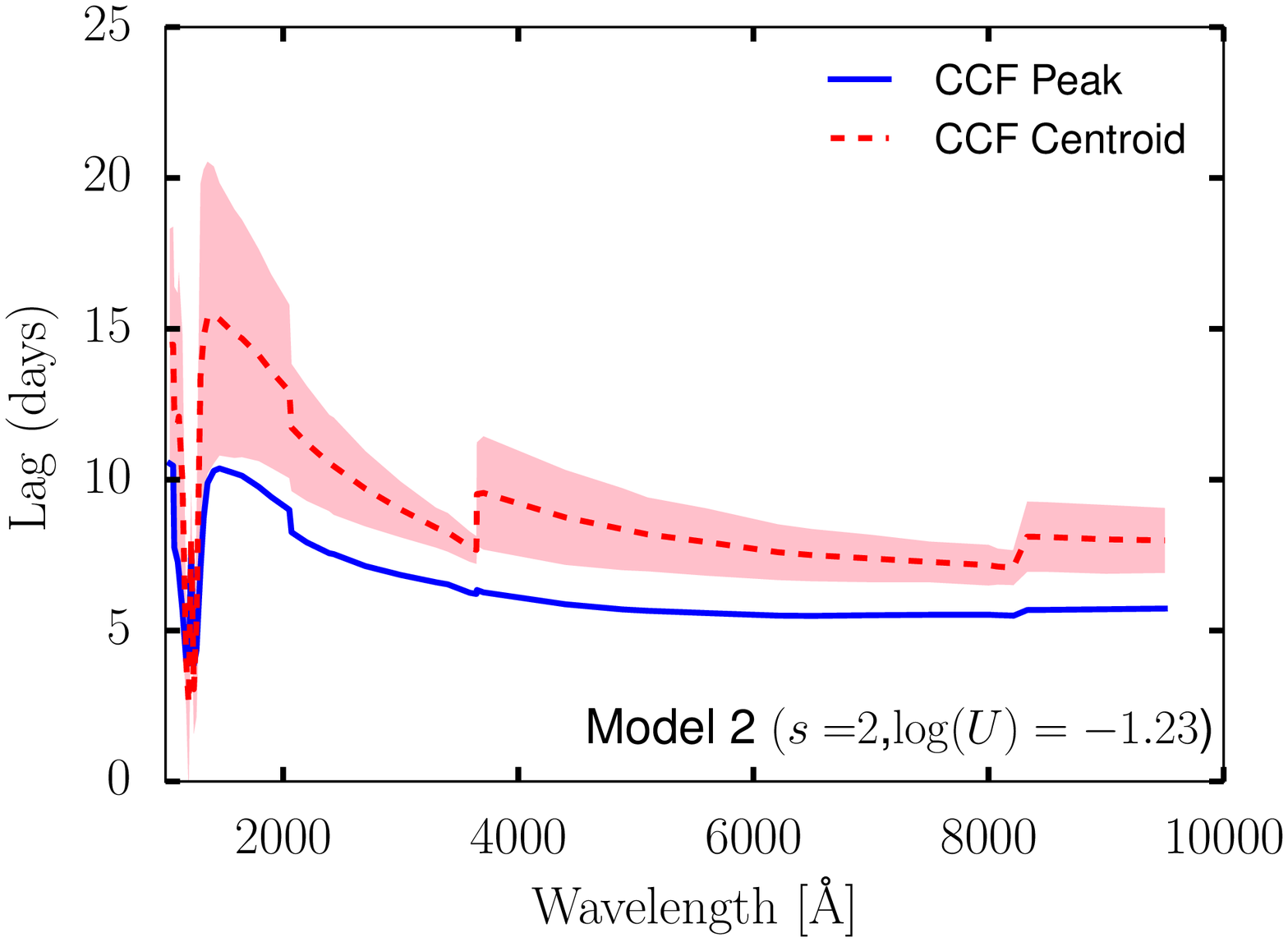}
	\includegraphics[width=0.49\linewidth]{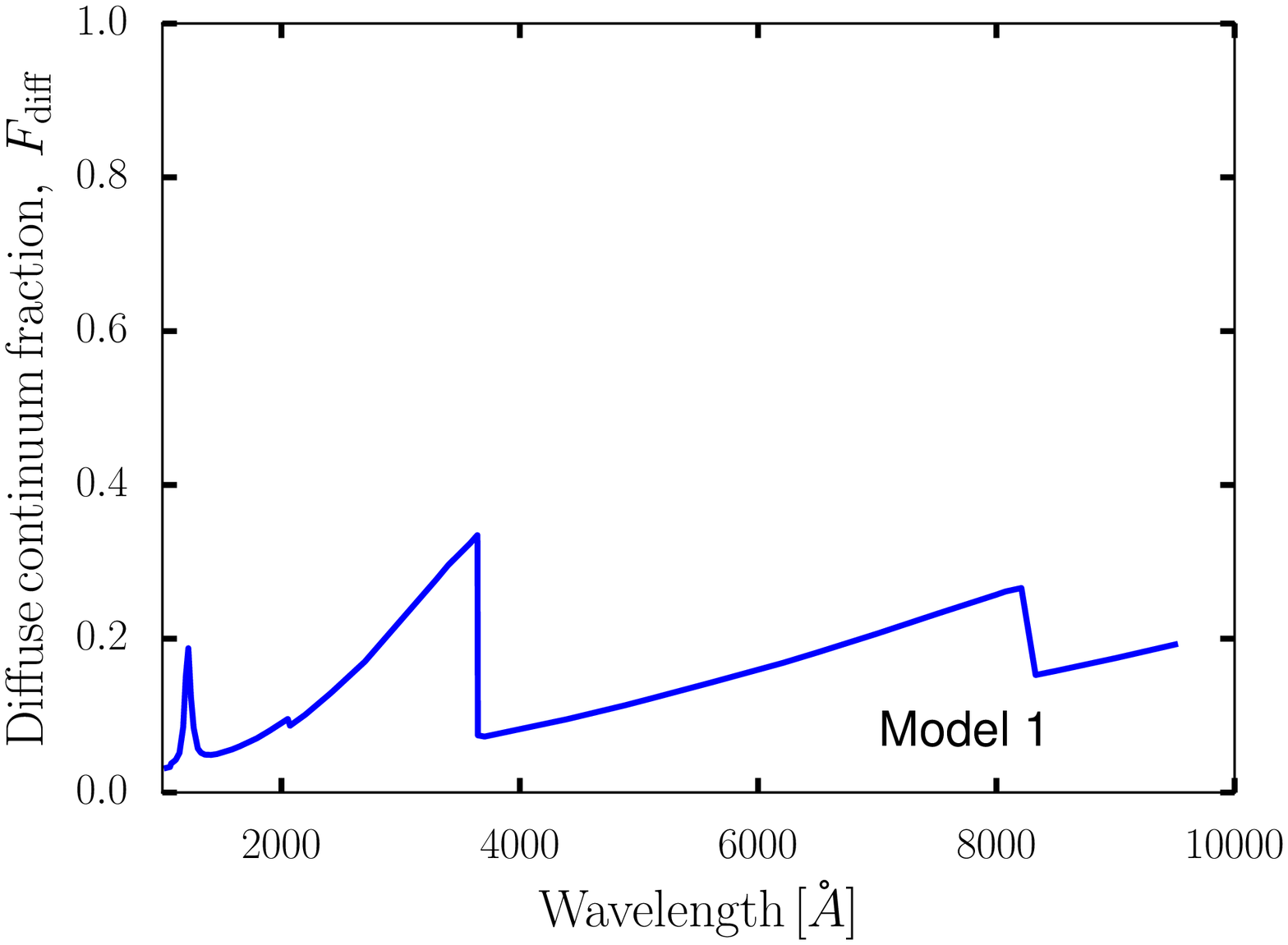}
	\includegraphics[width=0.49\linewidth]{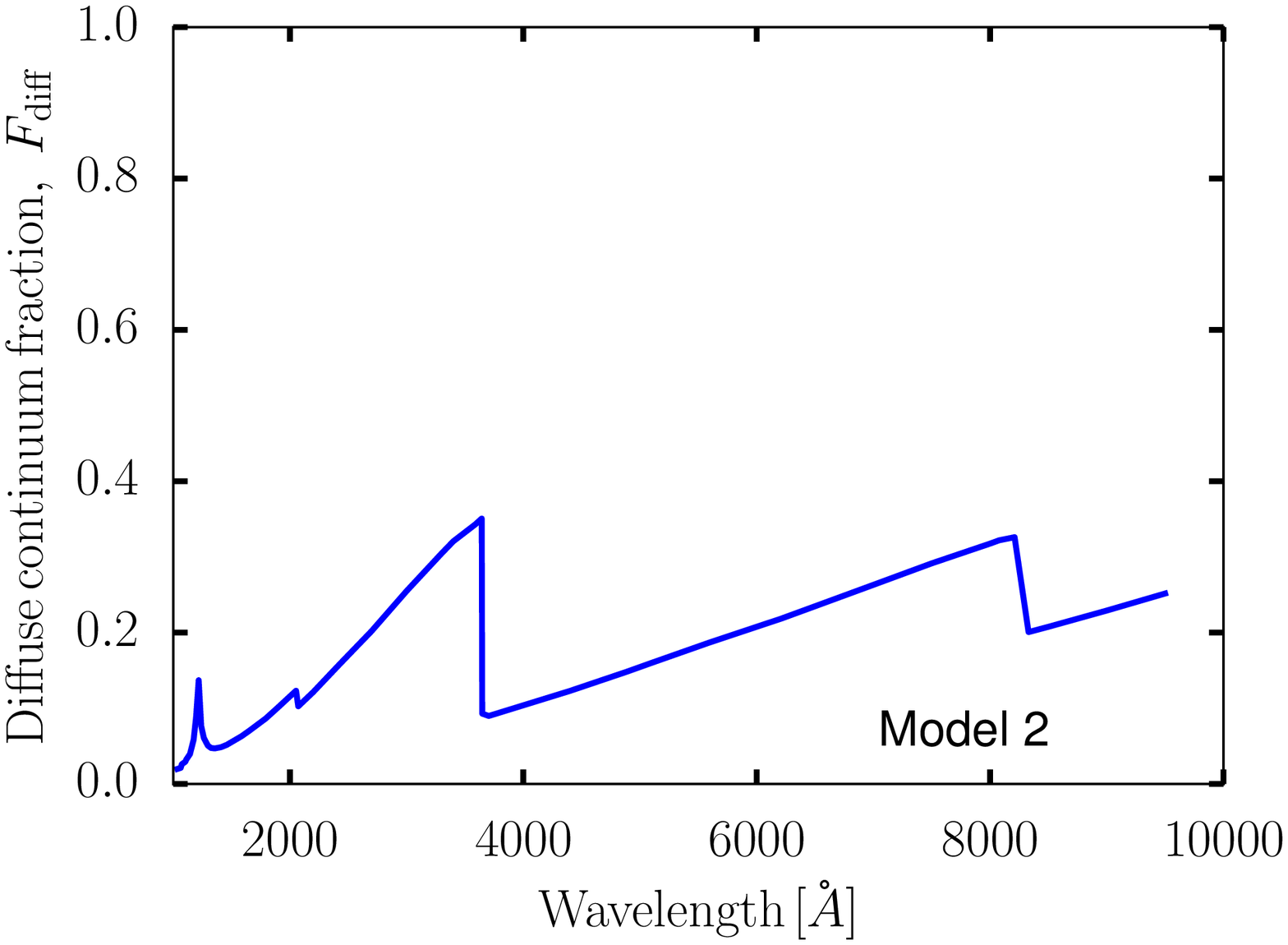}
	\caption{\emph{Top:} CCF peak lags and centroids for the BLR diffuse continuum component, in units of days, for Model 1 ($s=0$, left panel) and Model 2 ($s=2$, right panel). To obtain the lags, we drive the steady-state BLR models using a nuclear continuum light-curve with the logarithm of the flux following a damped random walk (DRW) with $T_\mathrm{DRW}=40$ days and $\sigma_{\mathrm{DRW}}=0.04$, similar to the continuum variation properties observed for NGC 5548 during 1989-1993. We repeat this simulation 10000 times; the lags presented are the mean values of the lag distribution at a given wavelength. The pink shaded region displays the 1$\sigma$ standard deviation of the CCF centroid distribution at each wavelength. \emph{Bottom:} $F_\mathrm{diff}$, the ratios of diffuse continuum $\mathrm{\lambda}L_{\mathrm{\lambda}}$ (DC), to total continuum $\mathrm{\lambda}L_{\mathrm{\lambda}}$(incident$+$DC), as a function of wavelength.} 
\label{fig:cont_lags}
\end{figure*}

For Models 1 and 2, the recovered CCF centroids for the DC ($\tau_\mathrm{CCF,cent}$, Figure \ref{fig:cont_lags}, top panels) are a factor $\sim2$ smaller than the corresponding response function centroids as derived in \S \ref{sec:results_cont}. This is the lag signal we would measure for the DC emission if it could be isolated from the total continuum light-curve (incident + diffuse + constant components). To first order, and assuming that the nuclear continuum has zero lag at the wavelength of interest, we can determine a rough estimate for the \emph{measured} continuum lag as the product of the diffuse continuum lag and the diffuse continuum fraction, $\tau_\mathrm{CCF,cent}\times F_{\mathrm{diff}}$, where $F_{\mathrm{diff}}=L_\nu(\mathrm{diff.})/(L_\nu(\mathrm{diff.})+L_\nu(\mathrm{nuc.}))$. For example, if the DC component dominates the measured continuum flux we expect to measure a delay for the measured continuum bands similar to that found for the DC. Conversely, for a weak DC component, the continuum flux is dominated by the (in this case lag-less) disk component, yielding zero delay. For intermediate DC contributions, the measured continuum delay will lie somewhere between these two extremes, depending on the exact details of the driving continuum variations.

Given values of $F_\mathrm{diff}$ in the range 0.1--0.3 (Figure \ref{fig:cont_lags}, bottom panels), the continuum for Models 1 and 2 will thus lag the incident continuum by several days at wavelengths short-ward of the Balmer and Paschen jumps. These estimates of the lag assume full coverage of the continuum source by the BLR. The ratio $F_{\mathrm{diff}}$, and thus the observed lag will, according to this prescription, scale linearly with source covering fraction. The measured continuum lag may therefore be smaller than that shown in Figure \ref{fig:cont_lags} by a factor $\sim$few. Note that here we neglect any intrinsic inter-band continuum lag due to the accretion disk geometry. We address both these issues in \S \ref{sec:discussion}.

\subsubsection{Dependence of CCF Lags on \nh\,for $s=0$:}\label{sec:CCF_nh_range}

\begin{figure*}
	\centering
	\includegraphics[width=0.99\linewidth]{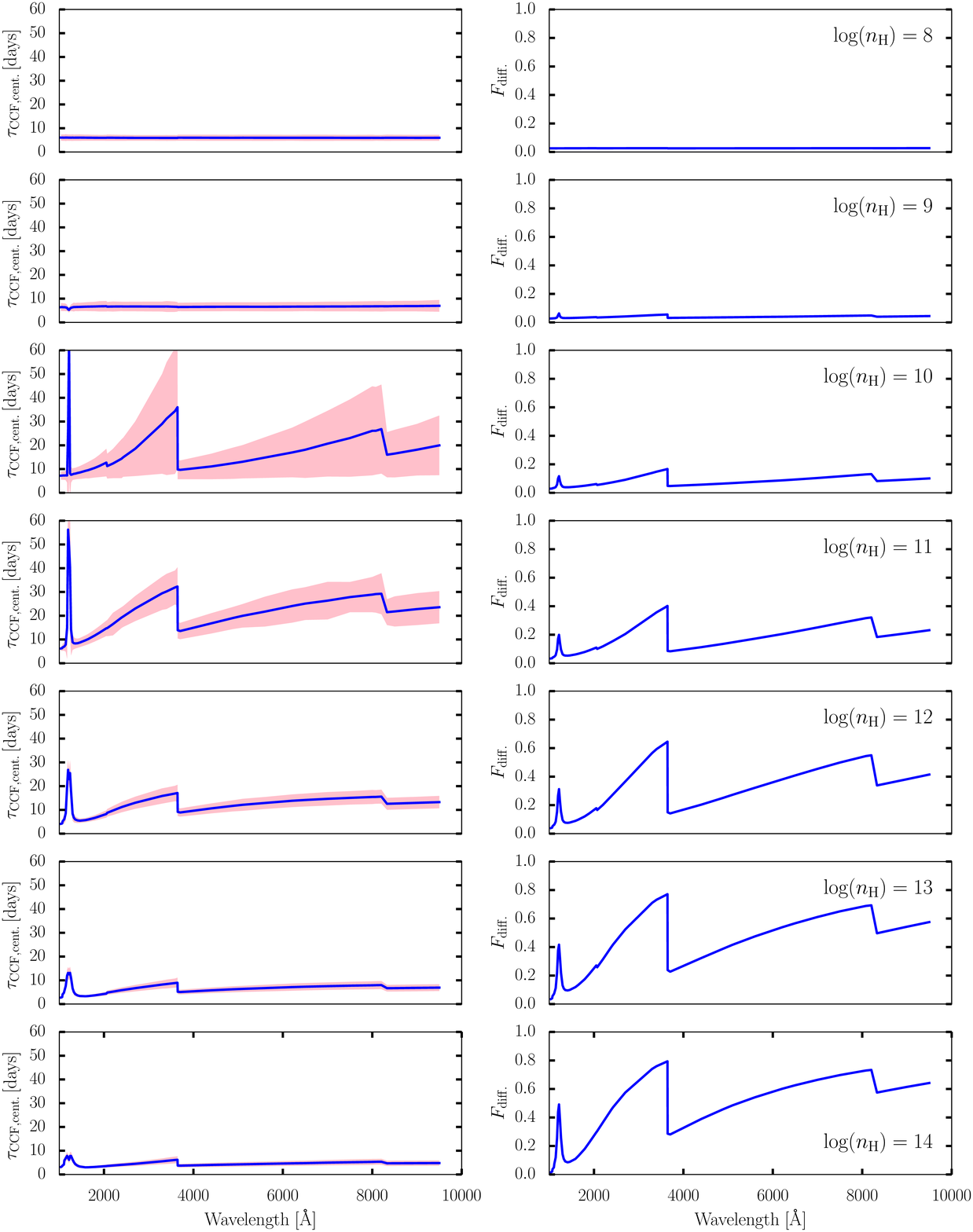}
	\caption{\emph{Left panels:} CCF centroids, in days, for the diffuse continuum component, for $s=0$ models spanning a range of $n_{\rm H}$, $8\le\lognh\le14$. The CCF is determined using the interpolated-CCF method (\S \ref{sec:method_measuring}), for 10000 realizations of a DRW driving continuum (\S \ref{sec:method_driving}). The pink shaded region displays the 1$\sigma$ standard deviation of the CCF centroid distribution at each wavelength. \emph{Right panels:} The diffuse continuum fraction, $F_{\mathrm{diff}}=L_\nu(\mathrm{diff.})/(L_\nu(\mathrm{diff.})+L_\nu(\mathrm{nuc.}))$, for each model.}
	\label{fig:cont_nh_lags}
\end{figure*}

At \lognh$<10$, $F_\mathrm{diff}$ is low, and thus the DC contributes little to the delay signal (Figure \ref{fig:cont_nh_lags}, top panels). As \nh\,increases, the luminosity of the DC approaches and eventually exceeds that of the nuclear continuum, i.e., $F_\mathrm{diff}$ increases with \nh (Figure \ref{fig:cont_nh_lags}, right panels). We note that, for \lognh$\geq10$, a substantial lag will be introduced into the observed continuum, as the product $\tau_\mathrm{CCF,cent}\times F_{\mathrm{diff}}$ becomes non-negligible. While the higher-density models produce DC emission primarily at smaller radii, this effect is balanced to some degree by the increase in $F_\mathrm{diff}$. In particular, for \lognh$>10$, the product $\tau_\mathrm{CCF,cent}\times F_{\mathrm{diff}}$ yields lags 
of between 4--12 days immediately blue-wards of the Balmer and Paschen jumps (assuming full source coverage).

\subsubsection{Dependence of CCF Lags on $\log(U)$ for $s=2$:}\label{sec:CCF_U_range}

\begin{figure*}
	\centering
	\includegraphics[width=0.99\linewidth]{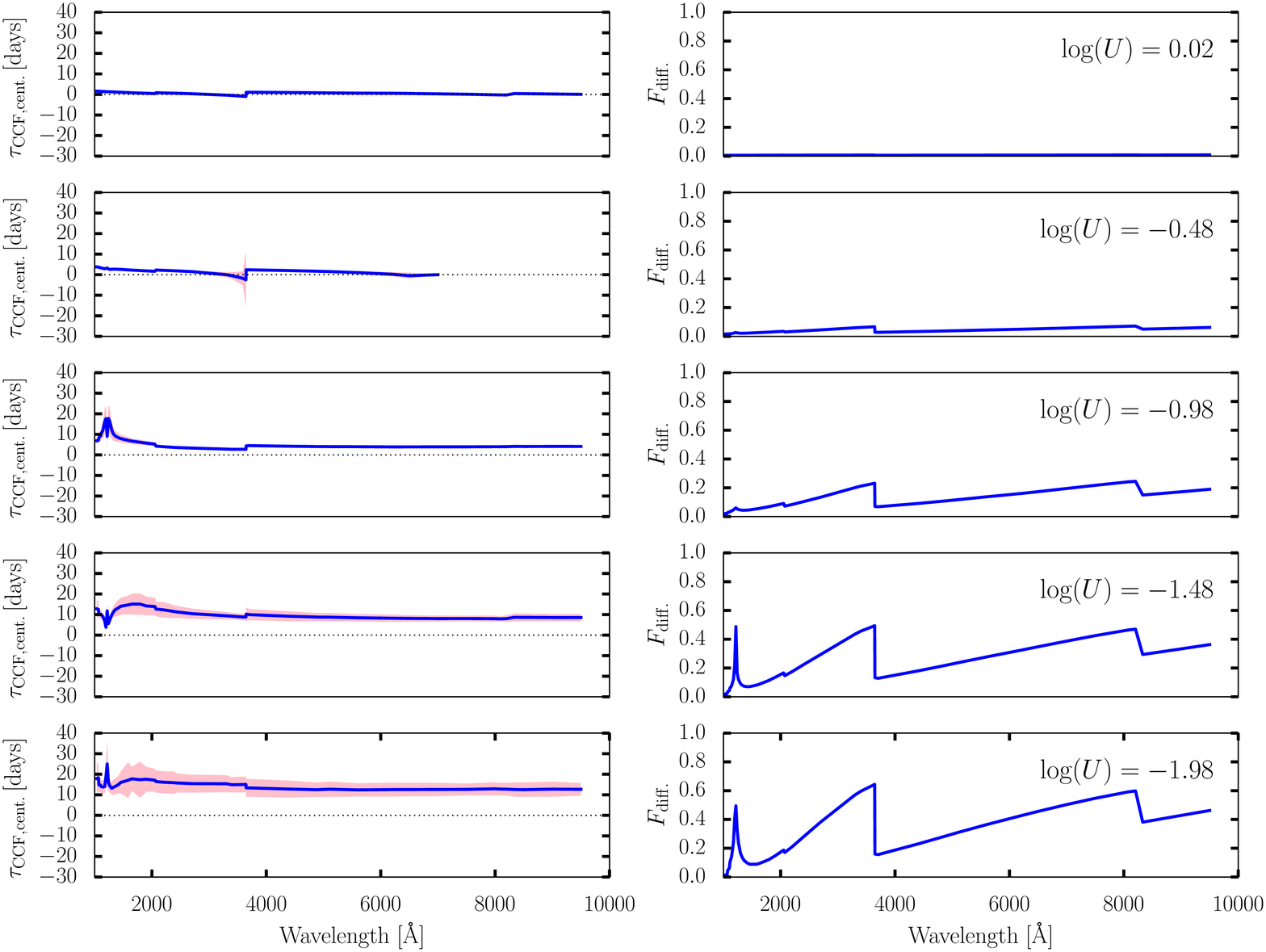}
	\caption{\emph{Left panels:} CCF centroids, in days, for the diffuse continuum component, for $s=2$ models spanning a range in ionization parameter, $0.52\ge \log U \ge-1.98$. The CCF is determined using the interpolated--CCF method (\S \ref{sec:method_measuring}), for 10000 realizations of a DRW driving continuum (\S \ref{sec:method_driving}). The pink shaded region displays the 1$\sigma$ standard deviation of the CCF centroid distribution at each wavelength. For the $\log(U)=-0.48$ model, the DC responds extremely weakly at wavelengths between 6500 \AA--8000 \AA, and the ICCF method fails to find a centroid for almost all DRW continuum realizations. We therefore do not include the $\tau_\mathrm{CCF}$ at wavelengths beyond than 6000 \AA\,for this model. \emph{Right panels:} The diffuse continuum fraction, $F_{\mathrm{diff}}=L_\nu(\mathrm{diff.})/(L_\nu(\mathrm{diff.})+L_\nu(\mathrm{nuc.}))$, for each model.} 
	\label{fig:cont_nh_lags_s2}
\end{figure*}

As noted in \S \ref{sec:results_cont_comparison}, the wavelength dependence of the response function centroids for $s=2$ models is in a sense `reversed' compared to that of the $s=0$ models, and tends to decrease with increasing wavelength. While the strength of the DC component is marginally larger for Model 2, the wavelength dependence of the DC delays is a relatively weak function of wavelength (Figure \ref{fig:cont_nh_lags_s2}, left panels), while $F_{\mathrm{diff}}$ depends strongly on wavelength. The scaled lags therefore will tend to show behaviour resembling that of Model 1, with delays generally increasing with increasing wavelength, but with less significant reductions in delay longward of the Balmer and Paschen jumps.

The DC responds on very short timescales for high ionization parameters $\log(U)\sim0$; however, very little diffuse continuum is produced at these ionization levels. On the density-flux plane ($n_{\rm H}$, $\Phi_{\rm H}$), the diffuse continuum contours are widely spaced for $\log[U]\apprle0$ (roughly the bottom-right regions for the continuum bands shown in Figure \ref{fig:cloudy_DC}) and thus the DC contribution is largely insensitive to $U$ over a broad range in ionization parameter, $-1\apprge\log(U)\apprge-2$. For ionization parameters small enough to produce sufficient BEL luminosity to match observations (i.e., $\log(U)\apprle -0.5$, Figure \ref{fig:line_lumins_s2}), the product $\tau_\mathrm{CCF,cent}\times F_{\mathrm{diff}}$ is $\sim2-8$ days blue-wards of the Balmer and Paschen jumps, assuming full BLR coverage.

\subsubsection{Dependence of the Diffuse Continuum Lags on  $T_{\mathrm{DRW}}$:}

\begin{figure*}
\centering
\includegraphics[width=0.99\linewidth]{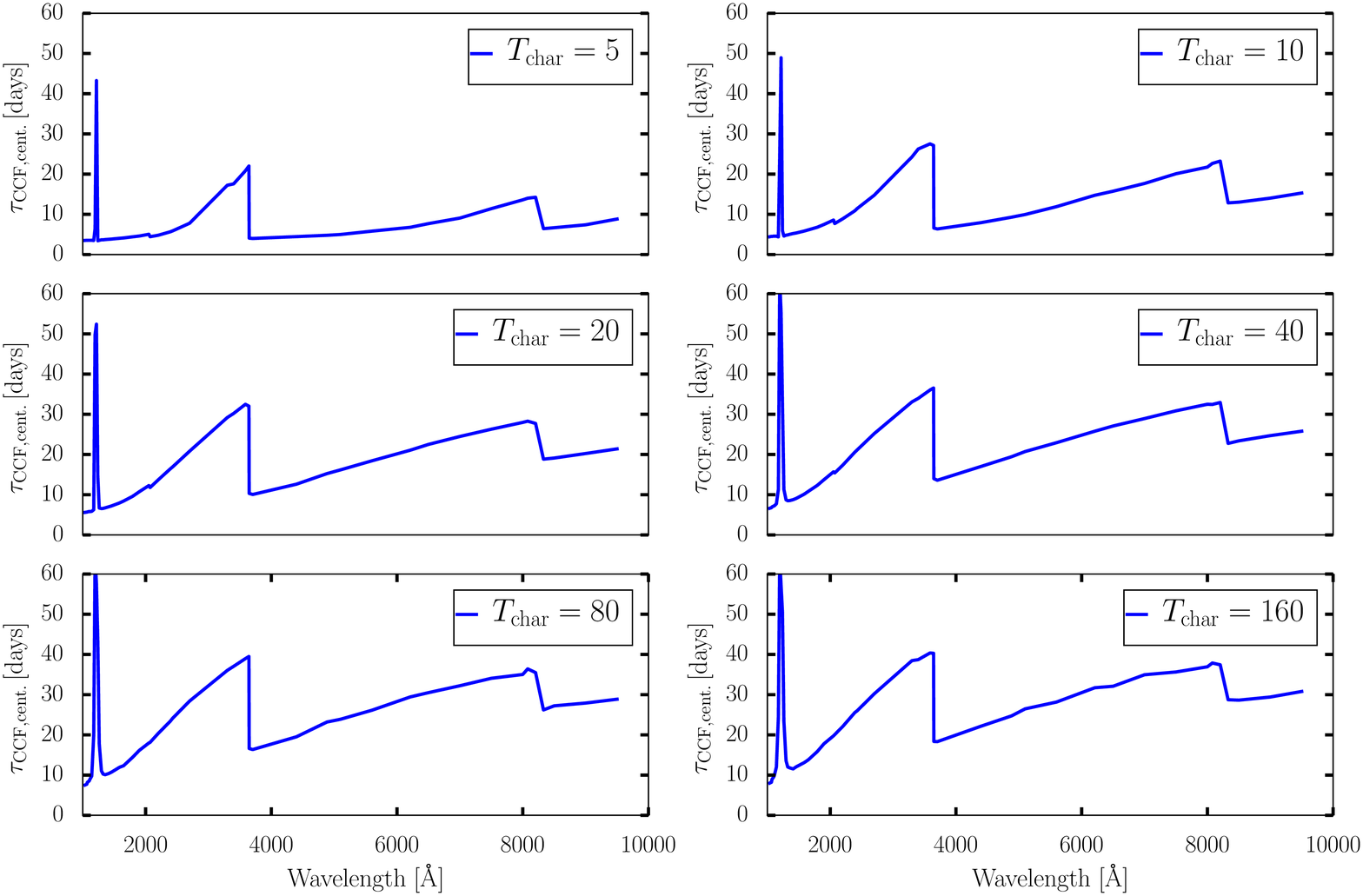}
\caption{Diffuse continuum CCF centroids, in days, for Model 1, but with driving continua spanning a range of damping timescales $T_\mathrm{DRW}$. The CCF is determined using the interpolated-CCF method (\S \ref{sec:method_measuring}), for 10000 realizations of each DRW driving continuum (\S \ref{sec:method_driving}).}
\label{fig:cont_tchar_lags}
\end{figure*}

To test the effect of the driving continuum behavior on the measured DC lags, we repeat our measurements of $\tau_{\mathrm{CCF,cent}}$ for Model 1, applying a range of characteristic DRW timescales, 5 days $\le T_\mathrm{DRW}\le160$ days. We keep the variability amplitude constant at $\sigma_{\mathrm{DRW}}=0.04$. For small values of $T_\mathrm{DRW}$, the DC lags we recover are between 50\%--100\% smaller immediately shortward of the Balmer jump, relative to the $T_\mathrm{DRW}=40$ day realizations (Figure \ref{fig:cont_tchar_lags}). Even so, we find $T_\mathrm{CCF,cent}\sim20$ days for this wavelength region even for $T_\mathrm{DRW}\sim5$ days. Thus, after scaling by the appropriate $F_{\rm diff}$ and source covering fraction, we do expect to observe some contamination of the observed lag spectrum due to DC, even assuming very short variability timescales for the driving nuclear continuum.

%% file: sections/sec_discussion.tex
\section{Discussion}\label{sec:discussion}

\begin{figure*}
\includegraphics[width=0.495\linewidth]{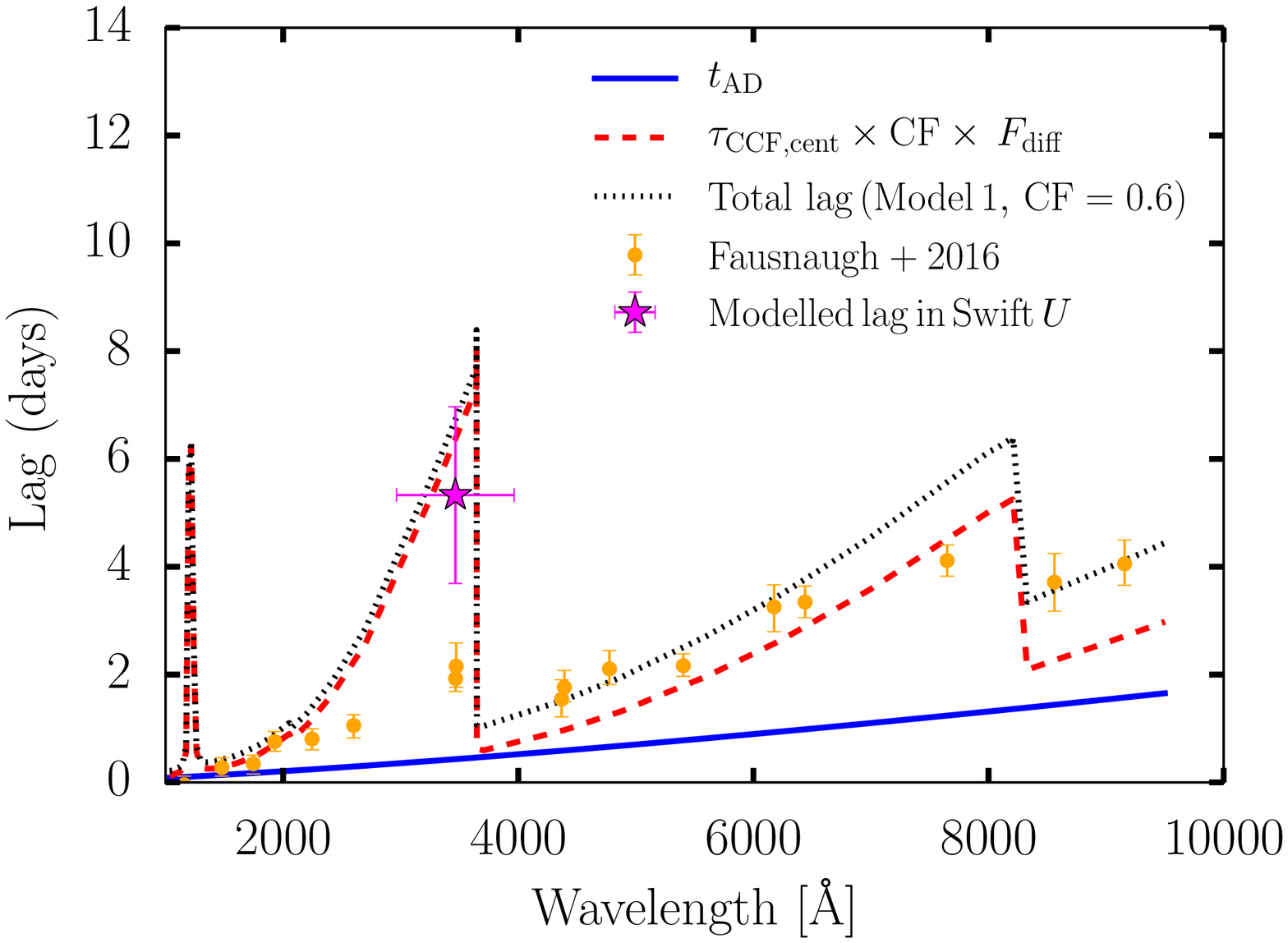}
\includegraphics[width=0.495\linewidth]{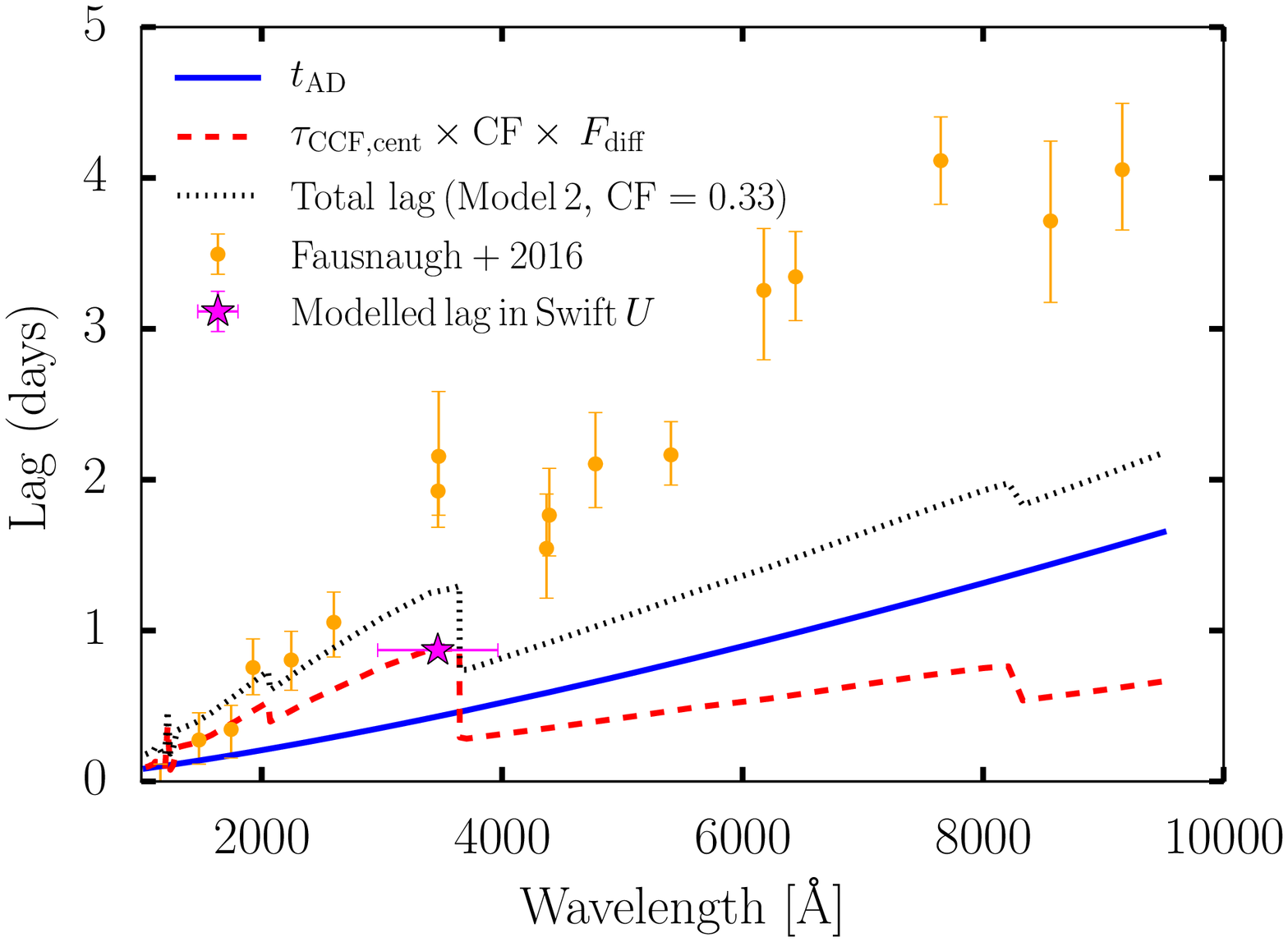}
\caption{An estimate of the total lag spectrum (black dotted curves) comprising the accretion disk lag $\tau_\mathrm{AD}$ (blue solid curves) and the DC lag (red dashed curves), for the minimum source covering fraction allowed by the line luminosities, i.e., 60\% for Model 1 (\emph{left panel}), and 33\% for Model 2 (\emph{right panel}). The accretion disk lag is calculated using Equation \ref{eq:disklags}, assuming an Eddington ratio of 0.1. The total lag is the weighted sum of the two components (Equation \ref{eq:lagsum}). Yellow circles indicate the measured total continuum lags for the 2014 AGN STORM (Peterson, PI) RM campaign of NGC~5548 \citep{Fausnaugh2016}. Clearly, for spherically symmetric pressure law models that generate sufficient BEL luminosity to match observations, the delay induced by DC contamination represents a significant contribution to the total observed delay at a given UV-optical wavelength. To illustrate the effect of measuring inter-band delays using broadband photometry, we estimate the observed lag in the \emph{Swift} UVOT \emph{U} bandpass for our Models 1 and 2 (magenta stars). The uncertainty on the \emph{U} band delay represents the standard deviation of the distribution of ICCF delay centroids obtained for 10000 realizations of the DRW continuum, while the horizontal `error-bar' represents the filter FWHM.}\label{fig:lagsum}
\end{figure*}

Our main finding is that for pressure-law models which roughly reproduce the observed BEL luminosities, a substantial DC component is also present. Contamination of the observed UV--optical continuum by this DC component introduces a delay signature distinct from that arising in the reverberating disk. This additional delay is of order $\sim$few days, as estimated by the product of $F_\mathrm{diff}\times\tau_\mathrm{CCF}$ at the Balmer continuum peak (Figures \ref{fig:cont_nh_lags}, \ref{fig:cont_nh_lags_s2}). The only pressure law models that do not introduce a substantial additional lag signal are those with low gas densities (for $s=0$ models) or high ionization parameters (for $s=2$ models); such models also fail to produce sufficient BEL flux to match the observed emission-line luminosities.

\subsection{The BLR Covering Fraction} 

The lags determined in \S\S \ref{sec:ccf_BEL}--\ref{sec:ccf_DC} assume that the BLR covers the entire continuum source, which is unphysical for AGN where we observe the continuum directly; realistic BLRs must have lower source coverage. The ratios of the model emission-line luminosities to the observed luminosities provide a rough lower limit on the BLR covering fraction allowed by the data for pressure-law density distributions. For Model 1, the \hbeta\,BEL is in any case under-luminous, possibly due to limitations of the photoionization modeling (\S \ref{sec:results_steadystate}). However, the observed \civ\,luminosity constrains the covering fraction (as seen from the continuum source) to be larger than $\sim 60$\% for Model 1. For Model 2, the covering fraction must exceed $\sim 32$\% in order to produce sufficient luminosity in all the BELs. Typical global BLR covering fractions for AGN are $\sim0.4$, as inferred from the statistics of intervening UV absorption lines \citep{Dunn2007}. Thus, the required covering fractions are not unreasonable, given our simple density distributions.

\subsection{Combined DC and Disk Lag Spectrum} 

The DC lag spectra (scaled by $F_\mathrm{diff}$) are shown in Figure \ref{fig:lagsum} (red dashed curves), assuming covering fractions of 60\%\,and 33\%\,for Models 1 and 2, respectively. A DC lag of this magnitude can be differentiated from the standard X-ray disk reprocessing models due to the prominent Balmer and Paschen features in the lag spectrum. \citet{Fausnaugh2016} find an additional lag of $\sim0.6-1.2$ days in their \emph{Swift u}-band data for NGC 5548, in comparison to the $\tau\propto\lambda^{4/3}$ model. In a qualitative sense, this resembles the signature of DC contamination predicted by our models. To facilitate a quantitative comparison with the observed lags, we now make a rough estimate of the lag spectra that are expected for our Models 1 and 2, including the effects of illumination of a temperature-stratified accretion disk by an X-ray corona. For the disk reprocessing lags, we employ the wavelength -- disk size scaling relation presented by \citet{Edelson2017} (their Equation 3):

\begin{equation}\label{eq:disklags}
r=ct_\mathrm{AD}\approx0.09\left(X\frac{\lambda}{1928\mathrm{\AA}}\right)^{4/3}M_8^{2/3}\left(\frac{\dot{m}_\mathrm{Edd}}{\eta}\right)^{1/3}
\end{equation}

\noindent where $\lambda$ denotes the wavelength observed, $M_8$ is the black hole mass in units of $10^8$ $M_{\astrosun}$, $\dot{m}_\mathrm{Edd}$ is the ratio of the accretion rate to the Eddington rate, and $\eta$ is the radiative efficiency. The accretion disk delay $t_{\mathrm{AD}}$ is in units of days. For NGC 5548, $M_8=0.52$ \citep{Bentz2015}; we assume $\eta=0.1$. The scaling factor $X$ encapsulates the mapping from disk surface temperature to effective wavelength at a given accretion disk radius $r$; we assume a flux-weighted radius based on the temperature stratification of the \citet{Shakura1973} $\alpha$-disk model, in which case $X=2.49$ \citep{Fausnaugh2016}. For each wavelength bin, we calculate the weighted sum of $t_{\mathrm{AD}}$ and the DC lags,

\begin{equation}\label{eq:lagsum}
\mathrm{Total\,\,lag}=(\mathrm{CF}\times F_{\mathrm{diff}})\tau_{\mathrm{CCF,cent}}+(1-\mathrm{CF}\times F_{\mathrm{diff}})t_\mathrm{AD},
\end{equation}

\noindent where CF denotes the total covering fraction. We note that this is merely a first-order approximation of the lag that would result from a combined accretion disk and DC lag signal; in particular, the assumption that the observed lag scales linearly with the relative luminosities may not be valid in practice. Nevertheless, this expression reduces to $t_{\mathrm{AD}}$ for small $F_{\mathrm{diff}}$, and to $\tau_\mathrm{CCF,cent}$ for large $F_{\mathrm{diff}}$ and covering fraction, as expected. 

We display the total lag spectra for Models 1 and 2, assuming the minimum covering fractions allowed by the observations, in Figure \ref{fig:lagsum} (black dotted curve). Comparison with the observed delays \citep{Fausnaugh2016} is complicated by the fact that most of these are derived from broadband photometry. Thus, each filter samples a range of delays. To illustrate this effect, we estimate the observed delay expected in the UVOT \emph{U} bandpass, given our Models 1 and 2 with covering fractions of 60\% and 33\%, respectively (Figure \ref{fig:lagsum}, magenta stars). We calculate these delays by making a weighted sum of the DC response functions at each discrete wavelength probed by our photoionization modeling, using weights corresponding to the normalized \emph{Swift} UVOT \emph{U} filter throughput function. This yields an approximate response function for the \emph{U} bandpass, for which we generate simulated \emph{U} band lightcurves using a DRW driving continuum, as per \S \ref{sec:ccf_DC}. We then estimate $F_\mathrm{diff}$ for the \emph{U} band as the weighted average $F_\mathrm{diff}$ for the individual continuum bands, again using weights corresponding to the normalized filter throughput function. We calculate the observed \emph{U} band delay using Equation \ref{eq:lagsum}, including the weighted average $t_\mathrm{AD}$ across the bandpass. For both models, our estimated \emph{U} band delay is somewhat shorter than the monochromatic delay just bluewards of the Balmer break, as expected given that the \emph{U} band also samples the region redwards of the Balmer break, for which the delay induced by the DC is short. We do not include the effects of BLR kinematics in this calculation. Rotational broadening will tend to smooth out any sharp features seen in the delay spectrum (for example, at the Balmer and Paschen jumps), but we expect this effect to be negligible for the \emph{U} band, for which the continuum longward of the Balmer jump contributes only a small fraction of the total light.

The accretion rate dependency of the underlying disk delay spectrum introduces an additional uncertainty when comparing to the observed lags. We assume $\dot{M}_\mathrm{Edd}=0.1$ for the disk spectrum displayed in Figure \ref{fig:lagsum}; higher Eddington ratios (or lower accretion efficiencies) produce steeper accretion disk lag spectra. \citet{Fausnaugh2016} argue that $\dot{M}_\mathrm{Edd}$ is unlikely to be much higher than 0.1 for NGC 5548 in 2014, unless the disk is seen very close to edge-on. Given these uncertainties, the simplicity of our pressure-law models, and choice of BLR geometry, we do not expect an exact correspondence between the predicted and observed delay spectra. In particular, Model 1 predicts total (disk plus DC) delays that roughly match the observed delays in the far-UV and the optical regimes, while the model lags near the Balmer continuum feature exceed the measured values. Model 2 matches the observed delays only in the far-UV. However, a significant delay contribution from DC emission, with a similar gross wavelength dependence, is likely ubiquitous for pressure law models capable of emitting strong broad-line flux (\S \ref{sec:ccf_DC}). These results strongly suggest that DC contamination is responsible for a substantial fraction of the observed delays \emph{across the entire UV-optical spectral region}, and not just at the Balmer continuum feature. Thus, DC contamination must be taken into account when interpreting the larger than expected accretion disk sizes inferred from continuum reverberation mapping (\S \ref{sec:intro_diskreverb}). Indeed, for the nearby Seyfert 1 galaxy NGC 4593, \citet{Cackett2017} find strong evidence for the DC Balmer continuum feature in their inter-band spectroscopic (\emph{HST}) and \emph{Swift} delay spectrum. Given the strength of this feature, they suggest that there may be a significant DC contamination of the delay spectrum at all UV-optical-NIR wavelengths. That is not to say that DC contamination is the \emph{only} significant cause of the observed long delays. In particular, scattering effects in the accretion disk atmosphere may influence the observed delays and their wavelength dependence, producing increased inter-band delays for the standard disk geometry \citep{Hall2018}. Non-standard accretion disks are also a distinct possibility. For example, \citet{Starkey2017} suggest that the observed delays for NGC 5548 may be due to a tilted inner-disk geometry \citep{Nealon2015}.

We emphasize that these single component pressure-law models are intended to grossly match the observed emission-line luminosities and not their detailed temporal behavior. \citet{Kaspi1999} study the temporal behavior of $s=~2$ pressure law BLR models. Their models reproduce the observed steady-state luminosities and typical delays for the strongest UV BEL (excluding H$\beta$), but do not reproduce the observed BEL lightcurves for NGC 5548 in detail. Thus, while pressure laws capture the gross properties of real BLRs, additional model complexity and/or deviations from spherical symmetry are required to reproduce all salient features. For this reason we do not attempt to `fit' our models (e.g., varying the normalizations of \nh\,and \ncol\,along with the covering factor) to the observed inter-band continuum delays.

\subsection{Atypically Small BEL Lags for NGC 5548} The BEL lags measured during the 2014 campaign are unusually small, given the continuum luminosity at the time \citep{Pei2017}. Our models cannot simultaneously produce sufficient BEL luminosity to match observations, and produce \hbeta\,lags of order $\sim5$ days. One possible explanation for the short observed BEL lags is that the nuclear continuum lacks variational power on sufficiently long timescales during this campaign; this may decrease the measured BEL lags \citep[e.g.,][]{Goad2014}. However, we do not find that adopting small values of $T_\mathrm{DRW}$ reduces the \hbeta\,lags produced by our model sufficiently to explain the lags observed in 2014. 

In early 2000 NGC~5548 went into an historic low-state during which the strong broad UV emission-lines, most notably C~{\sc iv} disappeared. We speculate that during this time the nuclear region became enshrouded with gas and dust. Large dust grains are robust to destruction by UV photons. Thus when climbing out of this low luminosity state, the destruction of UV photons on grains will reduce the strength of the emission-lines relative to what one would normally expect for the same continuum luminosity. The reduction in the strength of the C~{\sc iv} emission-line core relative to the emission-line wings in NGC~5548 during 2014 c.f. 1993, when the continuum luminosity was of similar strength, is consistent with this expectation.

Our models suggest that the diffuse continuum contamination induces some additional delay in the observed far-UV continuum. We note that \citet{Pei2017} determine the \hbeta\,lags relative to the continuum lightcurves at  1150 \AA\,and at 1367 \AA. At these wavelengths the additional lag induced by the diffuse continuum is less than 0.5 day for Model 2. For Model 1, the H\textsc{I} Rayleigh scattering feature in the DC produces significant additional lag in a narrow peak near these continuum bands (Figure \ref{fig:lagsum}). While we do not include kinematics in our models, this feature would be rotationally broadened in a real BLR. Even making the extreme assumption that the Rayleigh feature causes a $\sim2$-day shift in the adopted `zero lag' bandpass relative to the true ionizing continuum lightcurve, it is still insufficient to explain the measured $\sim5$-day lags in relation to the expected values at this continuum luminosity.

%% file: tables/line_lags_table.tex
\begin{deluxetable}{lccr}
	\tabletypesize{\small}
	\tablecaption{Observed Broad-Line Properties\label{tab:observedlags}}
	\tablewidth{0pt}
	\tablecaption{Observed \lline\,and line lags for NGC 5548\label{tab:observedlags}}
	\tablehead{
	\colhead{Line} & \colhead{$\log[\lline]$} & \colhead{Lag} & \colhead{Reference}\\
	\colhead{} & \colhead{} & \colhead{[days]} & \colhead{} \\
	\colhead{(1)} & \colhead{(2)} & \colhead{(3)} & \colhead{(4)}
	}
	\startdata
	\lya & 42.4 & $6.2^{+0.3}_{-0.3}$ & de Rosa+ \\
	\civ & 42.6 & $5.3^{+0.4}_{-0.5}$ & de Rosa+ \\
	\hbeta & 41.7 & $6.14^{+0.74}_{-0.98}$ & Pei+ \\
	\heuv & 41.7 & $2.5_{-0.3}^{+0.3}$ & de Rosa+ \\
	\heop & 40.7 & $2.46_{-0.25}^{+0.49}$ & Pei+ \\
	\enddata
	\tablecomments{(1) Broad emission line name. (2) Log of the line luminosity in units of erg s$^{-1}$, corrected for Galactic reddening and with narrow-line components subtracted; we use the decompositions presented by \citet{Korista2000} for the UV lines, and those of \citet{Peterson1991} for the Balmer lines. (3) Observed CCF centroid for the 2014 RM campaign, as measured using the ICCF method. (4) References: \citet{DeRosa2015}, \citet{Pei2017}. }
\end{deluxetable}

\clearpage

\begin{deluxetable}{lcccccccc}
	\tabletypesize{\small}
	\tablecaption{Steady-state Models\label{tab:modellags}}
	\tablewidth{0pt}
	\tablehead{
		\colhead{Line} & \colhead{$\log[\lline]$} & \colhead{$r_{\epsilon}$} & \colhead{$r_{\eta}$} & \colhead{$\tau_{\epsilon}$} & \colhead{$\tau_{\eta}$} & \colhead{$\tau_{\mathrm{CCF,peak}}$} & \colhead{$\tau_{\mathrm{CCF,cent}}$}  \\
		\colhead{} & \colhead{} & \colhead{[lightdays]} & \colhead{[lightdays]} & \colhead{[days]} & \colhead{[days]} & \colhead{[days]} & \colhead{[days]} \\
		\colhead{(1)} & \colhead{(2)} & \colhead{(3)} & \colhead{(4)} & \colhead{(5)} & \colhead{(6)} & \colhead{(7)} & \colhead{(8)}  \\
	}
	\startdata
	\multicolumn{8}{c}{\textbf{Model 1:} $s=0$, \lognh=10.75, \logncol=22.5} \\
	\multicolumn{8}{c}{} \\
	\lya & 42.8 & 36.2 & 46.5 & 47.7 & 61.6 & 35.7$\pm6.6$ & 39.4$\pm8.6$\\ 
	\civ & 42.7 & 20.9 & 35.0 & 24.6 & 40.9 & 29.7$\pm3.5$ & 33.6$\pm4.8$ \\ 
	\halpha & 41.9 & 58.9 & 70.2 & 68.3 & 81.4 & 41.2$\pm25.6$ & 44.9$\pm24.4$ \\ 
	\hbeta & 41.3 & 46.3 & 52.8 & 56.0 & 63.6 & 34.1$\pm4.7$ & 33.3$\pm7.2$ \\ 
	\heop & 41.1 & 21.4 & 30.1 & 22.3 & 31.2 & 17.6$\pm1.2$ & 20.5$\pm2.7$ \\ 
	\mgii & 42.0 & 74.4 & 101.3 & 92.7 & 126.8 & 76.8$\pm65.7$ & 77.4$\pm65.1$ \\ 
	\\
	\hline
	\\
	\multicolumn{8}{c}{\textbf{Model 2:} $s=2$, $\log(U)=-1.23$,  \logncol($r_{20}$)=22.5, \lognh($r_{20}$)=10.75} \\
	\multicolumn{8}{c}{} \\
	\lya & 43.1 & 50.5 & Cx & 50.5 & Cx & $28.6\pm3.6$ & $29.8\pm3.8$ \\ 
	\civ & 43.1 & 51.0 & Cx & 55.5 & Cx & $32.9\pm10.7$ & $36.1\pm10.6$ \\ 
	\halpha & 42.0 & 53.7 & Cx & 59.0 & Cx & $21.7\pm10.2$ & $22.4\pm10.6$ \\ 
	\hbeta & 41.4 & 53.7 & 32.4 & 57.9 & 34.3 & $23.3\pm10.1$ & $27.6\pm12.8$ \\ 
	\heuv & 42.2 & 56.7 & 39.4 & 63.9 & 47.8 & $36.0\pm19.8$ & $40.4\pm19.1$ \\ 
    \heop & 41.3 & 54.3 & 37.3 & 54.6 & 37.8 & $26.0\pm8.1$ & $30.6\pm9.7$ \\
	\mgii & 41.2 & 22.2 & 35.9 & 24.0 & 37.3 & $-25.2\pm89.8$ & $-25.2\pm89.6$ \\ 
	\enddata
	\tablecomments{The responsivity of the Mg~{\sc ii} BEL in our models is very low; thus, the CCF lags measured for that line are highly uncertain, and would not be accurately measurable in a real observing situation. (1) Broad emission line name. (2) Log of the line luminosity in units of erg s$^{-1}$, for a BLR covering $4\pi$ steradian of the continuum source. (3) Emissivity-weighted BLR radius, in units of lightdays. (4) Responsivity-weighted BLR radius, in units of lightdays. `Cx' denotes lines which have negative repsonsivity for a non-negligible fraction of the radial extent of the BLR; the integrated line luminosities may respond positively or negatively to continuum changes, and their response functions do not have a well-defined responsivity-weighted centroid. (5) Emissivity-weighted transfer function centroid, in days, allowing for anisotropic emission. (6) Responsivity-weighted transfer function centroid, allowing for anisotropic emission. `Cx' denotes lines which have negative repsonsivity for a non-negligible fraction of the radial extent of the BLR. (7) Mean CCF peak delay time for the continuum versus line response, using response functions that include the effects of local anisotropy and responsivity. The BEL are driven using a Damped Random Walk continuum with a 500-day baseline (\S \ref{sec:method_driving}). The measurement is repeated for 10000 realizations of the DRW continuum; for each realization, the CCF is obtained using the methodology of \citet{White1994} (\S \ref{sec:method_measuring}). The uncertainty shown is the standard deviation of the peak values. (7) Mean CCF centroid, and its standard deviation, for 10000 realizations of the DRW continuum. }
\end{deluxetable}